\documentclass[review]{elsarticle}
\pdfoutput=1
\usepackage{lineno,hyperref}
\usepackage{color}
\usepackage{amsmath, amssymb} 
\usepackage{rotating}
\modulolinenumbers[5]

\journal{Journal of \LaTeX\ Templates}

\definecolor{cinnamon}{rgb}{0.82, 0.41, 0.12}

\begin{document}

\begin{frontmatter}

\title{
Numerical study of the sedimentation of spheroidal particles}



%
%

\author[address1]{Mehdi Niazi Ardekani\corref{mycorrespondingauthor}}
\cortext[mycorrespondingauthor]{Corresponding author: mehd@mech.kth.se}

\author[address2]{Pedro Costa}

\author[address2]{Wim Paul Breugem}

\author[address1]{Luca Brandt}

\address[address1]{Linn$\acute{\textrm{e}}$ Flow Centre and SeRC (Swedish e-Science Research Centre),KTH Mechanics, S-100 44 Stockholm, Sweden}
\address[address2]{Laboratory for Aero $\&$ Hydrodynamics, TU-Delft, Delft, The Netherlands}

\begin{abstract}

The gravity-driven motion of rigid particles in a viscous fluid is relevant in many natural and industrial processes, yet this has mainly been investigated for spherical particles.  
We therefore consider the sedimentation of non-spherical (spheroidal) isolated and particle pairs in a viscous fluid via numerical simulations
using the Immersed Boundary Method.
The simulations performed here show that the critical Galileo number for the onset of secondary motions decreases as the spheroid aspect ratio departs from $1$. Above this critical threshold, oblate particles perform a zigzagging motion whereas prolate particles rotate around the vertical axis while having their broad side facing the falling direction. Instabilities of the vortices in the wake follow when farther increasing the Galileo number.
We also study the drafting-kissing-tumbling associated with the settling of particle pairs.
We find that the interaction time increases significantly for non-spherical particles and, more interestingly, spheroidal particles are attracted from larger lateral displacements. This has important implications for the estimation of collision kernels and can result in increasing clustering in suspensions of sedimenting spheroids.

\end{abstract}


\end{frontmatter}

\linenumbers

\section{Introduction}
\label{introduction}
The presence of solid rigid particles in a fluid alters the global transport and rheological properties of the mixture in complex and sometimes unpredictable ways.
 In recent years many efforts have therefore been devoted to develop numerical tools able to fully resolve the fluid-particle and particle-particle interactions and to allow us to investigate rigid particles immersed in an incompressible viscous fluid, see among others \cite{Yeo2010,Loisel2013,Lashgari2014,Picano2015,Lashgari2016,Yin2007}.  
Most of these previous studies consider spherical particles and indeed simulations of suspensions of non-spherical particles are relatively few despite the fact that these are more frequently found. Here we develop a numerical algorithm for spheroidal particles and use it to investigate the sedimentation of isolated and pairs of non-spherical particles. A spheroid, is an ellipsoid with two equal semi-diameters, existing in two shapes of prolate and oblate. For a prolate spheroid the symmetric axis is aligned with the major semi-diameter while for an oblate spheroid this axis is aligned with the minor semi-diameter of the spheroid.
     
\subsection{Sedimentation of isolated spheroids}
The gravity-driven motion of heavy particles in a viscous fluid has been a matter of interest among physicists and engineers for decades; it is, however, only recently that the progress in development of computational and experimental techniques has led to a better understanding of the physics behind it. 

The simple case of an isolated sphere, fixed in an uniform unbounded flow, has been considered first (see e.g.\ Johnson \& Patel \cite{Johnson1999}; Ghidersa \& Du¨ek \cite{Ghidersa2000}; Schouveiler \& Provensal \cite{Schouveiler2002}; Bouchet et al. \cite{Bouchet2006} ). These studies showed different wake structures in different Reynolds number regimes. Allowing the particle to move freely under the effect of gravity introduces new degrees of freedom as path instability can also occur (Jenny et al. \cite{Jenny2004}; Uhlmann \& Du\v{s}ek  \cite{Uhlmann2014}). Indeed, these authors reported first the appearance of an oblique wake and then vortex shedding and unsteady motions when increasing the settling speed.

Path and wake instability becomes more complicated in the case of a non-spherical particle as the orientation plays a role in the dynamics of the problem. 
Feng et al. (1994) \cite{Feng1994} performed two-dimensional numerical simulations of settling elliptic particles and revealed that, in stable conditions, an elliptic particle always falls with its long axis perpendicular to the gravity direction. 
For three-dimensional oblate particles, the symmetry axis is  also aligned with the falling direction in the steady motion at low settling speeds. Increasing the particle size or density, the system becomes unstable and disc-like particles are observed to oscillate horizontally.
The ensuing wake instability depends on the aspect ratio and the vortices in the wake are modified as soon as the particle symmetry axis has an angle with respect to the velocity direction, see the review in \cite{Ern2012}. 
The numerical  simulations of Mougin \& Magnaudet (2001) \cite{Mougin2001} and Magnaudet \& Mougin (2007) \cite{Magnaudet2007}, considering freely rising and fixed bubbles, revealed that the path instability is closely related to the wake instability. These authors reported a planar zigzagging motion, following a rectilinear path for a frozen oblate bubble with aspect ratio of $\mathcal{AR}=1/2.5$ (polar over equatorial radius), in agreement with the experimental observations of Ellingsen \& Risso (2001) \cite{Ellingsen2001}. Unlike the case of sedimenting discs and oblate  particles, little is known about prolate particles with finite aspect ratios. 
This study aims therefore to fill this gap by investigating the sedimentation of isolated  prolate and oblate particles in a viscous fluid and comparing the onset and characteristics of the unsteady motion as function of the Galileo number. (The latter quantifies the importance of buoyancy with respect to viscous forces). We find that oblate and prolate particles exhibit different secondary transversal motions with different vortical structures in the unsteady wake. The influence of the aspect ratio on the onset of these unsteady secondary motions is also discussed. 

\subsection{Pair interaction between settling spheroids} 
Joseph et al. (1987) \cite{Joseph1987} and Fortes et al. (1987) \cite {Fortes1987} report a peculiar particle pair interaction for two equal spherical particles.
This is the so-called drafting-kissing-tumbling (DKT) phenomenon: it is associated with wake effects and torques acting on two settling particles at sufficiently close vertical distance from each other \cite {Feng1994}. The trailing particle is attracted into the wake of the leading particle, forming a tall body which is unstable and turns. As a consequence, the trailing particle tumbles and falls ahead of that initially leading \cite{Prosperetti2007}, see the visualization in figure~\ref{fig:16}. This peculiar interaction has been studied by many both experimentally \cite{Feng1994,Fortes1987} and numerically \cite{Patankar2000,Glowinski2001,Breugem2012}. Fornari et al. (2016) \cite{Fornari2016} performed direct numerical simulations of a suspension of slightly-buoyant spherical particles in a quiescent and turbulent environment. They show that the DKT phenomenon induces a highly intermittent particle velocity distribution which counteracts the reduction of the mean settling velocity caused by the hindered settling effect in a quiescent flow.
 Pair interactions between settling non-spherical particles has not been studied before, although this is key to understand
 the collective dynamics of sedimenting non-spherical particles. We therefore examine the DKT of non-spherical particles in the second part of this work. Results of this study reveals that non-spherical particles are attracted towards each other from larger horizontal particle separations  and experience a significant increase in the duration of the kissing phase. 

\subsection{Immersed boundary method (IBM) for non-spherical particles}
Among the different approaches proposed to perform interface-resolved direct numerical simulations (DNS) of particle-laden flows, such as \textit{force coupling} \cite{Lomholt2003}, \textit{front tracking} \cite{Unverdi1992}, \textit{Physalis} \cite{Sierakowski2016,Zhang2005} and different algorithms based on the \textit{lattice Boltzmann} method for the fluid phase \cite{Ladd1994,Ladd19942}, we resort to the  \textit{Immersed boundary} method(IBM), which has gained popularity in recent years due to the possibility of using efficient computational methods for solving the Navier-Stokes equations on a Cartesian grid. The IBM was first developed by Peskin (1972) \cite{Peskin1972} and  numerous modifications and improvement have been suggested since then, see  \citep{Mittal2005}. Uhlmann (2005) \cite{Uhlmann2005} developed a computationally efficient IBM to fully resolve particle-laden flows. Breugem (2012) \cite{Breugem2012} improved this method by applying a multi-direct forcing scheme \cite{Luo2007} to better approximate the no-slip/no-penetration (ns/np) boundary condition on the surface of the particles and by introducing a slight retraction of the grid points on the surface towards the interior. The numerical stability of the code for mass density ratios (particle over fluid density ratio) near unity is also improved by a direct account of the inertia of the fluid contained within the particles \cite{Kempe2012JCP}. In this study the IBM method of Breugem (2012) \cite{Breugem2012}  is extended to ellipsoidal particles. A lubrication correction force based on the asymptotic solution of Jeffrey (1982) \cite{Jeffrey1982} is introduced when the gap width between the particles is less than a grid cell and the collision and friction model proposed by Costa et al. \cite{Costa2015} employed to calculate the normal and tangential collision forces. To this end, we approximate the interacting objects by two spheres with same mass and radius corresponding to the local curvature at the point of contact.

This paper is organised as follows. We discuss the governing equations and the details of the numerical method in section \ref{Governing equations and numerical method}, followed by a validation study in section \ref{Validation}. The results of the simulations are discussed in Section \ref{Results}, first considering isolated particles and then pair interactions. Main conclusions and final remarks are presented in Section \ref{Final remarks}.


\section{Governing equations and numerical method}
\label{Governing equations and numerical method}
\subsection{Governing equations}
The motion of rigid ellipsoidal particles is described by the Newton-Euler equations
\begin{subequations}
\label{eq:NewtonEuler}  
\begin{align}
\rho_p V_p \frac{ \mathrm{d} \textbf{U}_{p}}{\mathrm{d} t} = \textbf{F}_p \, , \\ 
\frac{ \mathrm{d} \left( \textbf{I}_p \, \pmb{\omega}_{p} \right) }{\mathrm{d} t} = \textbf{T}_p \, , 
\end{align}
\end{subequations}
where $\rho_p$, $V_p$ and $\textbf{I}_p$ are the mass density, volume and moment-of-inertia tensor of a particle. 
$\textbf{U}_p$ and  $\pmb{\omega}_{p}$ are the translational and  the angular velocity of the particle. The moment of Inertia ${\bf I}_p$ of a non-spherical particle 
changes with the particle orientation and is therefore kept in the time derivative. $\textbf{F}_p$ and $\textbf{T}_p$ are the net force and momentum resulting from hydrodynamic stresses on the particle surface, gravity and particle-particle interactions. These can be written as
\small
\begin{subequations}
\label{eq:Forces}  
\begin{align}
\textbf{F}_p =& \oint_{\partial {V}_p}  \left[ -p\textbf{I} + \mu_f \left( \nabla \textbf{u} + \nabla \textbf{u}^T \right) \right] \cdot  \textbf{n} \mathrm{d}A - V_p \nabla p_e + \left( \rho_p - \rho_f \right)V_p\textbf{g} + \textbf{F}_c , \, \\  
\textbf{T}_p =& \oint_{\partial {V}_p} \textbf{r} \times \left( \left[ -p\textbf{I} + \mu_f \left( \nabla \textbf{u} + \nabla \textbf{u}^T \right) \right] \cdot  \textbf{n} \right) \mathrm{d}A + \textbf{T}_c  \, , 
\end{align}   
\end{subequations}
\normalsize
where $\rho_f$ is the density of the fluid, $\textbf{g}$ the gravitational acceleration and $\textbf{r}$ indicates the distance from the surface to the center of the particle. 
The stress tensor is integrated over the surface of the particle, denoted $\partial {V}_p$. The out-ward pointing unit normal vector at the surface is denoted by $\textbf{n}$ and  the unit tensor by $\textbf{I}$. The terms  $- \rho_f V_p\textbf{g}$ and $V_p \nabla p_e$ account for the forces caused by the hydrostatic pressure and a constant pressure gradient $\nabla p_e$ or any external force that might be imposed to drive the flow \cite{Breugem2012}. The force and torque resulting from particle-particle (particle-wall) collisions are indicated by  $\textbf{F}_c$ and $\textbf{T}_c$. The fluid velocity $\textbf{u}$ and the  stress tensor $-p\textbf{I} + \mu_f \left( \nabla \textbf{u} + \nabla \textbf{u}^T \right)$  appearing in the Newton-Euler equations are obtained from solving the Navier-Stokes and continuity equations

\begin{subequations}
\label{eq:NS}  
\begin{align}
\rho_f (\frac{\partial \textbf{u}}{\partial t} + \nabla\cdot \textbf{u} \textbf{u}) =& -\nabla p_e -\nabla p + \mu_{f} \nabla^2 \textbf{u} + \rho_f \textbf{f} \, , \\ 
\nabla \cdot \textbf{u} =& \, 0 \, .
\end{align}
\end{subequations}

The extra term on the right hand side of the Navier-Stokes equations indicates the IBM force, active in the immediate vicinity of a particle surface to impose indirectly the no-slip / no-penetration (ns/np) boundary condition. In other words, a force distribution $\textbf{f}$ is imposed on the flow such that the fluid velocity at the surface  
 is equal to the particle surface velocity ($\textbf{U}_p + \pmb{\omega}_p \times \textbf{r}$). Eqs. (\ref{eq:NewtonEuler}) and (\ref{eq:NS}) are coupled through $\textbf{f}$ and they are solved together.
 
In the case of spheroidal particles sedimenting in a still fluid examined here, the two non-dimensional parameters defining the problem are the
Galileo number and the spheroid aspect ratio. The former is the ratio between gravitational and viscous forces, defined as
\begin{equation}
Ga \equiv \sqrt{\frac{|\rho_p/\rho_f - 1| g D_{eq}^3}{\nu^2} }, 
\label{eq:LuMo}  
\end{equation}
where $\rho_p/\rho_f$  is the particle to fluid density ratio, $D_{eq}$ the diameter of a sphere with the same volume as the ellipsoidal particle. The polar (symmetric semi-axis) and the equatorial radius, $a$ and $b$, respectively define the spheroid 
aspect ratio, $\mathcal{AR}=a/b$ (see figure~\ref{fig:1}).  The results will be  presented in terms of the Reynolds number $Re= U L/\nu$, with $L$ and $U$ the characteristic length and velocity scale (typically the particle equivalent diameter $D_{eq}$ and terminal velocity) and $\nu$ the kinematic viscosity.

\subsection{Numerical method}

\subsubsection{Grid geometry}

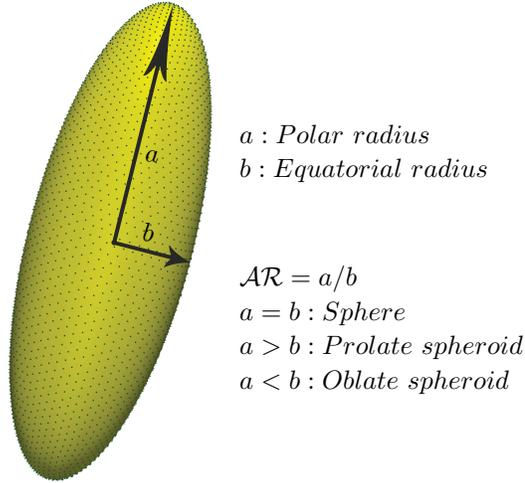
\begin{figure}[t]
\centering
\includegraphics[width=0.28\linewidth]{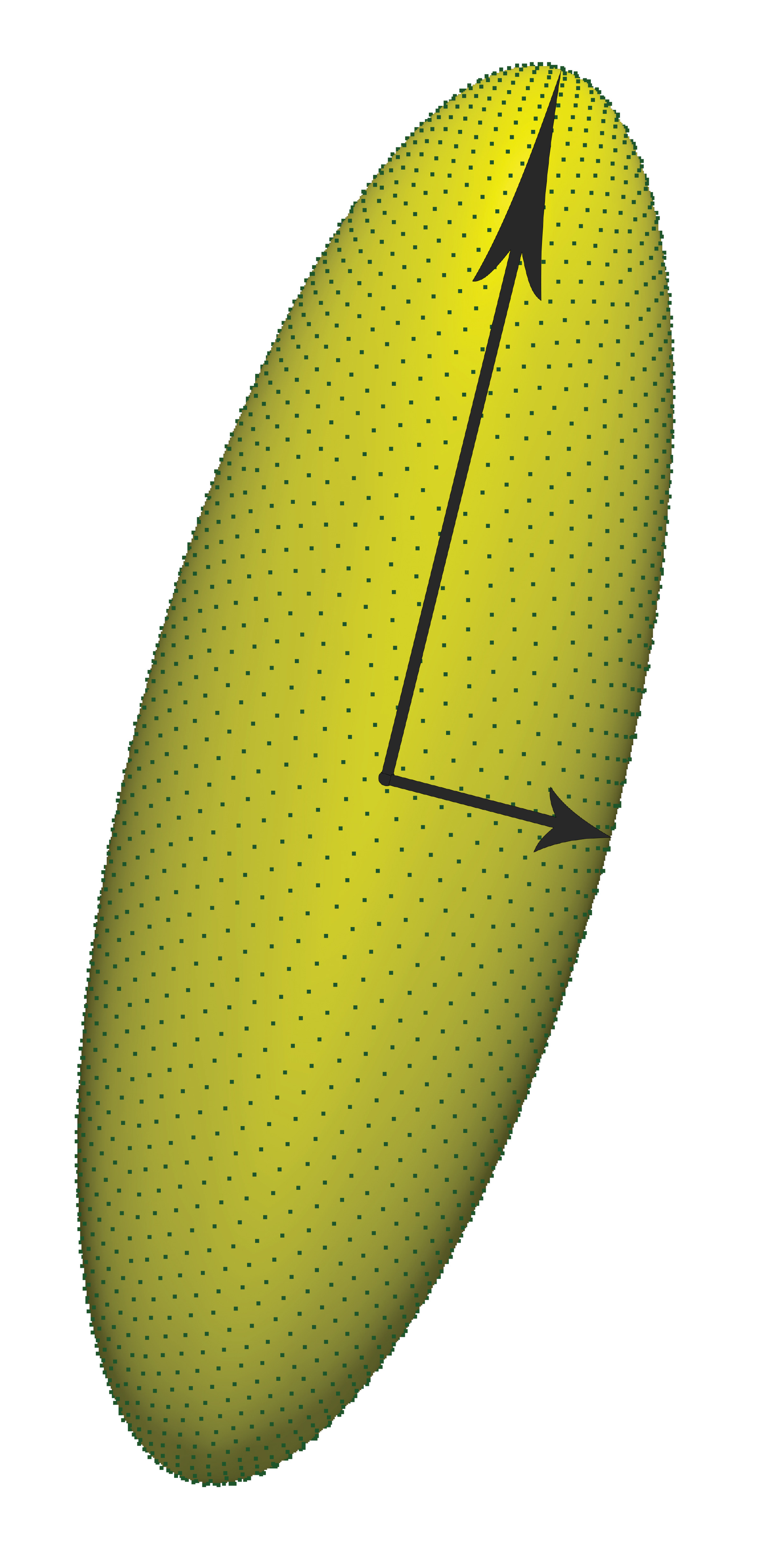}
\put(-36,131){\rotatebox{-5}{$a$}}
\put(-37,101){\rotatebox{-5}{$b$}}
\put(0,138){{$a : Polar \,\, radius$}}
\put(0,125){{$b : Equatorial \,\, radius$}}
\put(0,71){{$a = b: Sphere$}}
\put(0,58){{$a > b: Prolate \,\, spheroid$}}
\put(0,45){{$a < b: Oblate \,\, spheroid$}}
\put(0,84){{$\mathcal{AR}= a/b$}}
\vspace{-10pt}
\caption{\label{fig:1} 
Distribution of the Lagrangian grid points over the surface of a spheroidal particle.}
\end{figure}

A uniform ($ \Delta x = \Delta y = \Delta z$), staggered, Cartesian Eulerian grid is used for the flow and a Lagrangian grid is employed to represent the particles as shown in figure.~\ref{fig:1}. Uniform distribution of the Lagrangian points over the surface of the particles is obtained by an additional simulation of point charges moving on the 
surface of the spheroid we wish to simulate.
Driven by electrical forces, these charges reach an uniform equilibrium distribution after sufficiently long time. The number of Lagrangian points is defined such that the volume $\Delta V_l$ of the Lagrangian grid cells is as close as possible to the volume of the Eulerian grid cells, $ {\Delta x}^3$. Assuming that the Lagrangian grid corresponds to a thin shell of thickness $\Delta x$ the number of Lagrangian points can be calculated from
\small
\begin{equation}
N_l = \left[ \frac{\left( a + \Delta x /2 \right) {\left( b + \Delta x /2 \right)}^2  - \left( a - \Delta x /2 \right) {\left( b - \Delta x /2 \right)}^2 }{{ 3 \Delta x}^3 / \left(4 \pi \right)} \right]
\label{eq:eq:Nl} 
\end{equation}
\normalsize
where a, b are the polar (symmetric semi-axis) and the equatorial radii of the spheroidal particle .          

\subsubsection{Flow field solution}

The same \textit{pressure-correction} scheme used in Breugem (2012) \cite{Breugem2012} is employed to solve the flow field. 
Equations (\ref{eq:NS}a) and (\ref{eq:NS}b) are integrated in time using an explicit low-storage Runge-Kutta method. A first prediction velocity is used to approximate the IBM force, and a second one to compute the correction
pressure and update the pressure field.

\subsubsection{Solution of the particle motion}

Breugem (2012)  \cite{Breugem2012} shows that Eqs. (\ref{eq:Forces}) can be re-written in discrete form as
as
\begin{equation}\label{yyy}
\begin{split}
& \textbf{U}^q_p = \textbf{U}^{q-1}_p - \frac{\Delta t}{V_p}\frac{\rho_f}{\rho_p} \sum\limits_{l=1}^{N_L} \textbf{F}^{q-1/2}_l \Delta V_l + \frac{1}{V_p}\frac{\rho_f}{\rho_p} \left( {\left\lbrace \int_{V_p} \textbf{u} \mathrm{d} V \right\rbrace}^q - {\left\lbrace \int_{V_p} \textbf{u} \mathrm{d} V \right\rbrace}^{q-1} \right) \\[1em] 
  &+ \left( \alpha_q + \beta_q \right) \Delta t \left( 1 - \frac{\rho_f}{\rho_p} \right) \textbf{g} + \left( \frac{\alpha_q + \beta_q}{\rho_p V_p} \right) \Delta t \frac{\, \textbf{F}^q_c + \textbf{F}^{q-1}_c }{2} \, , 
\end{split}   
\end{equation}
for the linear momentum and 
\begin{equation}
\begin{split}
& \textbf{I}^q_p \, \pmb{\omega}^q_p = \textbf{I}^{q-1}_p \, \pmb{\omega}^{q-1}_p - \Delta t \rho_f \sum\limits_{l=1}^{N_L} \textbf{r}^{q-1}_l \times \textbf{F}^{q-1/2}_l \Delta V_l \\[1em] 
&+ \rho_f \left( {\left\lbrace \int_{V_p} \textbf{r} \times \textbf{u} \mathrm{d} V \right\rbrace}^q - {\left\lbrace \int_{V_p} \textbf{r} \times \textbf{u} \mathrm{d} V \right\rbrace}^{q-1} \right) + \left( \alpha_q + \beta_q \right) \Delta t  \frac{ \, \textbf{T}^q_c + \textbf{T}^{q-1}_c}{2}   \,   
\label{eq:Motions}  
\end{split}    
\end{equation}
for the angular momentum where $\textbf{r}$ is the position vector, $\textbf{x} - \textbf{x}_c$; 
these are integrated in time with the same Runge-Kutta method used for the flow.
$\textbf{I}_p \, \pmb{\omega}_p$ is obtained by solving Eq.( \ref{eq:Motions}) with the following iterative procedure.
\begin{enumerate}
\item As an initial guess $\textbf{I}_p$ is set equal to the moment-of-inertia tensor at the previous substep $\textbf{I}^{q-1}_p$.
\item $\pmb{\omega}_p$ is computed from the linear equations $\textbf{I}_p \, \pmb{\omega}_p = \textbf{B}$, where $\textbf{B}$ is the right hand side of eq. (\ref{eq:Motions}).
\item The particle rotation during the current substep is indicated by the rotation matrix $\textbf{A}$. This is associated to
an axis of rotation in the direction of  $ \left( \pmb{\omega}^q_p + \pmb{\omega}^{q-1}_p \right) /2$ and an angle of rotation $ \left| \left( \pmb{\omega}^q_p + \pmb{\omega}^{q-1}_p \right) /2 \right|  . \left( \alpha_q + \beta_q \right) . \, \Delta t $.
The rotation matrix is used to update the moment-of-inertia tensor from the previous substep,  $\textbf{I}_p = \textbf{A}\textbf{I}^{q-1}_p \textbf{A}^{-1}$.
\item The new $\textbf{I}_p$ is used as initial guess in step 1 until convergence within a threshold is obtained ($\textbf{I}_p^{new} - \textbf{I}_p^{old} < \epsilon$). 
\end{enumerate}

This procedure 
typically requires less than 5 iterations to converge.        
The orientation of the particle and the position of the Lagrangian points are updated  by means of the rotation matrix $\textbf{A}$. The position of the particle center and the velocity at the surface of particle are finally computed as
\hypertarget{eq:ctrpos}{}
\begin{subequations}
\begin{align}
&\textbf{x}^q_c  = \textbf{x}^{q-1}_c + \frac{\left(\alpha_q + \beta_q \right)}{2} \Delta t \left( \textbf{U}^q_p + \textbf{U}^{q-1}_p \right)  \, , \\[1em] 
&\textbf{U} \left( \textbf{X}^q_l \right) = \textbf{U}^q_p + \pmb{\omega}_p \times  \left( \textbf{X}^q_l - \textbf{x}^q_c \right)   \, .
\label{eq:ctrpos}  
\end{align}
\end{subequations}

\subsubsection{Lubrication and collision models}

\begin{figure}[t]
\centering
\includegraphics[width=0.8\linewidth]{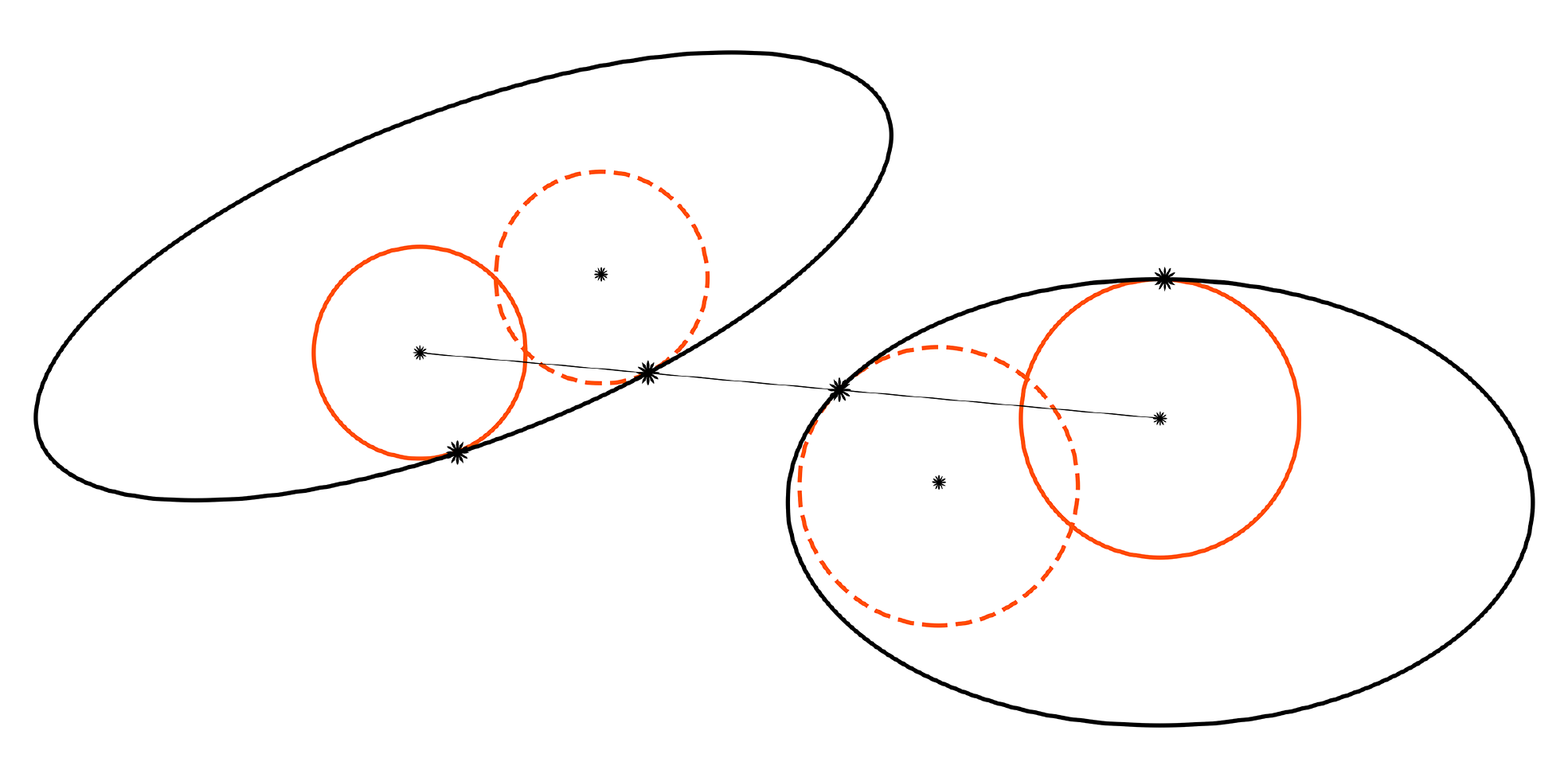}
\put(-70,60){{$c^k_y$}}
\put(-70,90){{$y^k$}}
\put(-123,48){{$c^{k+1}_y$}}
\put(-138,70){{$y^{k+1}$}}
\put(-194,45){{$x^k$}}
\put(-167,56){{$x^{k+1}$}}
\put(-213,70){{$c^k_x$}}
\put(-182,87){{$c^{k+1}_x$}}
\caption{\label{fig:2} 
Two-dimentional sketch of the iterative method used to find the nearest distance between two ellipsoids. $(x,y)^k$ are the current guesses for nearest points.} 
\end{figure}
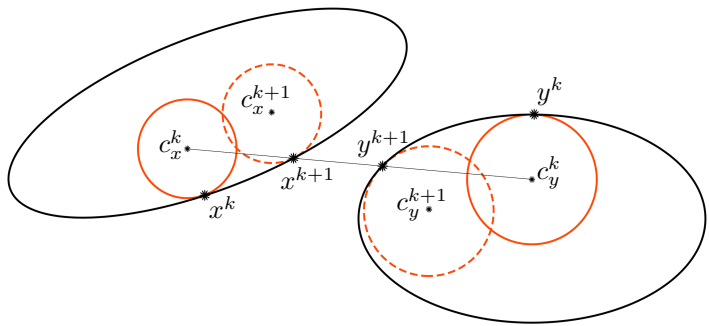

\paragraph{Lubrication model} $\,$ \\ 
A particle immersed in a viscous liquid  experiences lubrication forces while approaching a wall or another particle. These are due to the film drainage and have an analytical expression in the Stokes regime  
\cite{Brenner1961}. 
The lubrication force is well captured by the IBM method as long as the fluid in the thin gap between the two solid bodies
is well resolved.
However, for gaps smaller than the Eulerian mesh size, lubrication is under-predicted. To compensate for this inaccuracy and avoid computationally expensive grid refinements, a correction model based on asymptotic expansions of 
the lubrication force 
in the Stokes regime is used, see also \cite{Costa2015,Breugem2010}. 
Since lubrication is essentially a two-body problem dominated by the flow in the narrow gap separating two  surfaces \cite{Claeys1993}, spheroidal particles are represented as spheres with radius equal to the local radius of curvature of the spheroidal particle. 
In other words, we 
approximate the spheroidal particles near contact as spheres with the radius of curvature of the closest points of contact
and same mass as the original spheroid and
resort to an analytical solution for poly-disperse suspensions of spherical particles \cite{Jeffrey1982}. 
From a computational point of view, the two difficulties  are (i) to find the closest points on the surface of the two ellipsoids and (ii)  find the local radii of curvature. The efficient iterative method proposed by \cite{Lin2002} is employed here to find the closest distance between the two particles. 
The method can be summarised as follows, see Figure~\ref{fig:2} for a visual clarification in 2D.
\begin{enumerate}
\item The search algorithm starts from two arbitrary points on the surface of the two particles $(x,y)^k$, assumed initially as the nearest points.
\item Construct two {\it balls} completely inside the ellipsoids and tangent to the inner surface at the current guess for the nearest points.  
\item Find a new guess  $(x,y)^{k+1}$ by the intersection of the line of centres $(c_x,c_y)^k$ of the two balls and the surface of the two spheroids.
\item If not converged, go back to step 2. Convergence is obtained when the change of the slope of the line that connects the closest points is below a given threshold. 
\end{enumerate}
The procedure converges faster as the radius of the constructed balls increases, however these should fit entirely inside the spheroids.       

Once the nearest points are known, the Gaussian radius of curvature $R_i$ is calculated  based on the meridional and normal radii of curvature $M$ and $N$,
\begin{equation}
M = \frac{a^2b^2}{{\left( {\left( a sin \Phi \right)}^2 + {\left( b cos \Phi \right)}^2 \right) }^{3/2}}  \, ,  
N = \frac{b^2}{{\left( {\left( a sin \Phi \right)}^2 + {\left( b cos \Phi \right)}^2 \right) }^{1/2}} .
\label{eq:radii}  
\end{equation}
where the semi axes $a$ and $b$ are the polar and the equatorial radius of the spheroid and the angle $\Phi$ defines the latitude of the point at which the radius of curvature is being calculated.  
The Gaussian radius of curvature defines the radius of the best fitting sphere tangent to the given surface point 
\[
R_i = \sqrt{MN} = \frac{a^2b}{ {\left( a sin \Phi \right)}^2 + {\left( b cos \Phi \right)}^2  }  \,.
\]
Note that the best fitting sphere is different from the balls used in the algorithm above and its radius $R_i$  can be larger than $a$ and $b$.

\begin{figure}[t]
\centering
\includegraphics[width=0.6\linewidth]{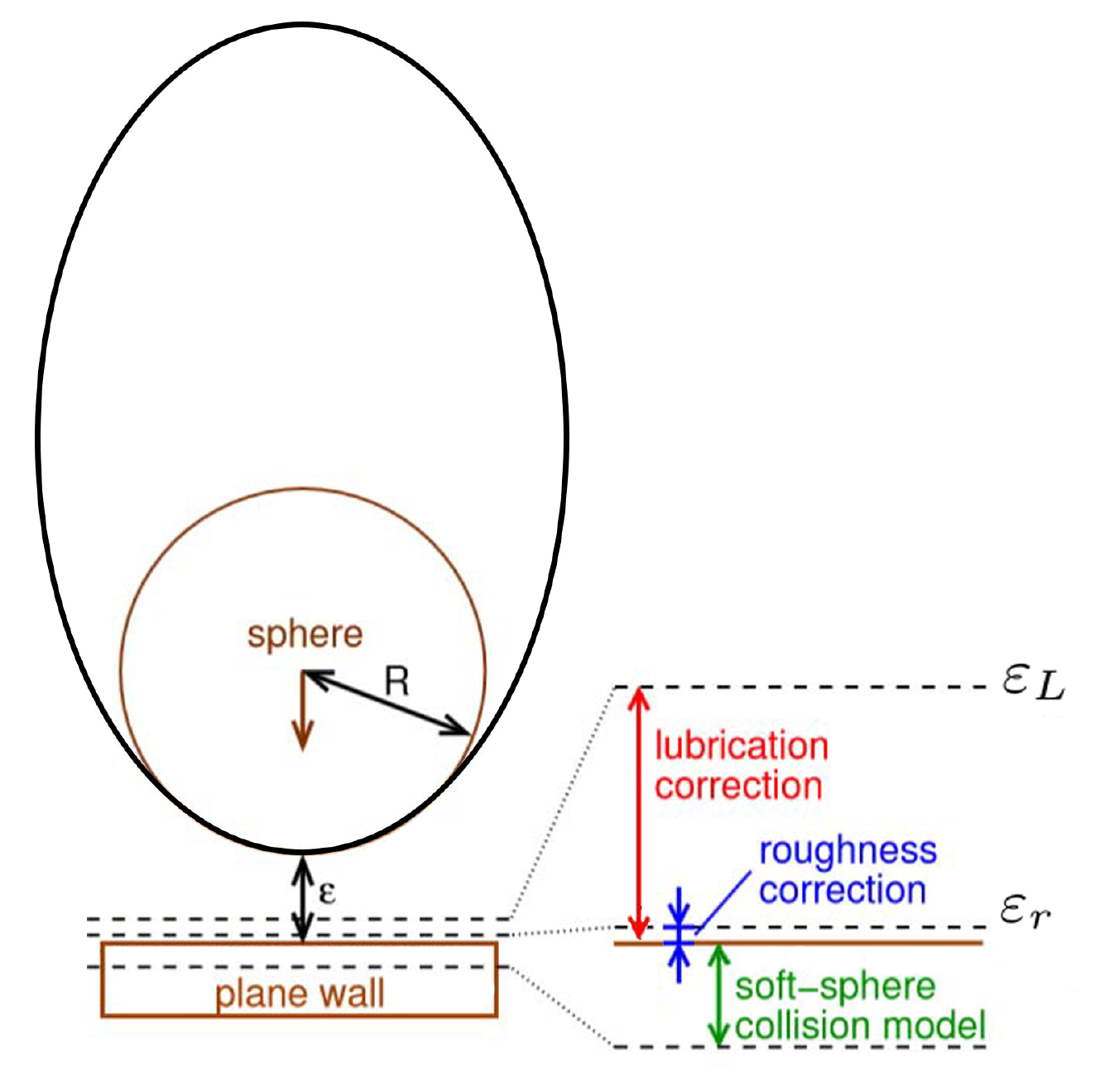}
\vspace{-10pt}
\caption{\label{fig:3} 
Schematic representation of the lubrication model applied to a sphere approaching a plane wall. A similar approach is used for particle-particle interactions.} 
\end{figure}

The lubrication model employed in this study is based on the asymptotic solution of Jeffrey (1982) \cite{Jeffrey1982} for spheres with different radii. This two-parameter solution considers normal lubrication effects given by 
\hypertarget{eq:LuMo}{}
\begin{equation}
\Delta F_{Lub} = -6 \pi \mu R_i \Delta U_n \left[ \lambda(\kappa,\varepsilon) -\lambda( \kappa, \varepsilon_{L}) \right]  \, ,
\label{eq:LuMo}  
\end{equation}
where $R_i$ is the radius of curvature, $\kappa$ the ratio between the radii of curvatures of the two spheres and $\varepsilon$ the gap width (closet distance) normalized by the larger radius of curvature. $\lambda$ is the Stokes amplification factor defined here as in Jeffrey (1982) \cite{Jeffrey1982}. $\varepsilon_L$ defines the normalized gap width at which the lubrication model becomes active. To account for the presence of surface roughness, and to limit the lubrication forces to finite values, a threshold width below which the value of the Stokes amplification factor becomes constant  ($\varepsilon \leq \varepsilon_r: \lambda(\kappa,\varepsilon) = \lambda(\kappa,\varepsilon_r)$) is introduced. 
We use here $\varepsilon_L=0.025$ and $\varepsilon_r=0.001$ in the case of particle-particle interactions 
and $\varepsilon_L=0.05$ and $\varepsilon_r=0.001$
for particle-wall interactions. A schematic representation of the lubrication model is given in figure~\ref{fig:3}. Lubrication corrections responsible for translational and rotational shearing are neglected in this study due to their slower divergence with the gap width: $  \Delta F \propto \ln \varepsilon$ versus $\Delta F \propto 1/ \varepsilon$ for normal lubrication.  

\begin{figure}[t!]
\centering
\includegraphics[width=0.495\textwidth]{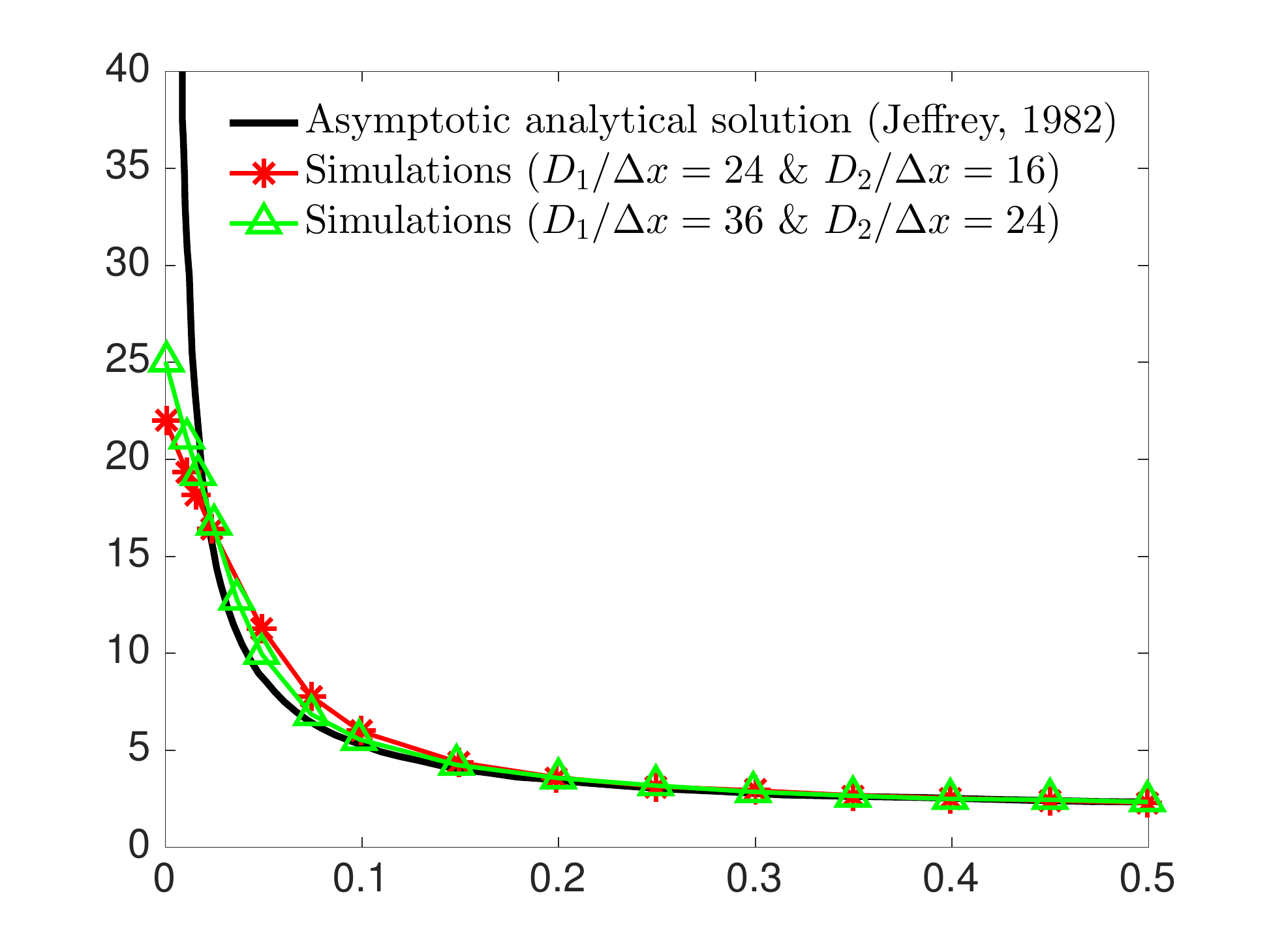}
\includegraphics[width=0.495\textwidth]{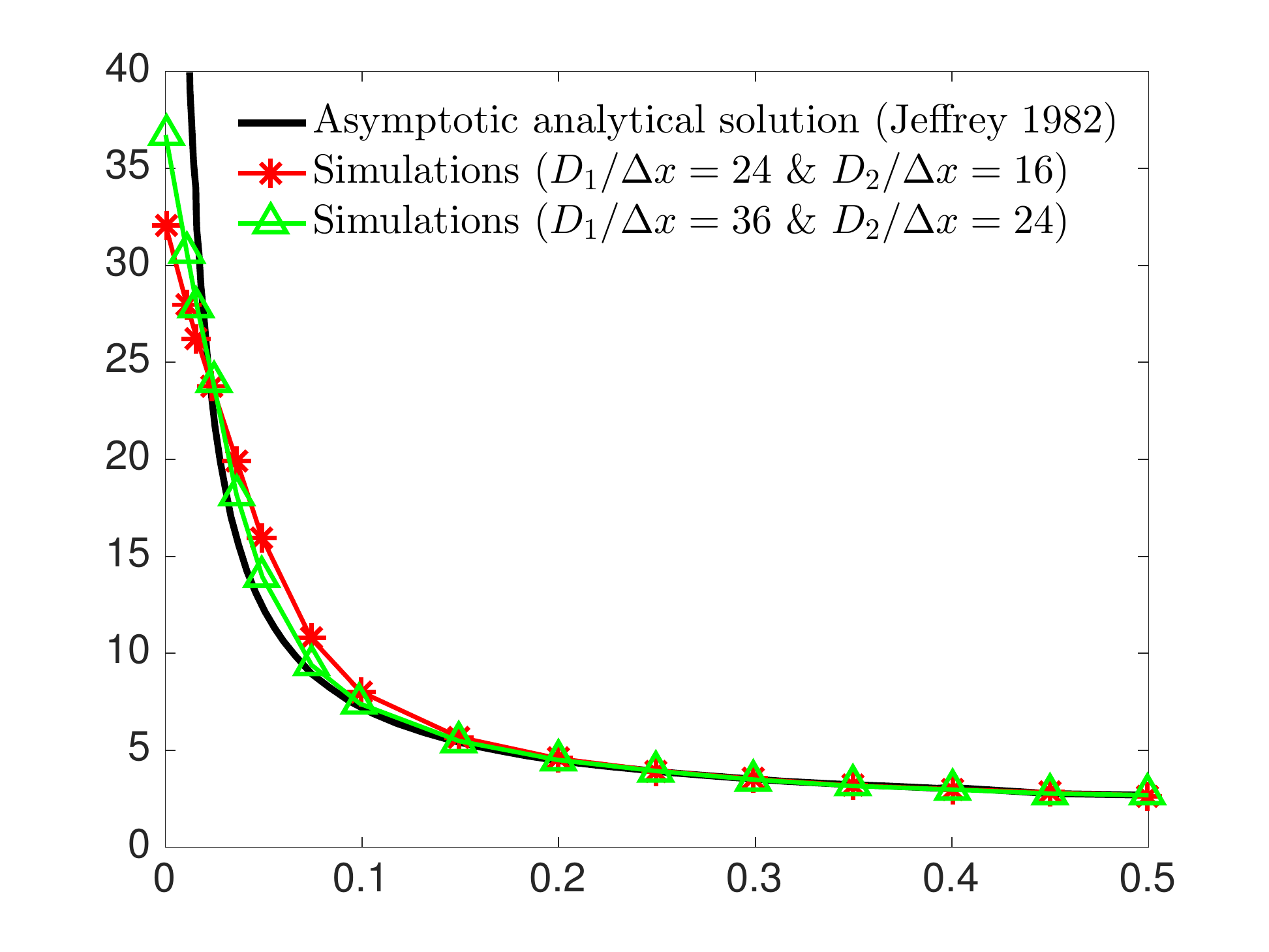}  
\put(-172,50){\rotatebox{90}{$F/ {\left(F_{sd}\right)}_{II}$}}
\put(-345,50){\rotatebox{90}{$F/ {\left(F_{sd}\right)}_I $}}
\put(-94,-5){{$\varepsilon/R_1$}}
\put(-267,-5){{$\varepsilon/R_1$}} 
\put(-338,-20){\footnotesize (a) Sphere I without lubrication correction}
\put(-166,-20){\footnotesize (b) Sphere II without lubrication correction}
\\
\includegraphics[width=0.495\textwidth]{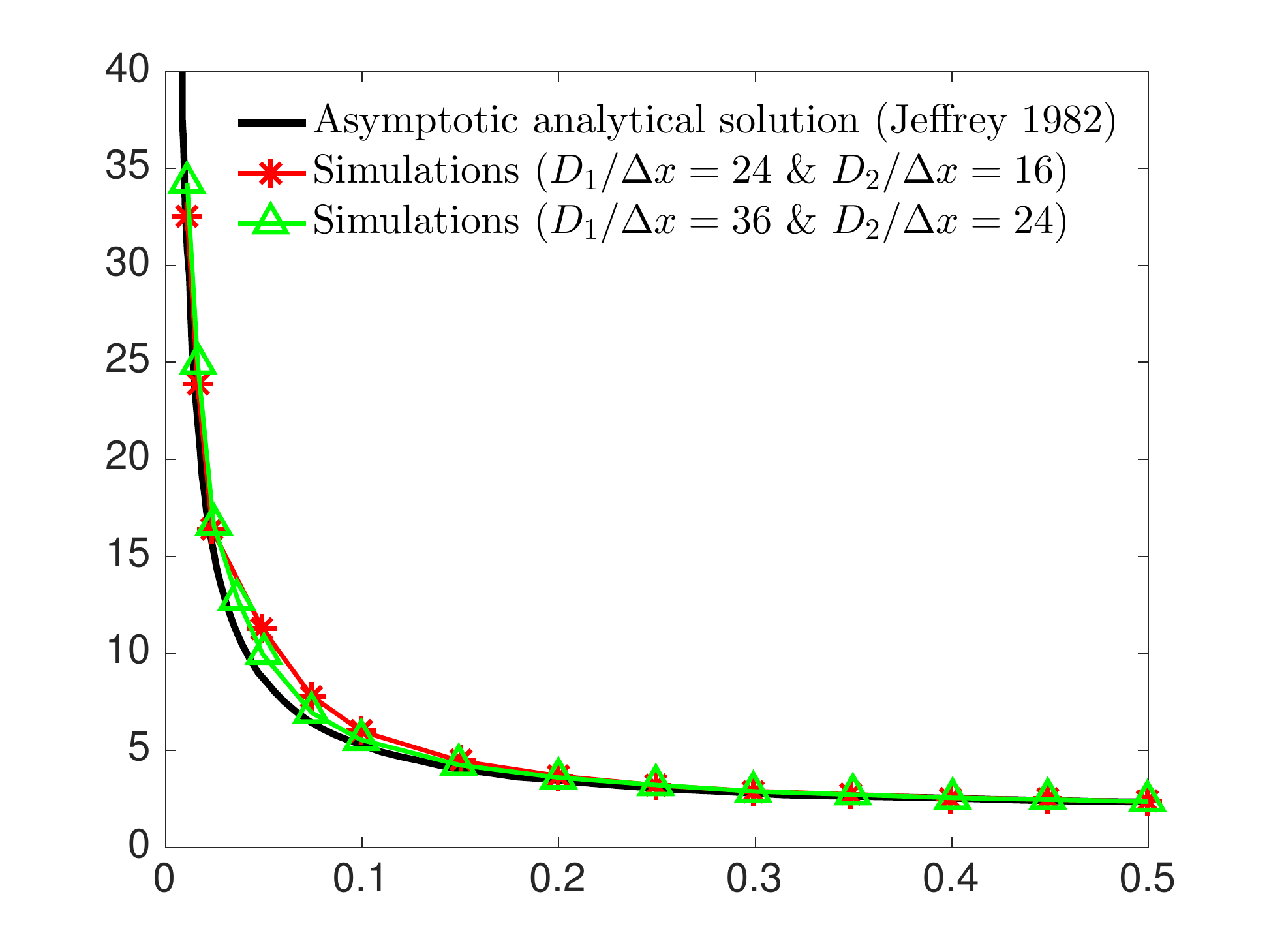}
\includegraphics[width=0.495\textwidth]{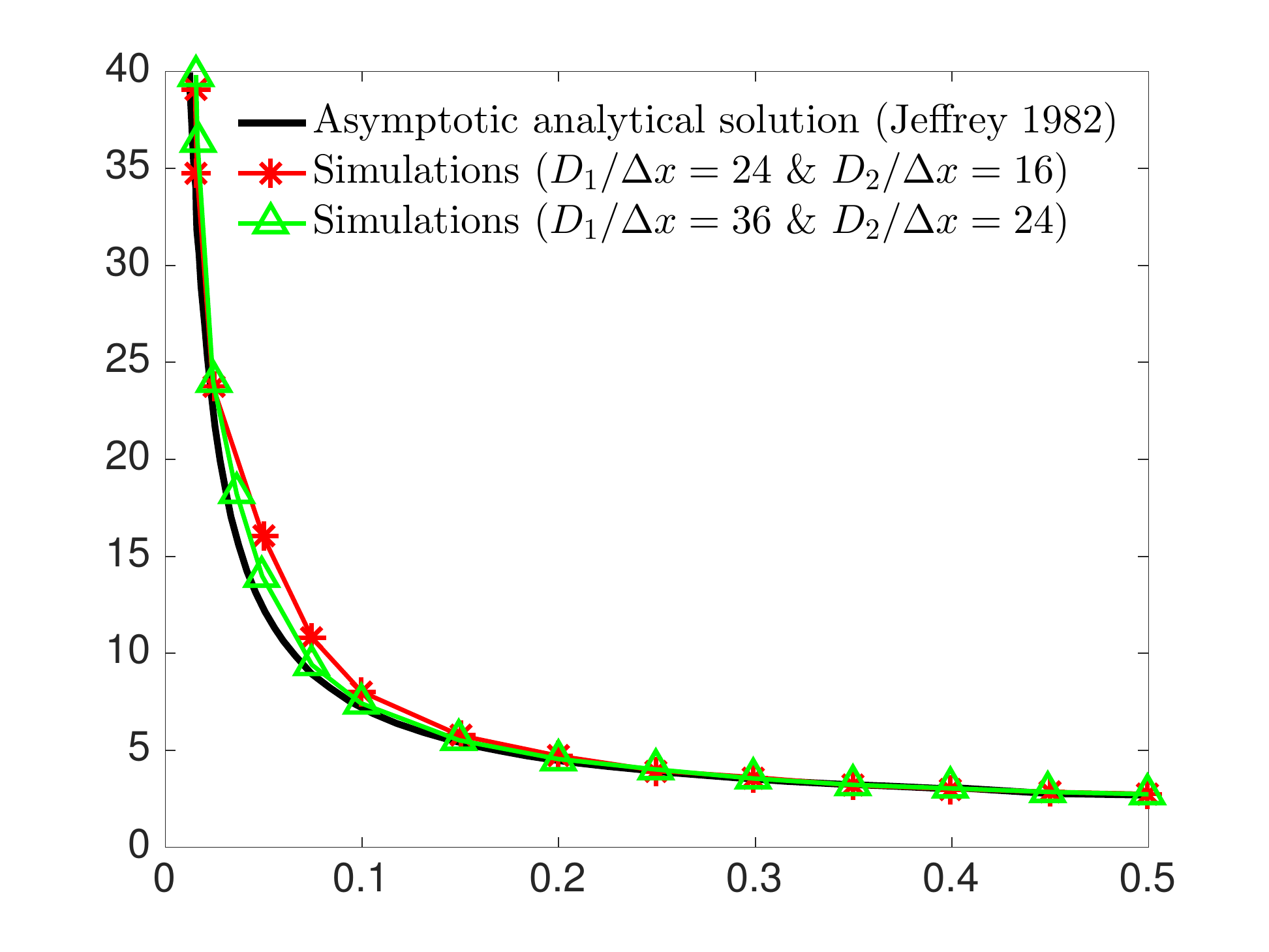}
\put(-172,50){\rotatebox{90}{$F/ {\left(F_{sd}\right)}_{II} $}}
\put(-345,50){\rotatebox{90}{$F/ {\left(F_{sd}\right)}_I $}}
\put(-94,-5){{$\varepsilon/R_1$}}
\put(-267,-5){{$\varepsilon/R_1$}}
\put(-332,-20){\footnotesize (c) Sphere I with lubrication correction}
\put(-162,-20){\footnotesize (d) Sphere II with lubrication correction}
\caption{\label{fig:4} 
Normal force between two unequal spheres ($R_1/R_2=1.5$) approaching at equal velocity with and without lubrication correction for two grid resolutions $24$ and $36$ grid points per larger diameter, compared to the analytical solution of Jeffrey (1982) \cite{Jeffrey1982}. The forces are normalized by the Stokes drag $F_{sd}$ in free space for each particle.} 
\end{figure}
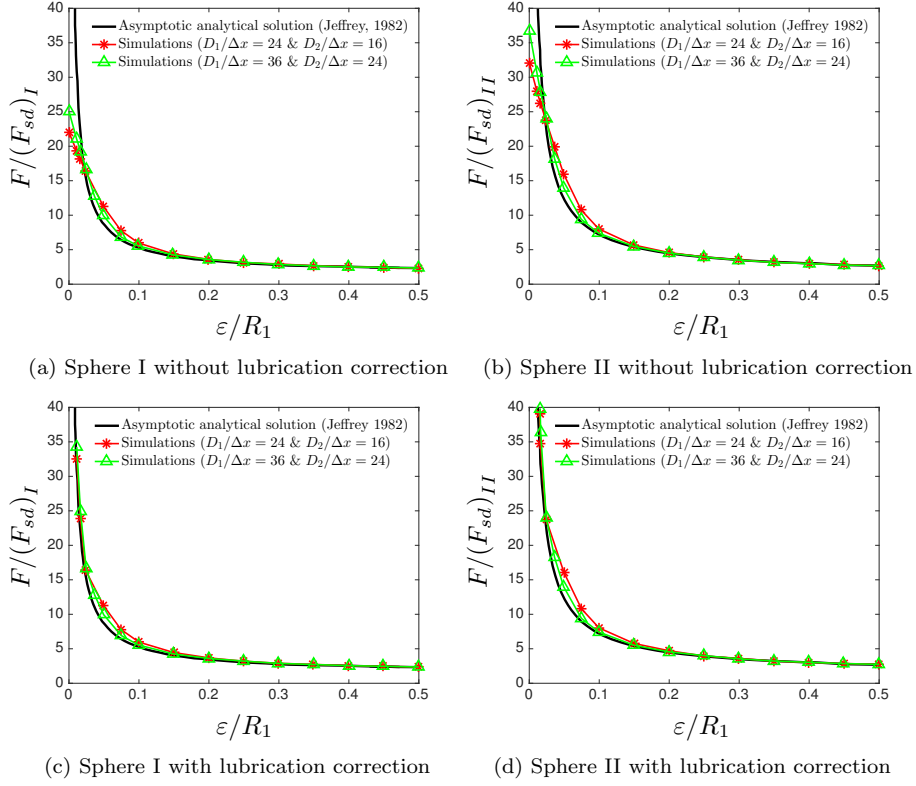


To validate the lubrication model, we compute the normal force between two spheres of different radii ($R_1/R_2=1.5$) approaching at equal velocity. Results for the interaction force with and without the lubrication correction are displayed in figure~\ref{fig:4} normalized by the Stokes drag in free space ($F/F_{sd}$). Without correction, the results are in good agreement with the analytical solution of Jeffrey (1982) \cite{Jeffrey1982} only when the grid can resolve the flow between the solid objects. For smaller gaps, eq.~(\ref{eq:LuMo}) correctly captures the increase in lubrication.


\paragraph{Collision model} $\,$ \\
When the gap width between two spheroids reduces to zero, the lubrication correction is switched off and a soft sphere collision model \cite{Costa2015} activated. 
To compute the collision forces we proceed as for the lubrication correction model, i.e.\ the spheroidal particles are approximated as spherical particles with  the same mass as the whole particle and with a radius corresponding to the local curvature at the contact points. 
The radii of the approximating spheres remain constant during the collision, simplifying the problem to that of the collision between two unequal spheres. 
The centres of the approximating colliding spheres are stored at the time step before the gap width becomes negative and updated during the collision using the particle velocity and the rotation matrix introduced above.  

\begin{figure}[h!]
\centering
\includegraphics[width=0.495\textwidth]{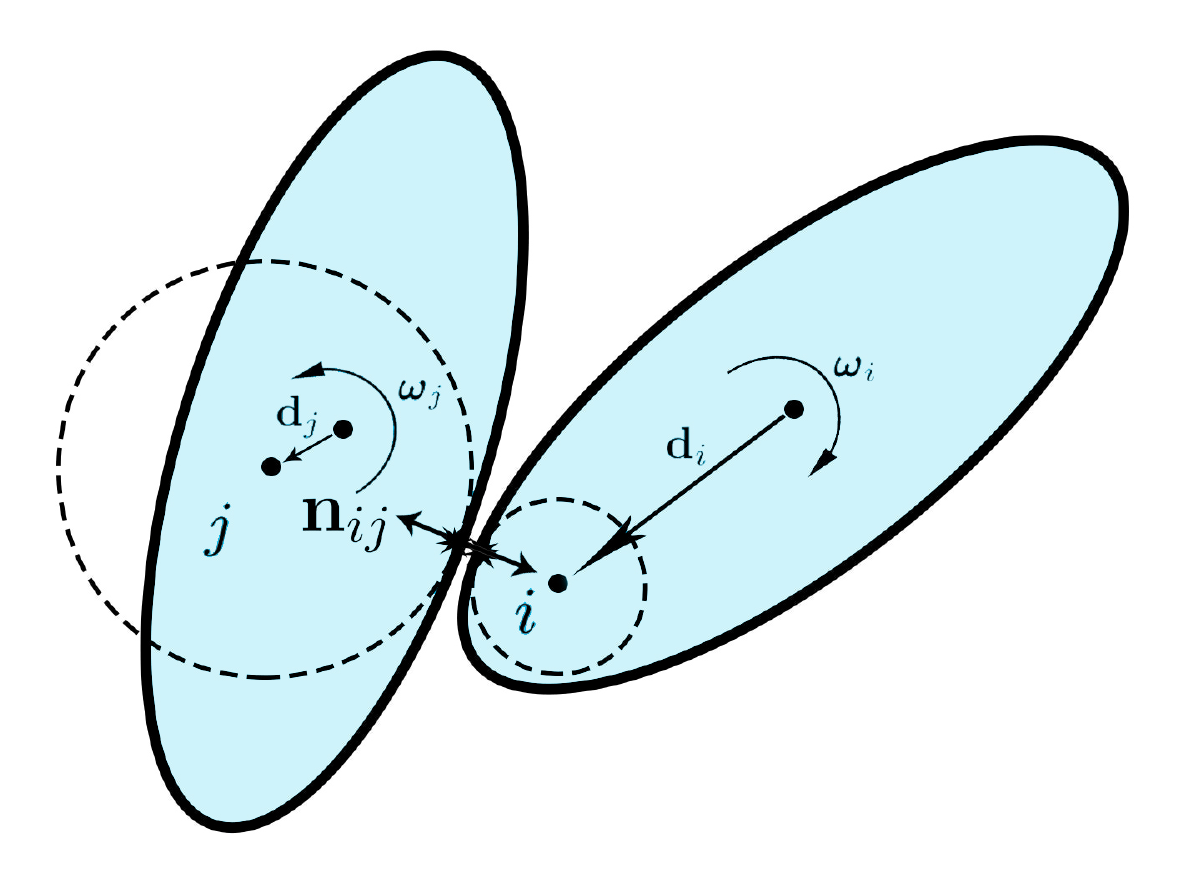}
\includegraphics[width=0.495\textwidth]{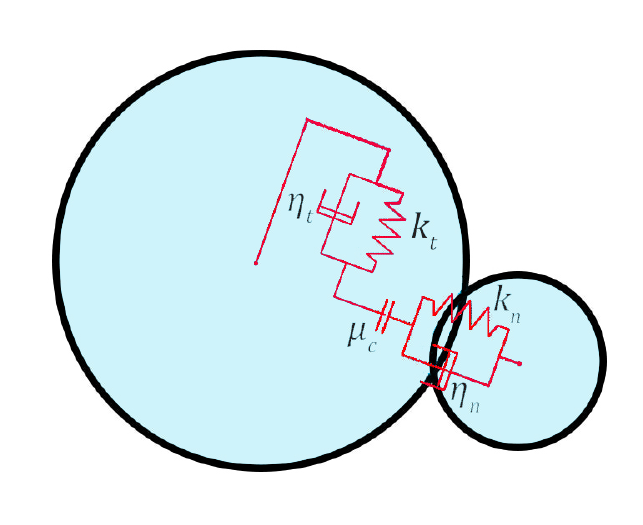}
\put(-316,-15){\footnotesize (a) Particles just before collision}
\put(-158,-15){\footnotesize (b) Approximating spheres in collision} 
\vspace{-1pt}
\caption{\label{fig:5} 
Collision model for spheroidal particles. Sketch of (a) the geometrical and kinetic parameters and (b) the spring-dashpot model used to compute normal and tangential forces.} 
\end{figure}

The soft sphere model used in \cite{Costa2015} is employed here to calculate the normal and tangential collision force. 
In this model, the forces  are computed using a linear spring-dashpot system in the normal and tangential directions, with an additional Coulomb friction slider to simulate friction as shown in Figure~\ref{fig:5}.

The collision time $T$ is allowed to stretch over $N$ time steps provided that the collision time is still much smaller than the characteristic time scale of the particle motion. This makes the numerical simulation of a wet collision more realistic since the fluid has enough time to adapt to the sudden change in the particle velocity as reported in Costa et al. 2015 \cite{Costa2015}.

In brief (more details can be found in \cite{Costa2015}), the normal collision force
 depends on the overlap between the two particles and on the normal relative velocity of the surface points located on the line-of-centers. The normal direction $\textbf{n}_{ij}$  is defined by the vector connecting the centres of the two colliding spheres,  
\begin{equation}
 \textbf{n}_{ij} = \frac{\textbf{x}_j - \textbf{x}_i}{||\textbf{x}_j - \textbf{x}_i||},
\end{equation}
the penetration as
\begin{equation}
\pmb{\delta}_{ij,n} = \left( R_i + R_j - ||\textbf{x}_j - \textbf{x}_i|| \right) \textbf{n}_{ij} ,
\end{equation}
and the normal relative velocity $\textbf{u}_{ij,n} = \left(\textbf{u}_{ij} \, \cdot \, \textbf{n}_{ij} \right)\textbf{n}_{ij}   $ with 
\begin{equation}
\textbf{u}_{ij} = \left(\textbf{u}_i +  \pmb{\omega}_i \times \textbf{d}_i + R_i \pmb{\omega}_i \times \textbf{n}_{ij} \right) - \left(\textbf{u}_j +  \pmb{\omega}_j \times \textbf{d}_j + R_j \pmb{\omega}_j \times \textbf{n}_{ji} \right) .
\end{equation}
where $\textbf{d}_i$ and $\textbf{d}_j$ are the vectors that connect the centres of spheroids to the centres of the approximated spheres, while $R_i$ and $R_j$ are the radii of the approximating spheres with centres at $\textbf{x}_i$ and $\textbf{x}_j$. The normal collision force acting on sphere $i$ when colliding with sphere $j$ is then expressed as
\begin{equation} \label{eq:fnormal}  
\textbf{F}_{ij,n} = -k_n \pmb{\delta}_{ij,n} - \eta_n \textbf{u}_{ij,n}
\end{equation}
with model coefficients
\begin{equation}
k_n = \frac{m_e \left( \pi ^2 + \ln^2 e_{n,d} \right)}{{\left(N \Delta t \right)}^2}  \, ,\eta_n = -\frac{2 m_e \ln e_{n,d}}{N \Delta t}  \, ,m_{e} = {\left( m_i^{-1} + m_i^{-1} \right)}^{-1}.
\end{equation}
$k_n$ and $\eta_n$ are the normal spring and dashpot coefficients, computed by solving the motion of a linear harmonic oscillator requiring that \cite{Van2004} 
(i) The magnitude of the normal relative velocity at the end of the collision is equal to the normal restitution coefficient $e_{n,d}$ times the normal velocity at the beginning of the collision. (ii) There is no overlap at the end of the collision ($t = N\Delta t$). 

The terms $m_i$ and $m_j$ in the expression above are the masses of the spheroidal particles and $N$ is the number of time steps over which the collision is stretched. Large values of $N$ cause a large overlap between particles and therefore an unrealistic delay of the particle rebound, while small values of $N$ result in a lack of accuracy  as the collision force may be very large. Here, we  use $N=8$.

The component of the collision force $\textbf{F}_{ij,t}$ in the tangential direction $\textbf{t}_{ij}$ is computed similarly with a Coulomb friction included to model the possibility of sliding motion.
The tangential force acting on sphere $i$ when colliding with sphere $j$ is expressed as:
\begin{equation}
\label{eq:ftang} 
\textbf{F}_{ij,t} = min\left( ||-k_t \pmb{\delta}_{ij,t} - \eta_t \textbf{u}_{ij,t}|| \,\,\, , \,\,\, ||-\mu_c \textbf{F}_{ij,n}|| \right) \textbf{t}_{ij},
\end{equation}
where the relative tangential velocity $\textbf{u}_{ij,t} = \textbf{u}_{ij} - \textbf{u}_{ij,n}$ and the tangential displacement is denoted $\pmb{\delta}_{ij,t}$. 
This is
computed during the collision by integration of the relative tangential velocity 
\[
\pmb{\delta}^{*^{n+1}}_{ij,t} = \textbf{A} \, \cdot \, \pmb{\delta}^n_{ij,t} +  \int_{t^n}^{t^{n+1}} \textbf{u}_{ij,t} \, \mathrm{d} t. 
\]
It should be noted that,  to comply with Coulomb's condition, the tangential displacement is saturated when the particles start sliding \cite{Luding2008},   
\small
\begin{equation}\label{eq:delta_Tang} 
\pmb{\delta}^{n+1}_{ij,t} = \left\lbrace
  \begin{array}{r@{}l}
    \pmb{\delta}^{*^{n+1}}_{ij,t} \,\,\,\,\,\,\,\,\,\,\,\,\,\,\,\,\,\,\,\,\,\,\,\,\,\,\,\,\,\,\,\,\,\,\,\,\,\,\,\,\,\,\,\,\,\,\,\,\,\,\,\,\,\,\,\,\,\,\,\,\,\,\,\,\,\,\,\,\,\,\, , \,\,\, if \,\,\,  ||\textbf{F}_{ij,t}|| \leq \mu_c ||\textbf{F}_{ij,n}|| \, , \\
    \left( 1/k_t \right) \left( -\mu_c ||\textbf{F}_{ij,n}|| \textbf{t}_{ij} - \eta_t \textbf{u}_{ij,t} \right) \,\,\, , \,\,\, if \,\,\,  ||\textbf{F}_{ij,t}|| > \mu_c ||\textbf{F}_{ij,n}||  .
  \end{array}
  \right.
\end{equation}
\normalsize
The coefficients in Eq. (\ref{eq:ftang}) are defined as 
\[
k_t = \frac{m_{e,t} \left( \pi ^2 + \ln^2 e_{t,d} \right)}{{\left(N \Delta t \right)}^2}  \, ,\eta_t = -\frac{2 m_{e,t} \ln e_{n,t}}{N \Delta t}  \, , m_{e,t} = {\left( 1 + 1/K^2 \right)}^{-1} m_e.
\]
where $k_t$ and $\eta_t$ are the tangential spring and dashpot coefficients and $K$ is the normalized particle radius of gyration for the approximating spheres ($\sqrt{2/5}$). 

The normal and tangential collision forces at the points of contact are finally transferred to the spheroids centres
\hypertarget{eq:fcol}{}
\begin{subequations}
\begin{align}
 & \textbf{F}^c_{ij} = \textbf{F}_{ij,n} + \textbf{F}_{ij,t}   \, , \\[1em]
 & \textbf{T}^c_{ij} = \left(\textbf{d}_{i} + R_i \textbf{n}_{ij}\right) \times \textbf{F}_{ij,t} + \textbf{d}_i \times \textbf{F}_{ij,n}     \, .
\label{eq:fcol}  
\end{align}
\end{subequations}
\subsubsection{Parallelization}
The numerical algorithm detailed in the previous subsections is implemented in Fortran with MPI libraries 
 for parallel execution on multi-processor machines with distributed memory. 
 For the parallelization of the Navier--Stokes equations we adopt a standard domain decomposition in two dimensions (streamwise and spanwise)
 since a 
 3D parallelization might result in an unbalanced distribution of  the computational load among the processors when a preferential direction exists, e.g.\ in the case of sediments. 
 %
 The particle-related computations follow a master-slave parallelization similar to that used in Breugem (2012) \cite{Breugem2012}. 
 %
 The processor where the center of a particle is located is denoted as master while the neighbours  containing at least one Lagrangian point as slaves. 
 The method requires that a particle fit entirely inside one processor domain, so that it cannot belong to more than 3 slaves. 
 To find the slave neighbors, the ellipsoid is projected in the plane of parallelization. From the equation of the projected ellipse, 
 we compute the intersections with the boundaries of the master domain and thus identify
 the slave neighbours.
 The slave processor communicates the data to the master processor, which is the one responsible for the computations of the particle motion.

\section{Validation}
\label{Validation}

\subsection{Spheroids in uniform shear flow} 

\begin{figure}[t]
\centering
\includegraphics[width=0.4\linewidth]{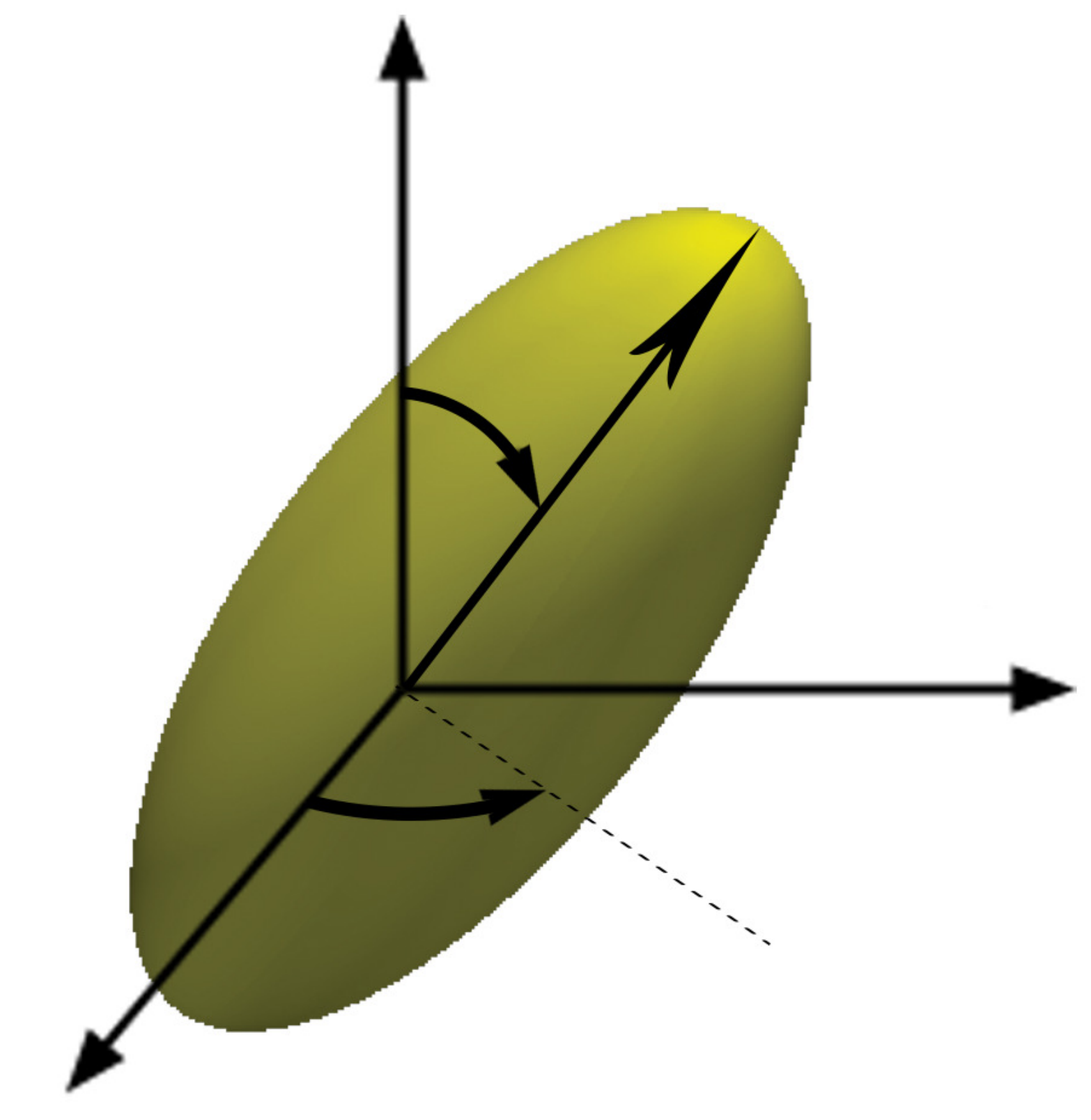}
\put(-5,62){{\large $y$}}
\put(-125,-3){{\large $x$}}
\put(-82,135){{\large $z$}}
\put(-90,28){{\large $\phi$}}
\put(-75,91){{\large $\theta$}}
\caption{\label{fig:6} 
Angles defining the orientation of the spheroid: $\theta$ defines the angle between the symmetric axis of the spheroid and the $z$-axis while $\phi$ indicates the angle between the projected symmetric axis in the $xy$ plane and the $x$-axis.} 
\end{figure}

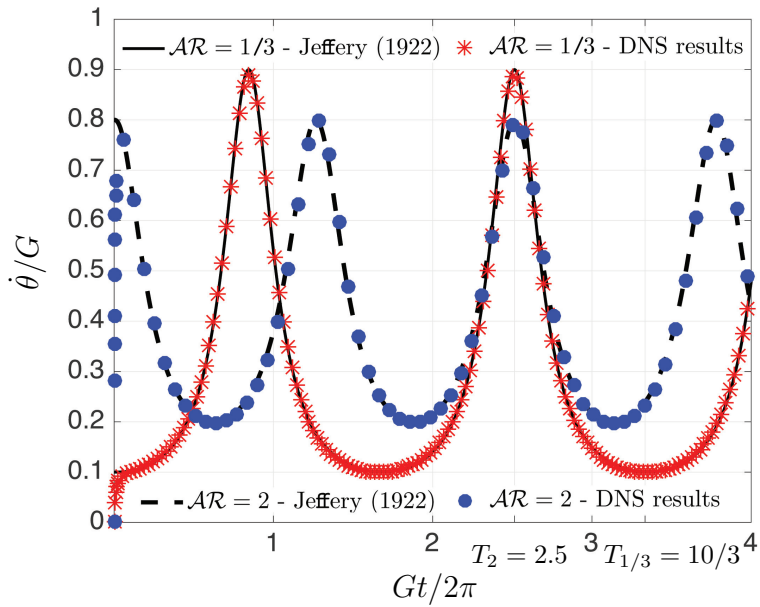
\begin{figure}[t]
\centering
\includegraphics[width=0.9\linewidth]{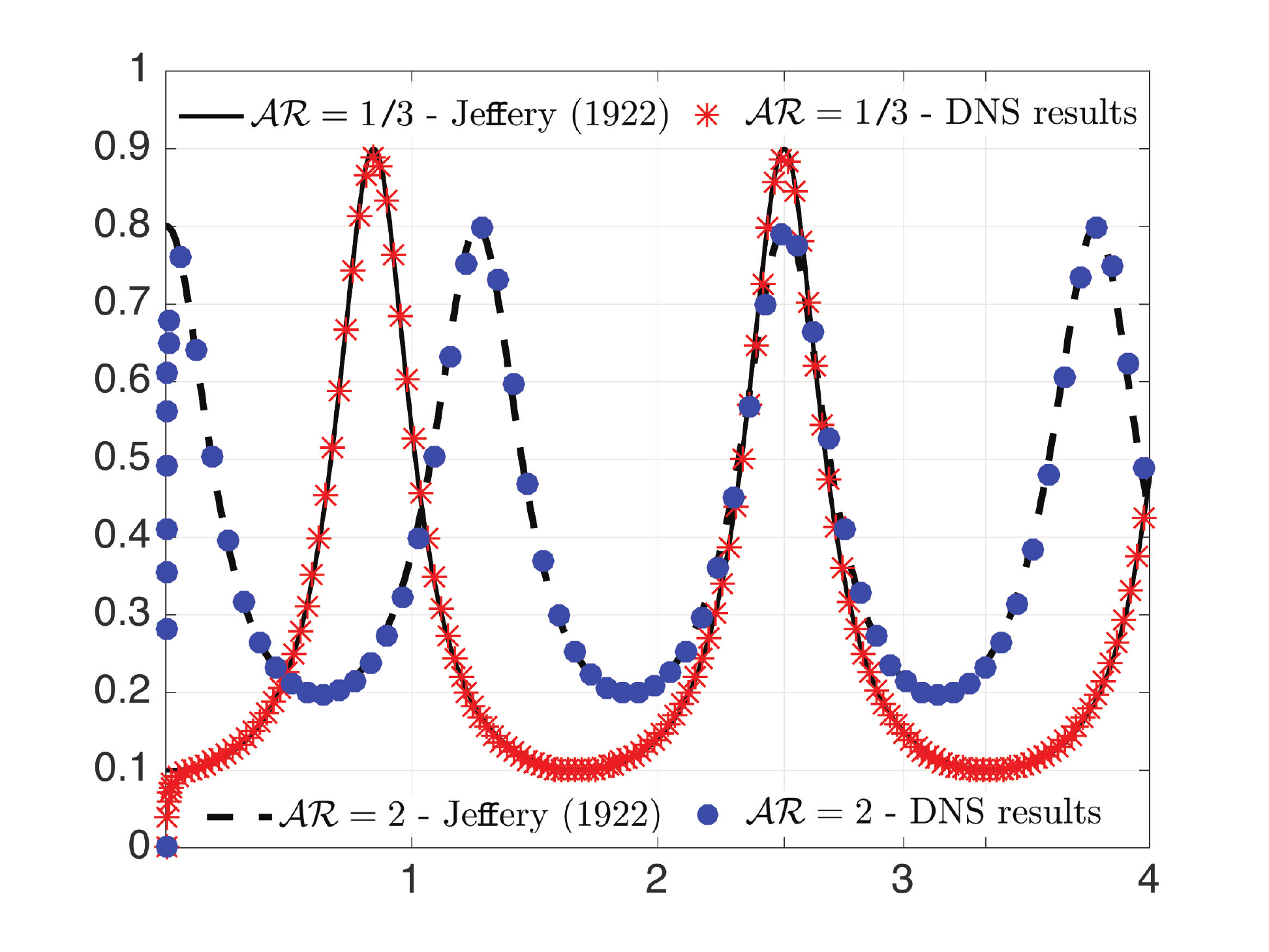}
\put(-310,110){\rotatebox{90}{ \large $\dot{\theta}/G$}}
\put(-165,-3){{\large $Gt/2\pi$}}
\put(-135,10){{$\tiny T_{2} = 2.5$}}
\put(-85,10){{$\tiny T_{1/3} = 10/3$}}
\caption{\label{fig:7} 
Spanwise component of the angular velocity
of spheroids with $\mathcal{AR}= 2$ and $1/3$ against the analytical solution by Jeffery (1922) \cite{Jeffery1922}. Time and angular velocity are non-dimensionalized with $2\pi/G$ and the shear rate $G$, respectively.} 
\end{figure}

The equations of motion of spheroidal particles derived by Jeffery (1922) \cite{Jeffery1922} have been widely used in the literature to track the motion of point particles, particles smaller than the smallest flow scale,
at vanishing particle Reynolds number $Re_p$ \cite{Zhao2014,Marchioli2013}. Jeffery (1922) \cite{Jeffery1922} also derived the analytical solution for the angular velocities $\dot{\theta}$ and $\dot{\phi}$ in the inertialess regime, $Re_p = 0$, in a simple shear flow, 
\hypertarget{eq:jeffery}{}
\begin{subequations}
\begin{align}
 & \dot{\theta} = -\frac{G}{a^2+b^2} \left( a^2 cos^2 \theta + b^2 sin^2 \theta \right)  \, , \\[1em]
 & \dot{\phi} = \frac{G \, | a^2-b^2 |}{4 \left( a^2+b^2 \right)} sin 2 \theta sin 2 \phi     \, ,
\label{eq:jeffery}  
\end{align}
\end{subequations}
where the semi axes $a$ and $b$ are the polar (symmetric semi-axis) and the equatorial radius of the spheroid, $G$ the imposed shear rate, $\theta$ ($ 0 \leq \theta < \pi$) and $\phi$ ($ 0 \leq \phi < 2 \pi$)  the angles defining the orientation of the spheroid, see figure~\ref{fig:6}. 

In this study, we simulate two spheroids with aspect ratios of $\mathcal{AR}= 2$ and $1/3$  in a plane Couette flow at $Re_p = 0.1$.
$Re_p$ is defined by shear rate $G$ and the equivalent particle diameter $D_{eq}$, i.e. the diameter of a sphere with the same volume of the original spheroid:
\hypertarget{eq:ReP}{}
\begin{equation}
 Re_{p} \equiv  \frac{G D_{eq}^2}{\nu} \, \, ,  D_{eq} = 2{\left( ab^2 \right) }^{1/3}.
\end{equation}
Simulations are performed in a domain of size $10D_{eq} \times 10D_{eq} \times 10D_{eq} $ with 32 grid points per $D_{eq}$ and periodic boundary conditions in the direction perpendicular to the velocity gradient. The initial particle orientation is set to $\phi = \theta = 0$ with no initial angular velocity. The particles tumble around the spanwise (normal to the shear plane) axis, as deduced by the analytical solution reported above with period
 $T = \frac{2 \pi}{G} \left( \mathcal{AR} + 1/\mathcal{AR} \right) $.
  The results, shown in figure~\ref{fig:7}, exhibit excellent agreement with the analytical solution. 
  
\subsection{Oblate ellipsoid in cross flow}

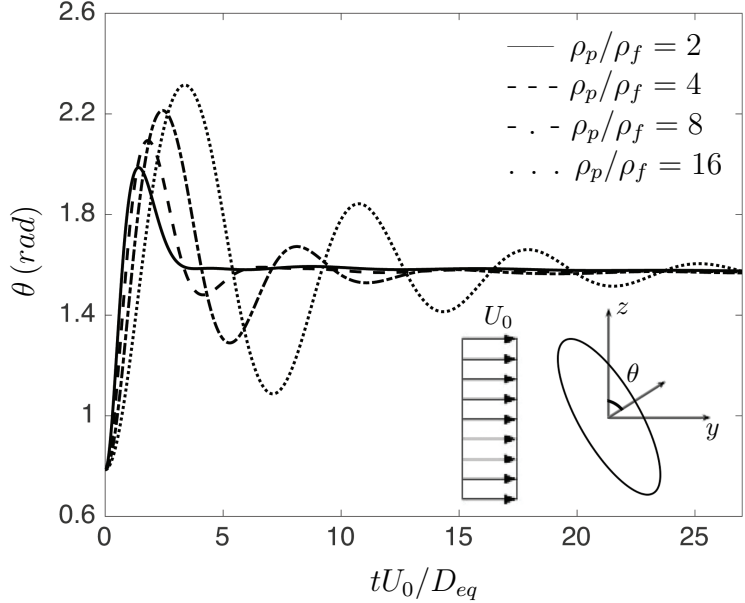
\begin{figure}[t]
\centering
\includegraphics[width=0.9\linewidth]{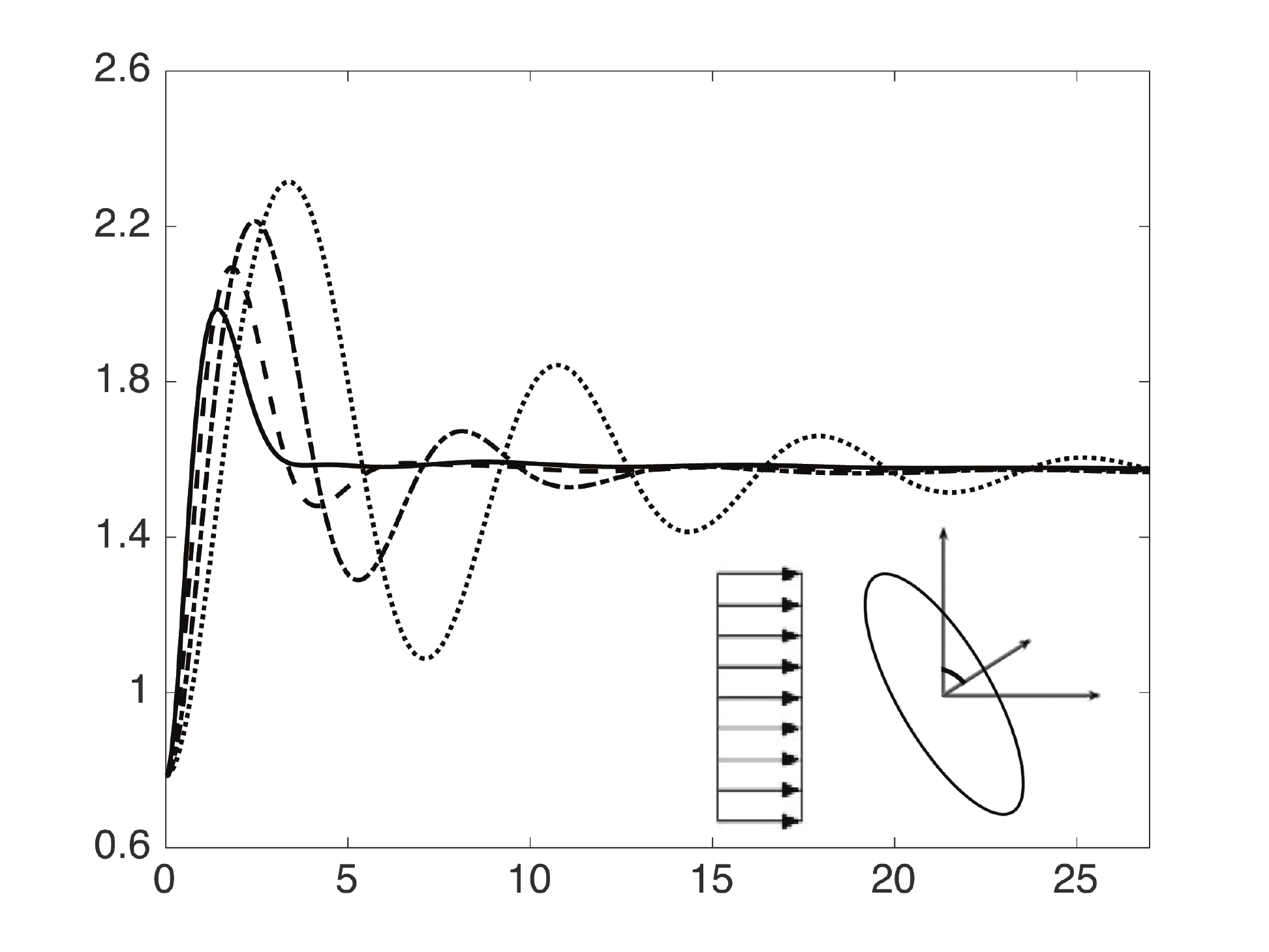}
\put(-306,105){\rotatebox{90}{ \large $\theta \, (rad)$}}
\put(-170,-3){{\large $t U_0 / D_{eq}$}}
\put(-118,200){{\large ----- $\,\rho_p/\rho_f = 2$}}
\put(-118,185){{\large - - - $\rho_p/\rho_f = 4$}}
\put(-118,170){{\large - . - $\rho_p/\rho_f = 8$}}
\put(-118,155){{\large . . . $\rho_p/\rho_f = 16$}}
\put(-73,76){{$\theta$}}
\put(-77,102){{$z$}}
\put(-43,56){{$y$}}
\put(-127,98){{$U_0$}}
\caption{\label{fig:8} 
Oscillating oblate in crossflow. Time evolution of the angle $\phi$  between the particle minor (symmetric) axis and the $z$ axis for different density ratios $\rho_p/\rho_f = 2$, $4$, $8$ and $16$, $Re_p = 100$ and the aspect ratio $\mathcal{AR}= 1/2.5$.} 
\end{figure}  

In this test case, the position of the centre is fixed while the particle is allowed to rotate freely around all three axes. Independent of the initial orientation, oblate particles align their semi-minor axis  with the flow direction. This is consistent with the findings by Feng et al. (1994) \cite{Feng1994} that elliptic particles fall with their major axis perpendicular to the gravity direction. As observed by Clift et al. (1978) \cite{Clift2005} and Kempe et al. (2009) \cite{Kempe2009}, given an initial deflection, the particle oscillates around one of its major axes (depending on the plane of deflection) and reaches a final equilibrium with its minor axis aligned with the flow. Moreover, by increasing the ratio of particle to fluid density, the period and the magnitude of the oscillations increase. 

The simulations are performed here for an oblate particle with $\mathcal{AR}= 1/2.5$ and different density ratios $\rho_p/\rho_f = 2$, $4$, $8$ and $16$ in a numerical domain of $15D_{eq} \times 100D_{eq} \times 15D_{eq}$ in the spanwise $x$, streamwise $y$, and wall-normal $z$ directions. The domain is periodic in the wall-parallel directions with two walls moving at same speed in the $y$ direction to create a uniform cross flow. The resolution is $32$ grid point per $D_{eq}$ and the particles Reynolds number, defined by the incoming flow velocity $U_0$ and the equivalent particle diameter $D_{eq}$,  $Re_p=100$. 
The initial deflection $\theta=\pi/4$, with $\theta$ the angle  between the particle minor (symmetric) axis and the $z$-axis. 
The evolution of  $\theta$ is reported in figure~\ref{fig:8} for different density ratios. It is observed, as expected, that the particle reaches an equilibrium with its major axis perpendicular to the flow direction after oscillations of the minor axis around the $x$-axis. The magnitude of oscillations increases with the density ratio.


\section{Results} 
\label{Results} 
We study the sedimentation of isolated and particle pairs in a viscous fluid. The results focus on the effect of shape and density ratio on the particle motion. 

\subsection{Sedimentation of isolated spheroids}

The sedimentation of isolated spheroids is simulated in
a domain of $15D_{eq} \times 15D_{eq} \times 125D_{eq}$ in the $x,y$ and $z$ directions, with gravity acting in the negative $z$ direction.  
Periodic boundary conditions are imposed in the horizontal directions whereas a free surface and a rigid wall are used at the upper and  bottom boundary. 
The equivalent particle diameter $D_{eq}=1.67 \times 10^{-3}m$, corresponding to the diameter of the spherical particles used in the numerical studies of Glowinski et al. (2001) \cite{Glowinski2001}, Sharma and Patankar (2005) \cite{Sharma2005} and Breugem (2012) \cite{Breugem2012}, and water is considered as the fluid. The volume of the particles is kept equal to $(1/6) \pi D_{eq}^3$, while the Galileo number $Ga$, varied by considering different density ratios ($\rho_p/\rho_f$).  We investigate spheroids with aspect ratios $\mathcal{AR}= 1/5,1/3,1,3$ and $5$ with a resolution of $32$ grid points per $D_{eq}$ for all cases except for the particles with aspect ratios $\mathcal{AR}= 1/5$ and $5$ where $48$ grid points per $D_{eq}$ are used. 
The resolution is higher than what is typically used for spherical particles \cite{Fornari2016,Picano2015,Uhlmann20142} to keep an adequate number points per semi-minor axis of the spheroid, which decreases with the aspect ratio. High grid resolution is also needed to capture the flow structures in the unsteady particle wake, especially at the highest settling speed. The spheroidal particle starts falling from rest with its major axis perpendicular to the falling direction. This orientation is chosen because other initial orientations are not stable. It is observed here, in agreement with findings in the literature \cite{Feng1994,Ern2012}, that a spheroidal particle eventually falls
 with its major axis perpendicular to the gravity direction independent of its initial orientation. 
  
\begin{figure}[t!]
\centering
\includegraphics[width=0.31\textwidth]{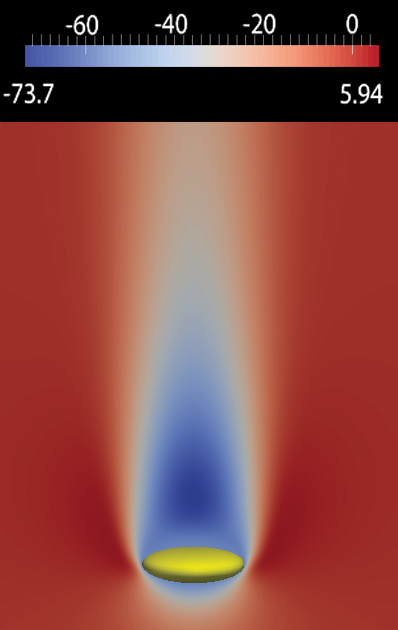} 
\put(-93,-15){\footnotesize (a) Oblate - $Ga=80$} \hspace{8pt}
\includegraphics[width=0.31\textwidth]{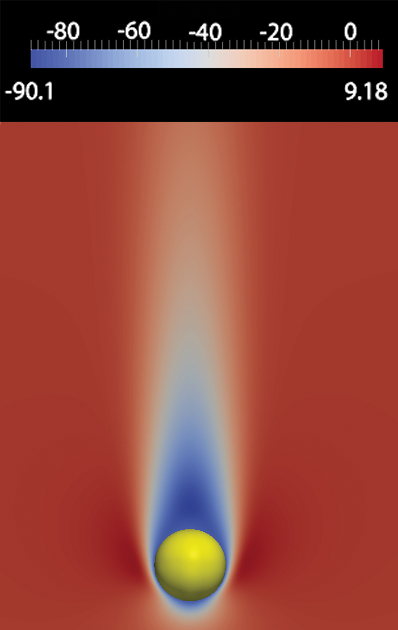} 
\put(-93,-15){\footnotesize (b) Sphere - $Ga=80$} \hspace{8pt}
\includegraphics[width=0.31\textwidth]{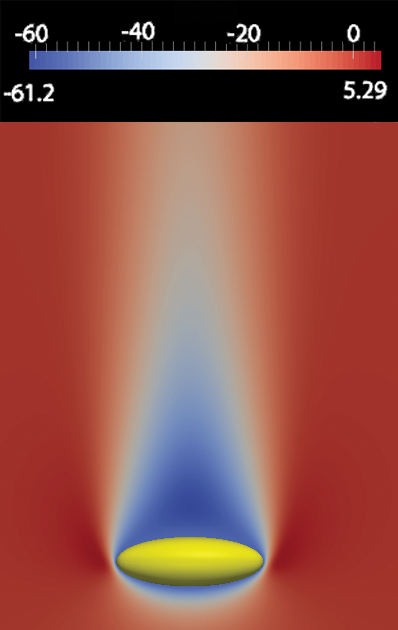} 
\put(-93,-15){\footnotesize (c) Prolate - $Ga=70$} \\ \vspace{15pt}
\includegraphics[width=0.31\textwidth]{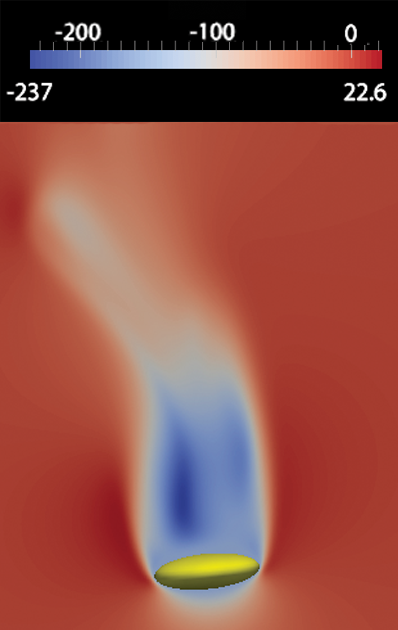} 
\put(-94,-15){\footnotesize (d) Oblate - $Ga=250$} \hspace{8pt}
\includegraphics[width=0.31\textwidth]{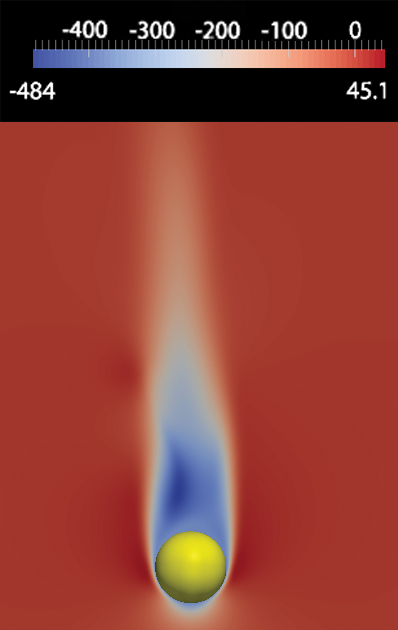} 
\put(-94,-15){\footnotesize (e) Sphere - $Ga=250$} \hspace{8pt}
\includegraphics[width=0.31\textwidth]{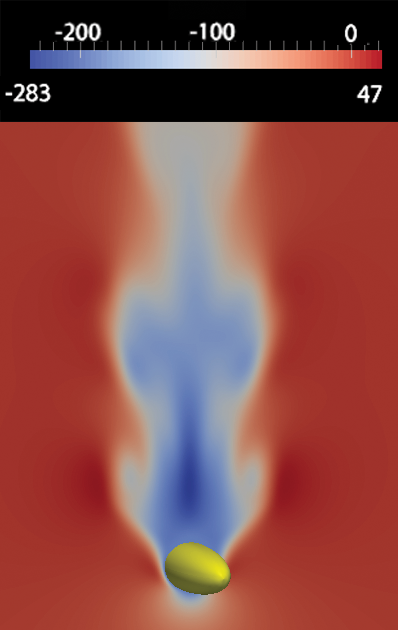} 
\put(-94,-15){\footnotesize (f) Prolate - $Ga=250$} 
\caption{\label{fig:9} Iso contours of vertical velocity, divided by $\nu/D_{eq}$, for stable and unstable wake behind settling spheroids of aspect ratios $\mathcal{AR}=1/3,1$ and $3$. The Galileo number is indicated in each plot.} 
\end{figure}

For isolated spheres, the steady-state settling velocity $u_t$ is often expressed in terms of a terminal Reynolds number, $Re_t \equiv u_t D_{eq} / \nu$. Empirical relations can be found in the literature to express $Re_t$ as function of  $Ga$. Yin \& Koch (2007) \cite{Yin2007}, among others, report the drag coefficient for isolated spheres as a function of $Re_t$, from which the relation between $Ga$ and $Re_t$ can be obtained as shown in Fornari et al. (2016) \cite{Fornari2016}:
\small
\hypertarget{eq:repga}{}
\begin{equation}
 Ga^2 = \left\lbrace
  \begin{array}{r@{}l}
    18 Re_t \left[ 1 + 0.1315 Re_t^{\left( 0.82 - 0.05 \log Re_t\right)} \right] \,\,\,\,\, , \,\,\, if \,\,\,  0.01 < Re_t \leq 20 \, , \\
    18 Re_t \left[ 1 + 0.1935 Re_t^{0.6305} \right] \,\,\, \,\,\, \,\,\, \,\,\, \,\,\, \,\,\, \,\,\, \,\,\, \,\,\, \,\,\, \, \, \, , \,\,\, if \,\,\,\,\,   20 < Re_t \leq 260 \, .
  \end{array}
  \right.
\label{eq:repga}  
\end{equation}
\normalsize

These relations are used to justify the length of our computational domain in the gravity direction ($125D_{eq}$).  Indeed the terminal velocity, $Re_t$, obtained at $Ga=80$ and $180$ differs by approximately $2 \%$ from the predictions using Eqs. (\ref{eq:repga})  ($Re_t=83$ and $243$, compared to the predicted values of $85$ and $248$). 

Jenny et al. (2004) \cite{Jenny2004} and Uhlmann \& Du\v{s}ek (2014) \cite{Uhlmann2014} study the sedimentation of a sphere in a viscous fluid
and find four different regimes. Below $Ga \approx 155$ a spherical particle settles steadily on a straight vertical path with an axisymmetric wake consisting of a single toroidal vortex. The wake becomes oblique (with planar symmetry) as the Galileo number increases above $155$, and the particle experiences a finite horizontal drift. A pair of thread-like quasi-axial vortices appear in this regime. For $Ga$ from approximately $185$ to $215$ the particle exhibits periodic oscillations and the wake becomes time-dependent, still preserving the planar symmetry;  the wake vortices evolve into a hairpin structure. Finally as $Ga$ further increases, the planar symmetry of the wake is broken and the particle follows a chaotic motion. 
Our results for spheres are consistent with the findings of \cite{Jenny2004,Uhlmann2014} and will not be reported here.

When considering the settling of isolated spheroids, we observe two different types of unsteady motion, different for oblates and prolates as $Ga$ exceeds the critical threshold for the first bifurcation.  Steady and unsteady wakes of spheroids with aspect ratios $\mathcal{AR}=1/3,1$ and $3$ are depicted in figure~\ref{fig:9}. The prolate particle rotates around the vertical ($z$) axis, while the oblate particle performs the so called zigzagging motion \cite{Mougin2006}. The details of the particle motions when increasing the Galileo number are discussed next for the oblate, $\mathcal{AR}=1/3$, and the prolate particle, $\mathcal{AR}=3$.

\subsubsection{Oblate particles}

The oblate particle, $\mathcal{AR}=1/3$, falls along a straight vertical path with an axisymmetric wake for $Ga \lesssim 130$ (corresponding to $Re_t \approx 103$).  For $Ga \gtrsim130$ the particle path is not vertical anymore, exhibiting an oscillatory motion. Fernandes et al. (2007) \cite{Fernandes2007} report a critical value of $Re_t=150$ for the onset of oscillatory motion of a circular disc based on the diameter of the disc.   
Using the major diameter to define $Re_t$, and not on the equivalent diameter, we obtain a value of $148.5$,  close to the findings in \cite{Fernandes2007} for discs with same $\mathcal{AR}=1/3$, and thus an additional validation for our numerical code. 

\begin{figure}[t]
\centering
\includegraphics[width=0.9\linewidth]{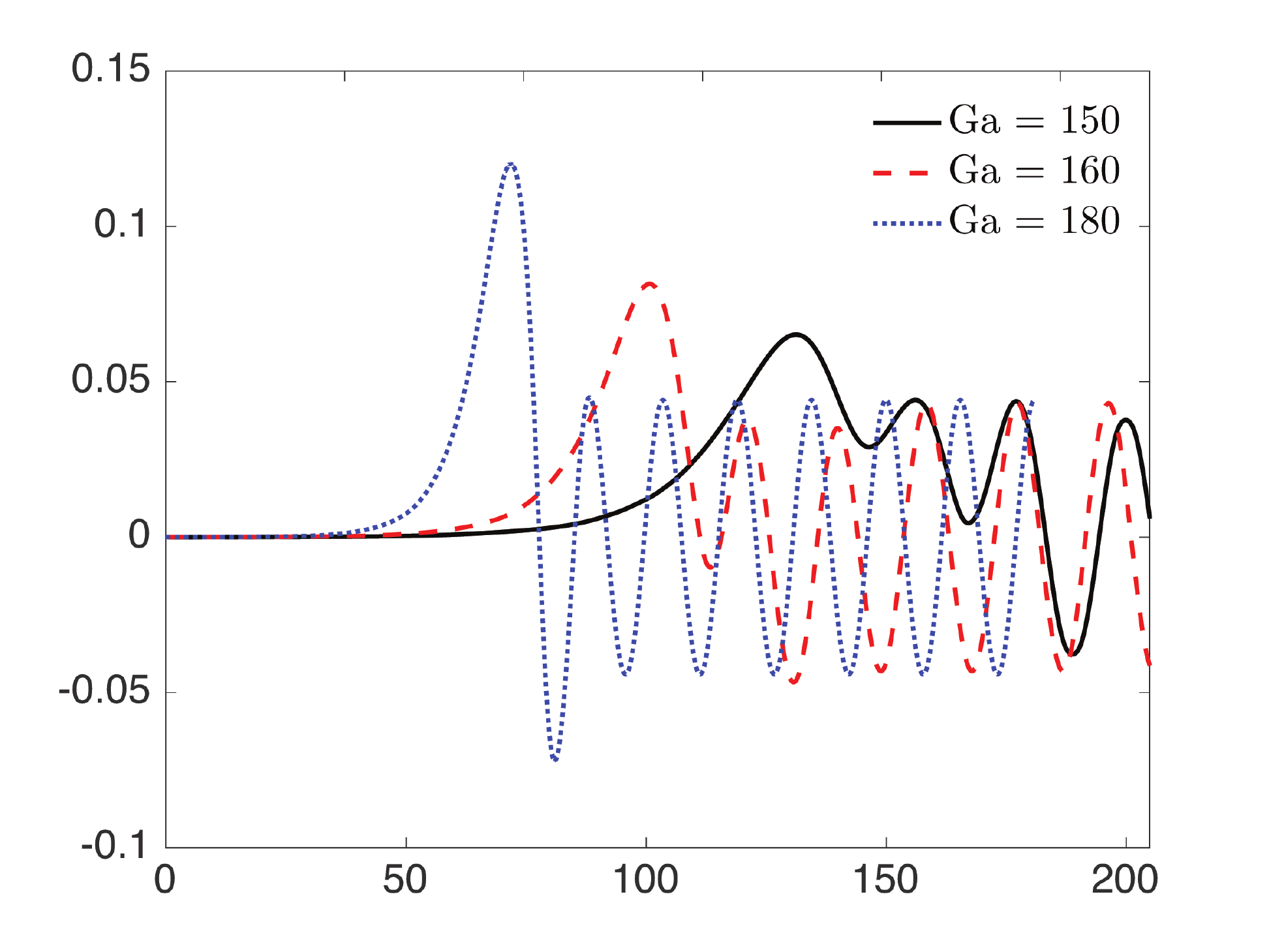}
\put(-310,105){\rotatebox{90}{ \large $V_n/u_{\,t}$}}
\put(-165,-3){{\large $t \, \sqrt{ g/D_{eq} } $}}
\vspace{-5pt}
\caption{\label{fig:10} 
Time history of the horizontal velocity $V_n$, normalized by the corresponding terminal settling velocity, for an oblate particle of aspect ratio 1/3 and three Galileo numbers, $Ga =150, 160$ and $180$.} 
\end{figure}

As $Ga$ exceeds the critical value of  $130$, the oblate particle experiences a horizontal drift in a random direction $n$. 
%
For $Ga$ numbers larger than $160$, the horizontal velocity soon oscillates 
around $0$, settling into a periodic motion. 
Interestingly, for $Ga$ numbers between $130$ and $160$ the oblate particle experiences an oscillation of the drift velocity around a positive value for some significant time interval before initiating the periodic motion. The length of this drift motion 
decreases in the presence of noise in the flow.
The time history of the horizontal velocity $V_n$ is shown in figure~\ref{fig:10} for different $Ga$  to document the different transients and the importance of the first shed vortex \cite{Ern2012}.

\begin{figure}[t]
\centering
\includegraphics[width=0.94\linewidth]{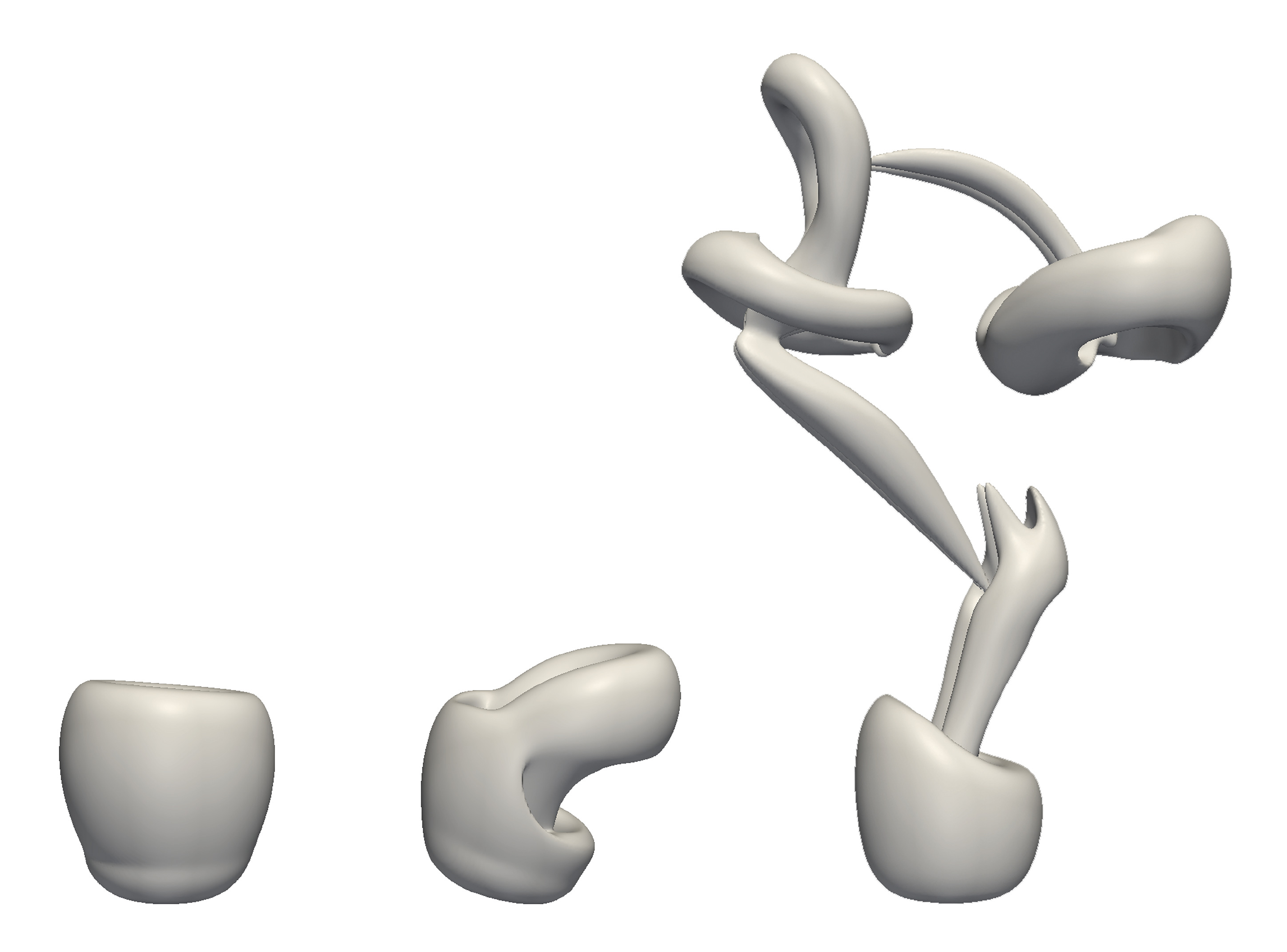}
\put(-230,205){{\large $(b)$}}
\put(-153,205){{\large $(c)$}}
\put(-320,205){{\large $(a)$}}
\vspace{-5pt}
\caption{\label{fig:11} 
Development of vortices in the wake of an oblate particle with $\mathcal{AR}=1/3$ and $Ga=180$. Iso-surfaces of Q-criterion equal to 5\% of its maximum are used to identify the vortices.} 
\end{figure}

A thorough discussion on the oscillatory paths of the disc-like cylinders and oblate spheroids can be found in Ern et al. (2012) \cite{Ern2012}. Magnaudet \& Mougin (2007)\cite{Magnaudet2007} and Yang \& Prosperetti (2007)\cite{Yang2007} relate the path instability to wake instabilities.
We therefore analyze the wake vortices to understand their relation to the particle motion. 
As shown in figure~\ref{fig:11}a), 
initially the wake of an oblate particle consists of a single toroidal vortex, attached to the particle, similar to that of spherical particles in the steady vertical regime.
As the instability develops, the particle rotates around one of its major-axes, perpendicular to gravity and to the horizontal direction in which it is drifting.  
When the angle with respect to the horizontal direction increases, a part of the toroidal vortex detaches forming the head of a hairpin vortex (see figure~\ref{fig:11}b); this soon develops  further  into a full hairpin structure. 
This vortex pushes the flow near and around the particle upwards, forming a low pressure region that generates a torque on the particle in the opposite direction. Owing to inertia, the oblate particle eventually reaches the opposite inclination. 
New hairpin vortices then detaches on the other side and so on each time the particle changes orientation (see figure~\ref{fig:11}c). The formation of these vortices is also discussed by Auguste et al. (2010)\cite{Auguste2010}. 
 For $Ga$ between $130$ and $160$ the first hairpin vortex detaching is significantly weaker than those observed for $Ga \gtrsim160$: the particle does not change its orientation and continues to drift in the same direction for some time, see figure~\ref{fig:10}. This confirms the importance of the strength of the first vortex, as discussed in Ern et al. (2012) \cite{Ern2012}.

\begin{figure}[t!]
\centering
\includegraphics[width=0.9\linewidth]{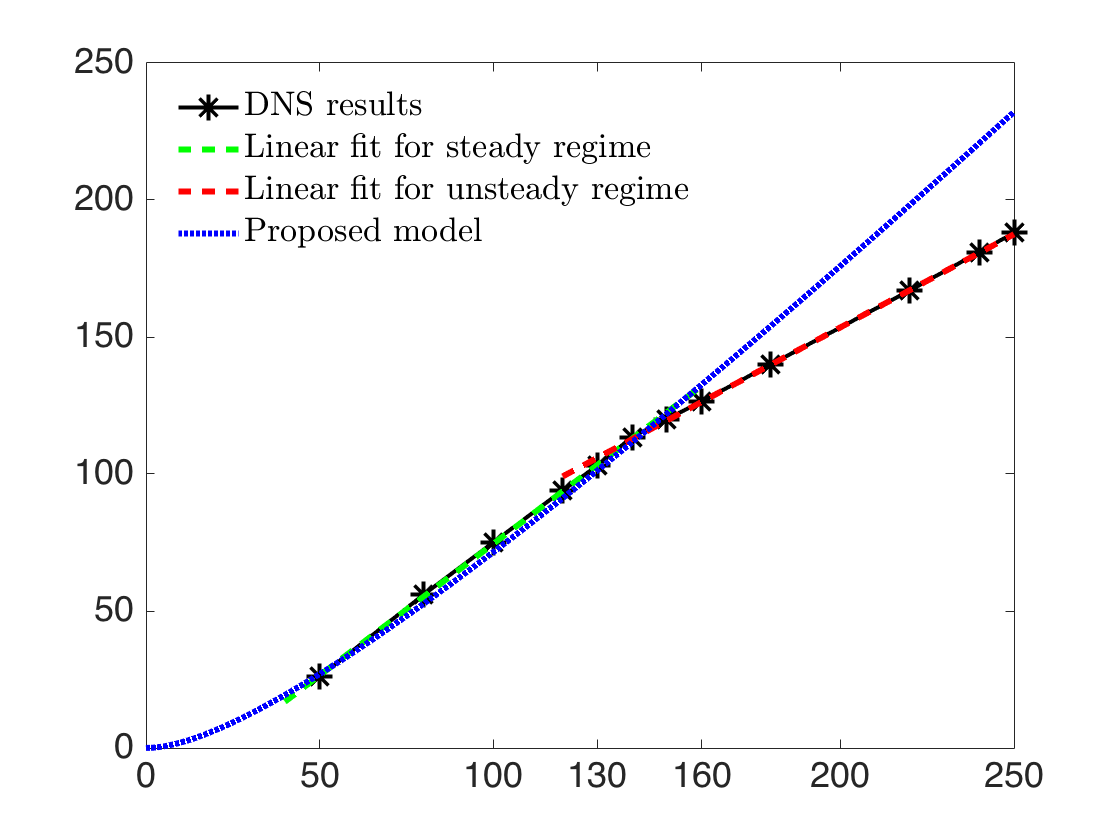}
\put(-310,110){\rotatebox{90}{ \large $Re_t$}}
\put(-155,-3){{\large $Ga$}}
\put(-245,40){\rotatebox{37}{ \large $\textcolor{black}{Re_t = 0.96Ga - 21.5}$}}
\put(-135,98){\rotatebox{27}{ \large $\textcolor{black}{Re_t = 0.68Ga + 17.5}$}}
\vspace{-5pt}
\caption{\label{fig:12} 
Terminal Reynolds number $Re_t$, versus $Ga$ for oblate spheroids with $\mathcal{AR}=1/3$. Linear fitting are also reported for the steady and unsteady regimes. The blue dotted line in the figure indicates the predictions from the proposed model.} 
\end{figure}

As the oblate particle experiences these oscillations, its vertical velocity decreases during the transient, which can be explained by conservation of energy of the system. Since the first and the second hairpin vortex are the two strongest in terms of their magnitude, the particle experiences two sudden decelerations before settling to the final regime, characterised by  oscillation of its terminal velocity of the order of $1-2\%$ of its settling speed. The terminal Reynolds number, $Re_t$, based on the averaged settling speed, is depicted for different $Ga$ numbers in figure~\ref{fig:12} where we also report linear fitting of the data in the steady and unsteady regimes.  
The slope  observed in the steady regime ($Ga \lesssim 130$) smoothly changes and reduces in the unsteady configurations ($Ga \gtrsim130$). A simple model is proposed to predict the terminal Reynolds number for spheroidal particles at low Galileo numbers based on the assumption that for oblates, spheres and prolate particles the steady flow (wake) regime is similar and only the frontal surface area differs. As shown in figure~\ref{fig:12} the model provides a good estimate up till $Ga \approx 130$ where the flow regime is steady and the particle path is vertical. 
The details of the suggested model can be found in appendix \hyperlink{appA}{A}. 

\subsubsection{Prolate particles}

\begin{figure}[t]
\centering
\includegraphics[width=0.9\linewidth]{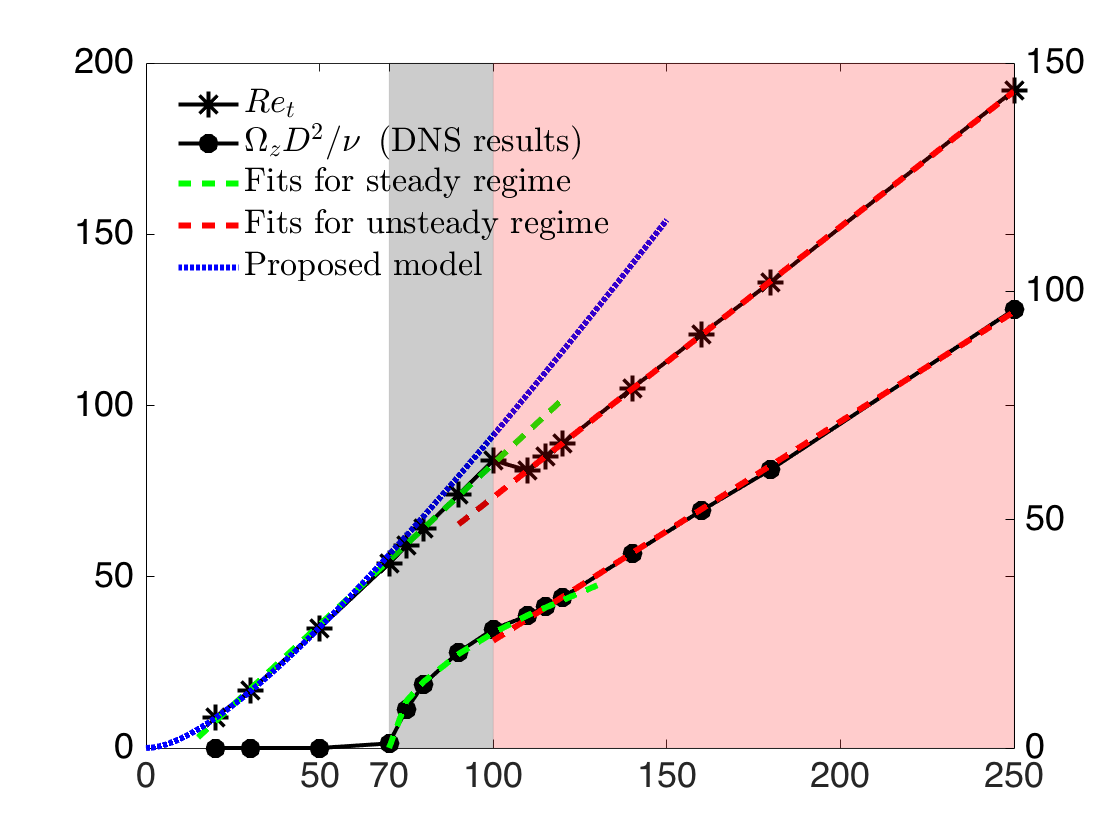}
\put(-305,110){\rotatebox{90}{ \Large $Re_t$}}
\put(-15,145){\rotatebox{-90}{ \Large $\Omega_z D^2 /\nu$}}
\put(-155,-3){{\Large $Ga$}}
\put(-257,40){\rotatebox{43}{ \large $\textcolor{black}{Re_t = 0.94Ga - 11}$}}
\put(-145,103){\rotatebox{38}{ \large $\textcolor{black}{Re_t = 0.79Ga - 5.8}$}}
\put(-150,55){\rotatebox{33}{ \large $\textcolor{black}{\Omega_z D^2 /\nu= 0.48Ga - 24.5}$}}
\put(-225,30){\rotatebox{38}{ \large $\textcolor{black}{4.6\sqrt{Ga - 70}}$}}
\vspace{-5pt}
\caption{\label{fig:13} 
Terminal Reynolds number $Re_t$ and terminal angular velocity $\Omega_z$, divided by $\nu/D^2$, versus $Ga$ for prolate spheroids with $\mathcal{AR}=3$. The results of proposed model is shown by the blue dotted line. White background indicates the steady regime, grey refers to the regime where the particle rotates around the vertical axis and the wake consists of four thread-like quasi-axial vortices and pink displays the regime with spiral wake structures (see figure~\ref{fig:14}).} 
\end{figure}

\begin{figure}[t]
\centering
\includegraphics[width=0.86\linewidth]{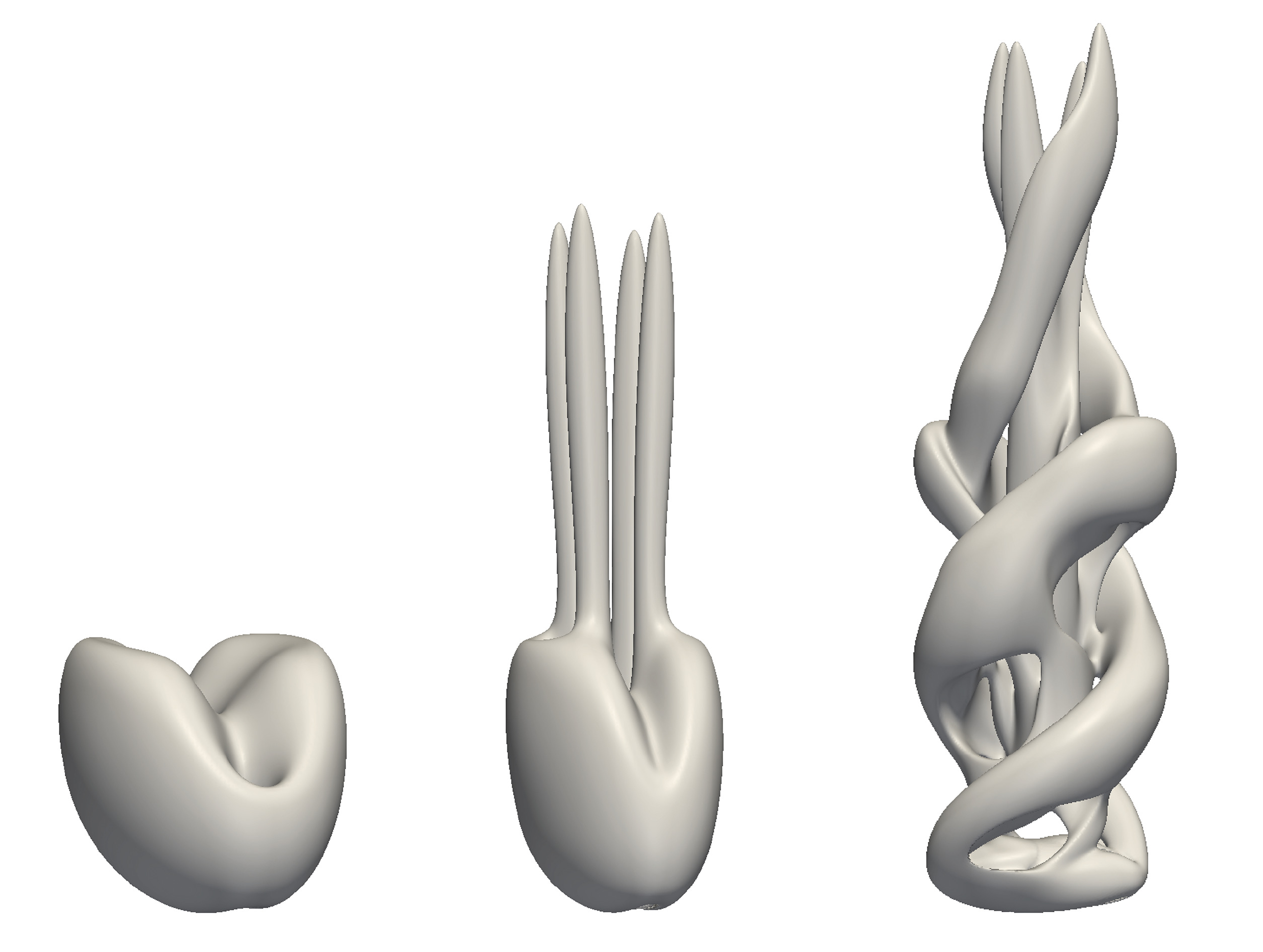}
\put(-202,185){{\large $(b)$}}
\put(-99,185){{\large $(c)$}}
\put(-300,185){{\large $(a)$}}
\vspace{-15pt}
\caption{\label{fig:14} 
Vortical structures in the wake of a prolate particle with $\mathcal{AR}=3$ for different Galileo numbers corresponding to the different regimes presented in figure~\ref{fig:13}. Iso-surfaces of Q-criterion equal to 5\% of its maximum are used to identify the vortices at Galileo numbers $60$, $80$ and $180$, respectively.} 
\end{figure}

The onset of secondary motions for the prolate particle with $\mathcal{AR}=3$ is observed at considerably lower Galileo numbers. The settling particle is found to rotate around the vertical direction, $z$-axis, for $Ga$ exceeding the critical value of $70$. 
The terminal Reynolds number, $Re_t$,  and the $z$--component of the angular velocity, $\Omega_z$, are depicted in figure~\ref{fig:13} as function of $Ga$. We report here the regime velocities, reached after an initial transient corresponding to a falling distance of about $50$ $D_{eq}$. 
The dependence of the  terminal velocity on the Galileo number can be approximated by a line for $Ga <100$, even when the particle undergoes rotation. 
The slope is found to decrease for $Ga>100$: here $Re_t$ displays a sudden decrease, which we will explain below by looking at the flow in the particle wake. 
As for the oblate, assuming the drag depends only on the frontal area fits the data nicely at lower $Ga$, see appendix \hyperlink{appA}{A}. 
For $Ga\in [70,100]$, the angular velocity increases from zero as $\sqrt{Ga-70}$ before settling to a linear law as $Ga$ exceeds $100$.

To better understand the sudden drop of the vertical and angular velocity at $Ga \approx100$, we study the structure of the wake
behind the prolate particle for the regimes indicated by the different background colours in figure~\ref{fig:13}. 
As shown in figure~\ref{fig:14} a) the wake is steady and symmetric for $Ga < 70$,  consisting of two recirculation regions at the sides. This regime corresponds to the white colour in figure~\ref{fig:13}.
As soon as the particle rotates around the gravity direction four thread-like quasi-axial vortices appear in its wake (figure~\ref{fig:14}b); these vortices are observed for $70<Ga<100$. This regime, indicated by grey colour in figure~\ref{fig:13}, is very 
sensitive to external perturbations and indeed any small noise such as the presence of another particle or the vicinity of a wall can trigger an instability. This is eventually observed for $Ga>100$ in the form of helical vortices (see figure~\ref{fig:14}c representing the flow at $Ga=180$). 
The flow kinetic energy increases suddenly when the wake becomes helical, resulting in the reduction of  the particle vertical velocity shown above. The drop in $Re_t$ ( cf.\ figure~\ref{fig:13}) can therefore be explained by the instability of the vortices in the wake. This
occurs at $Ga=100$ in the simulations presented here, without any additional external noise.  
As mentioned above, however, the exact value of $Ga$ at which the helical vortices become unstable depends on the ambient noise, suggesting that we are in the presence of a subcritical instability.

\begin{figure}[t]
\centering
\includegraphics[width=0.85\linewidth]{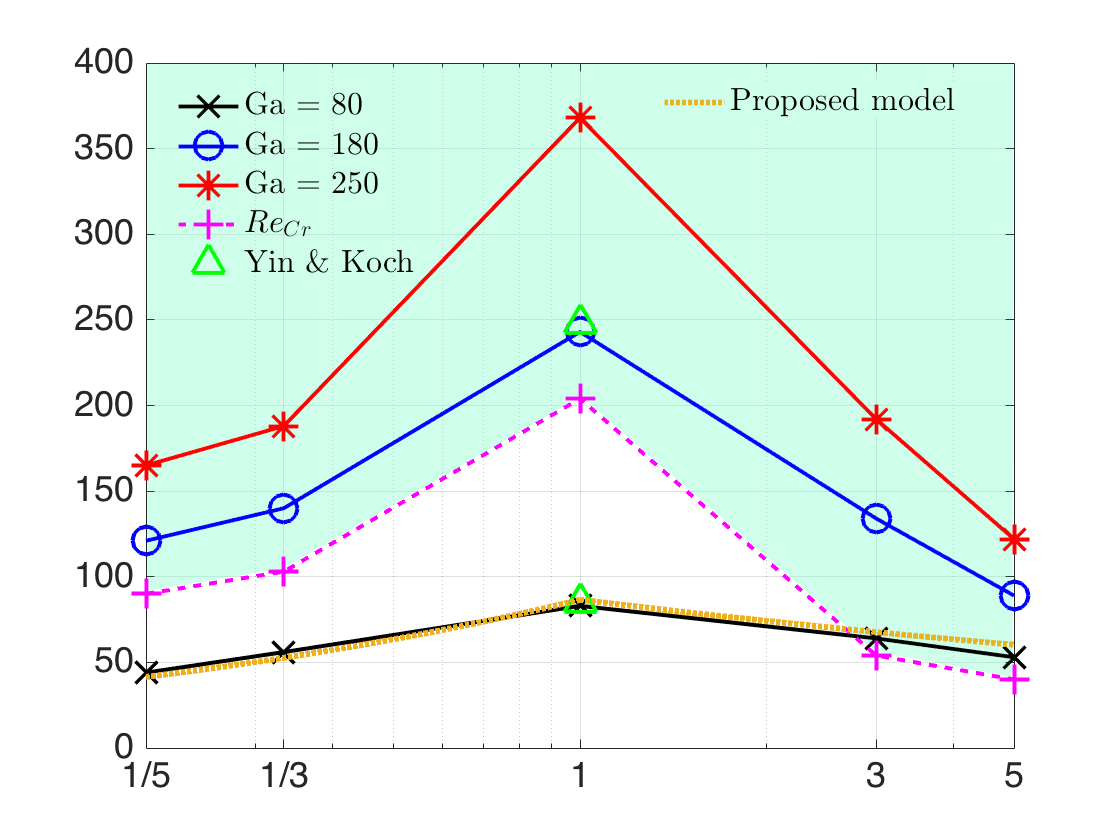}
\put(-295,100){\rotatebox{90}{ \Large $Re_t$}}
\put(-154,0){{\Large $\mathcal{AR}$}}
\vspace{-15pt}
\caption{\label{fig:15} 
Terminal Reynolds number $Re_t$, versus the aspect ratio  $\mathcal{AR}$ for three different Galileo numbers.  The region where the particles experience an unsteady  motion is indicated by the light green background.} 
\end{figure}

The effect of the particle shape on the terminal velocity and on the onset of secondary motions is also investigated for particles with aspect ratios $\mathcal{AR}=1/5$ and $5$. The terminal Reynolds numbers at $Ga= 80, 180$ and 250 are depicted in figure~\ref{fig:15} as function of the spheroid aspect ratio; the $Re_t$ predicted by the empirical relation in Yin \& Koch (2007) \cite{Yin2007} for spheres is also indicated in the figure for the two lowest $Ga$. 
The critical Reynolds number $Re_{cr}$ above which the particles undergo an unsteady motion is determined and indicated by the light green background. 
Finally, we also report the prediction of the simple model 
assuming the settling speed can be directly related to the frontal area for $Ga=80$, when the wake is quasi-steady. 
We notice good agreement for oblate particles and a slightly lower accuracy 
for prolate spheroids, which can be explained by the rotational motion they already experience at $Ga=80$. 
Spherical particles have the largest settling speed since the sphere corresponds to the object of minimum area perpendicular to the settling direction for a given volume.
If the wakes are quasi-steady for all aspect ratios, $Ga=80$, the minimum cross-section can therefore explain the maximum terminal velocity.
This explanation, however, does not hold at higher $Ga$ when the particle motion and the flow become unsteady.


\subsection{Drafting-Kissing-Tumbling (DKT) of spheroids}

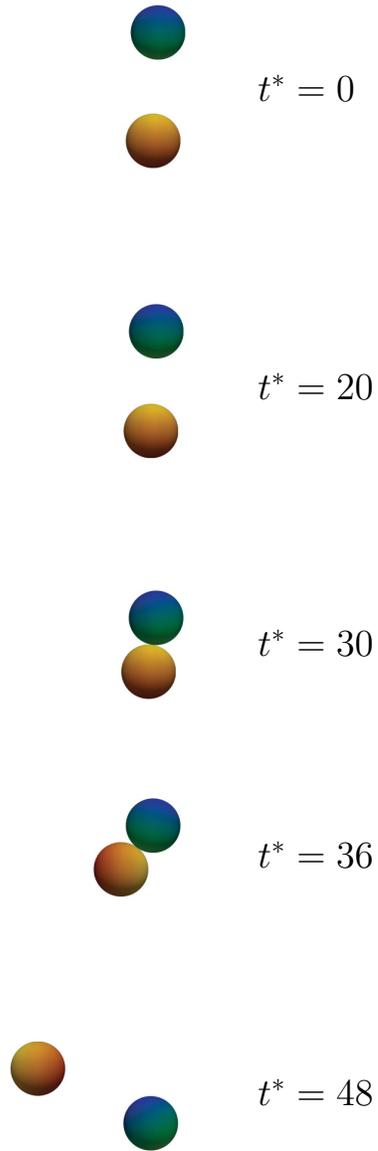
\begin{figure}[p]
\centering
\includegraphics[width=0.4\linewidth]{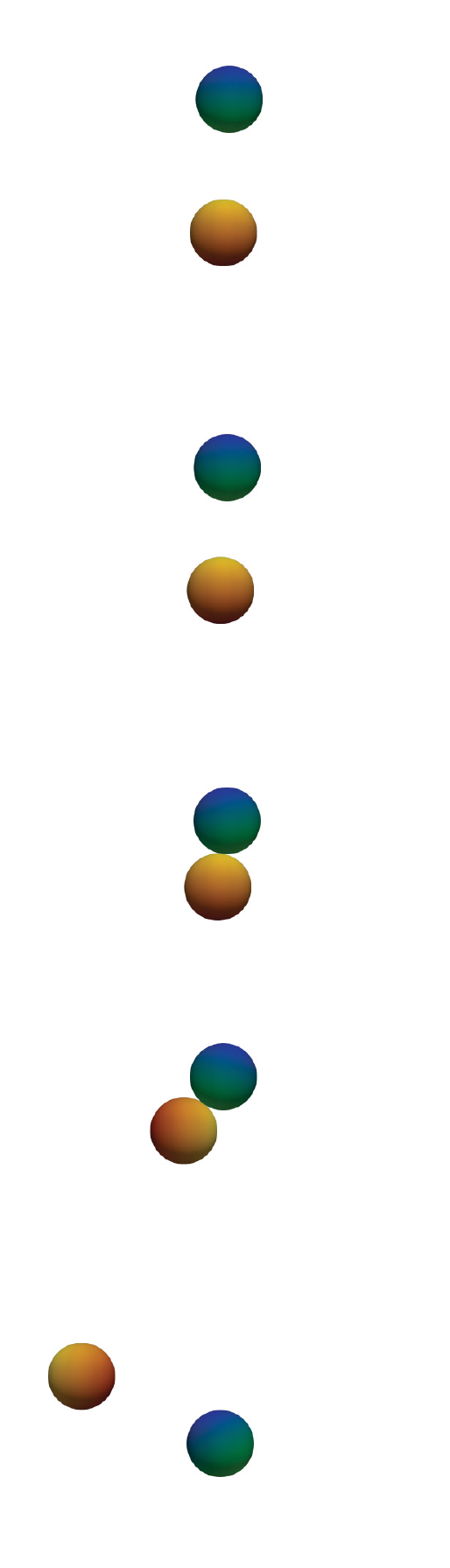}
\put(-30,45){{\Large $t^* = 48$}}
\put(-30,135){{\Large $t^* = 36$}}
\put(-30,215){{\Large $t^* = 30$}}
\put(-30,312){{\Large $t^* = 20$}}
\put(-30,425){{\Large $t^* = 0$}}
\vspace{-15pt}
\caption{\label{fig:16} 
Sequence of Drafting-Kissing-Tumbling (DKT) of two equal spheres from visualizations at non-dimensional times $t^* = 0$, $20$, $30$, $36$ and $48$.}
\end{figure}

Next, we study \textit{Drafting-Kissing-Tumbling} (DKT) of spheroidal particles with different aspect ratios. This peculiar pair interaction has been studied for two equal spheres both experimentally \cite{Feng1994,Fortes1987} and numerically \cite{Breugem2012,Glowinski2001,Patankar2000}.
The process is reproduced here in figure~\ref{fig:16} from our simulations with two equal spheres. The results are reported in non-dimensional time $t^* = t \sqrt{D_{eq}/g}$.
 Initially, the trailing particle is attracted into the wake of the leading one and drafted towards it with increasing velocity (drafting phase), until they 
are in contact (kissing phase). The particles in contact form a long body with its major axis parallel to gravity. As discussed previously, this orientation is unstable, as a long body tends to fall with its major axis perpendicular to the falling direction. The two particles therefore tumble \cite{Prosperetti2007} (tumbling phase).

We consider now non-spherical particles with the same volume as those in the numerical studies of Glowinski et al. (2001)  \cite{Glowinski2001}.
 Simulations are performed at $Ga = 80$, corresponding to a density ratio of $1.14$ and $D_{eq}=$  1.67 mm for spheroids with aspect ratios $1$, $3$ and $1/3$. 
 The corresponding terminal Reynolds number are $83$, $64$ and $53$ respectively, see figure \ref{fig:15}.

\begin{table}[t]
\centering
\begin{tabular}{l | *{6}{c}}
Case & $1$ & $1/3$ & $3$-$0^{\circ}$ & $3$-$45^{\circ}$ & $3$-$90^{\circ}$ \\
\hline
Aspect ratio & $1$ & $1/3$ & $3$ & $3$ & $3$ \\ 
Initial orientation of $P_1$ & --- & $(0,0,1)$ & $(1,0,0)$ & $(1,0,0)$ & $(1,0,0)$\\ 
Initial orientation of $P_2$ & --- & $(0,0,1)$ & $(1,0,0)$ & $(\sqrt{2}/2,\sqrt{2}/2,0)$ & $(0,1,0)$
\end{tabular} 
\caption{\label{table:1}Initial orientations of the spheroids at $t^*=0$ for all studied cases. The orientation vector is defined by the direction of the particle symmetry axis.} 
\end{table}

\begin{figure}[t]
\centering
\includegraphics[width=0.2\textwidth]{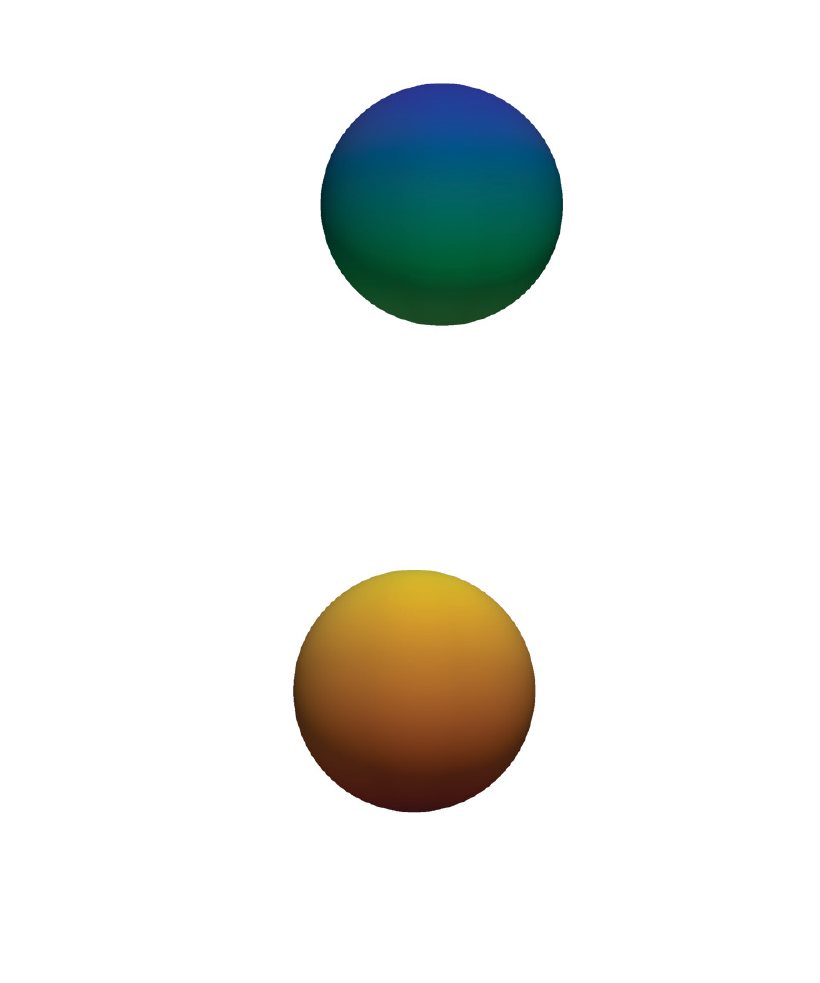} \hspace{-10pt} 
\put(-33,-18){\footnotesize (a)} 
\includegraphics[width=0.2\textwidth]{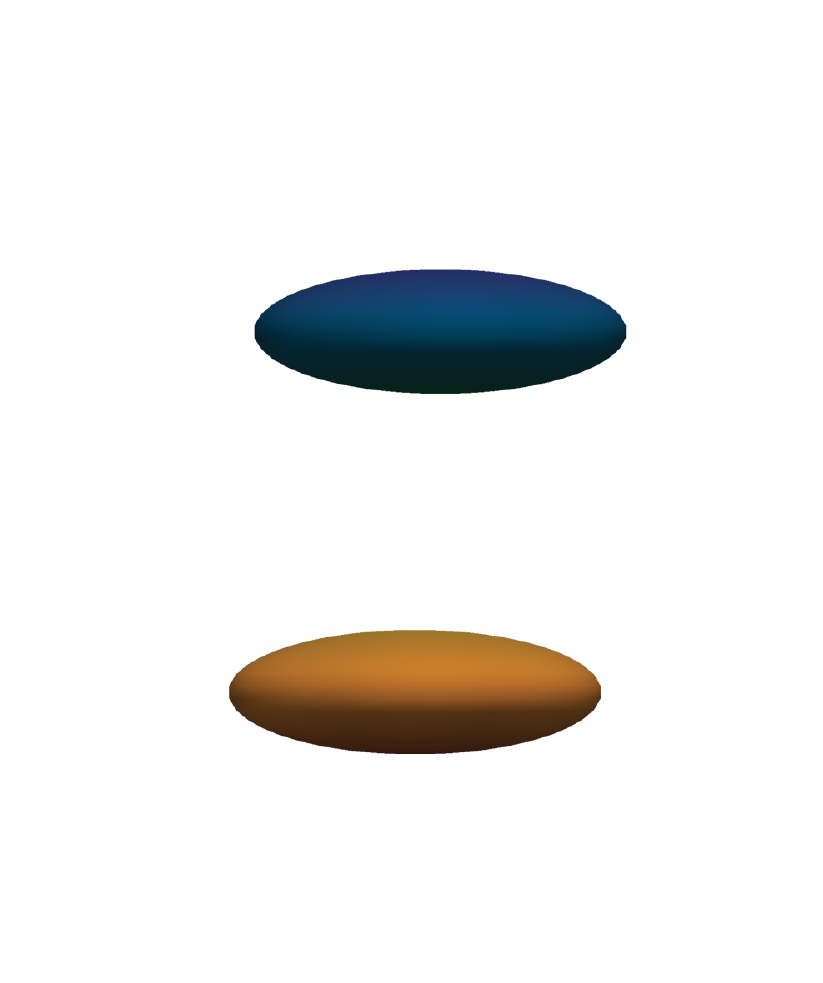} \hspace{-10pt} 
\put(-33,-18){\footnotesize (b)}
\includegraphics[width=0.2\textwidth]{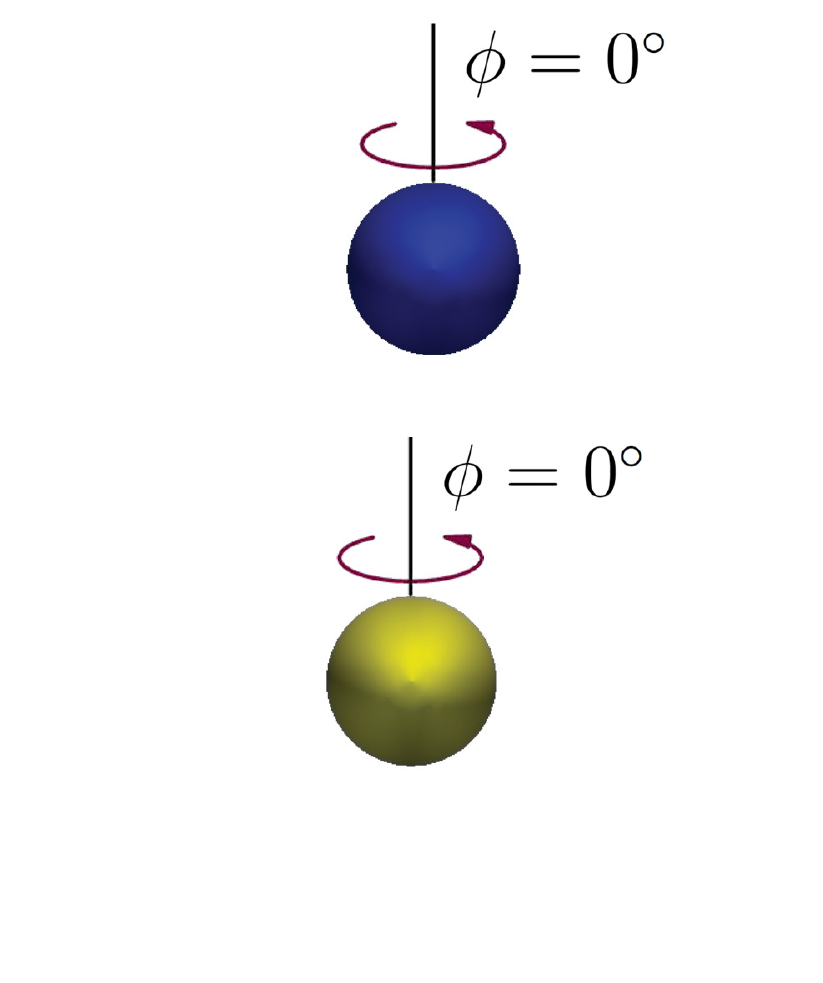} \hspace{-10pt} 
\put(-33,-18){\footnotesize (c)}
\includegraphics[width=0.2\textwidth]{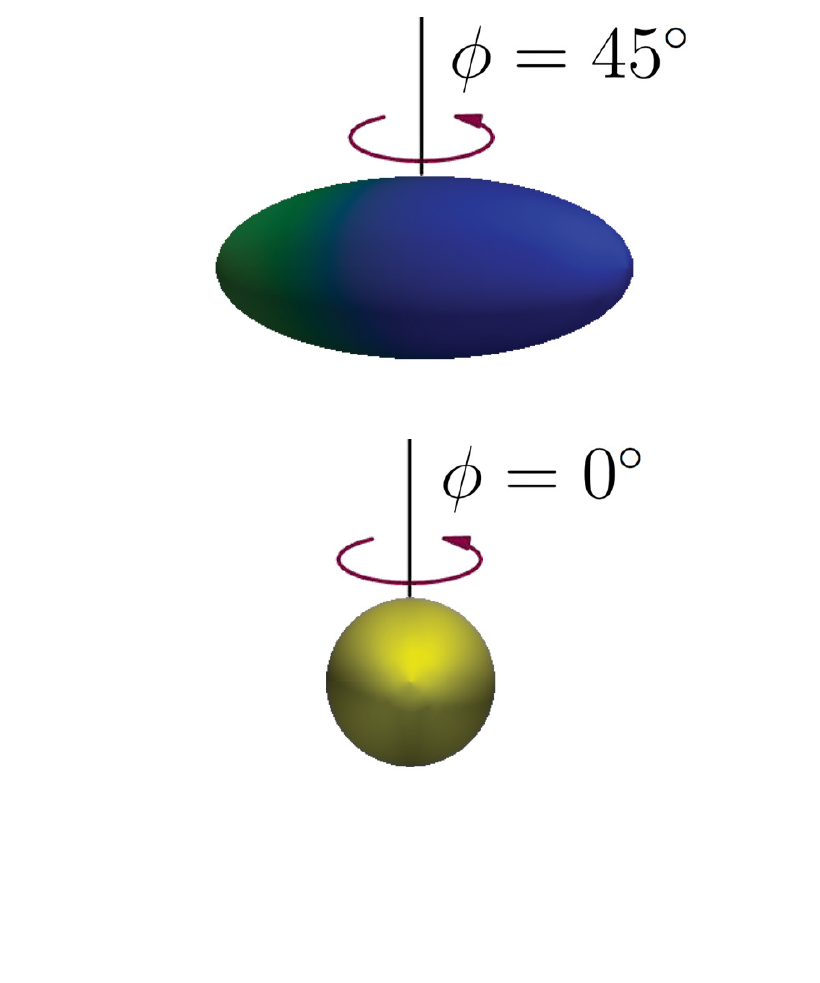} \hspace{-10pt} 
\put(-33,-18){\footnotesize (d)}
\includegraphics[width=0.2\textwidth]{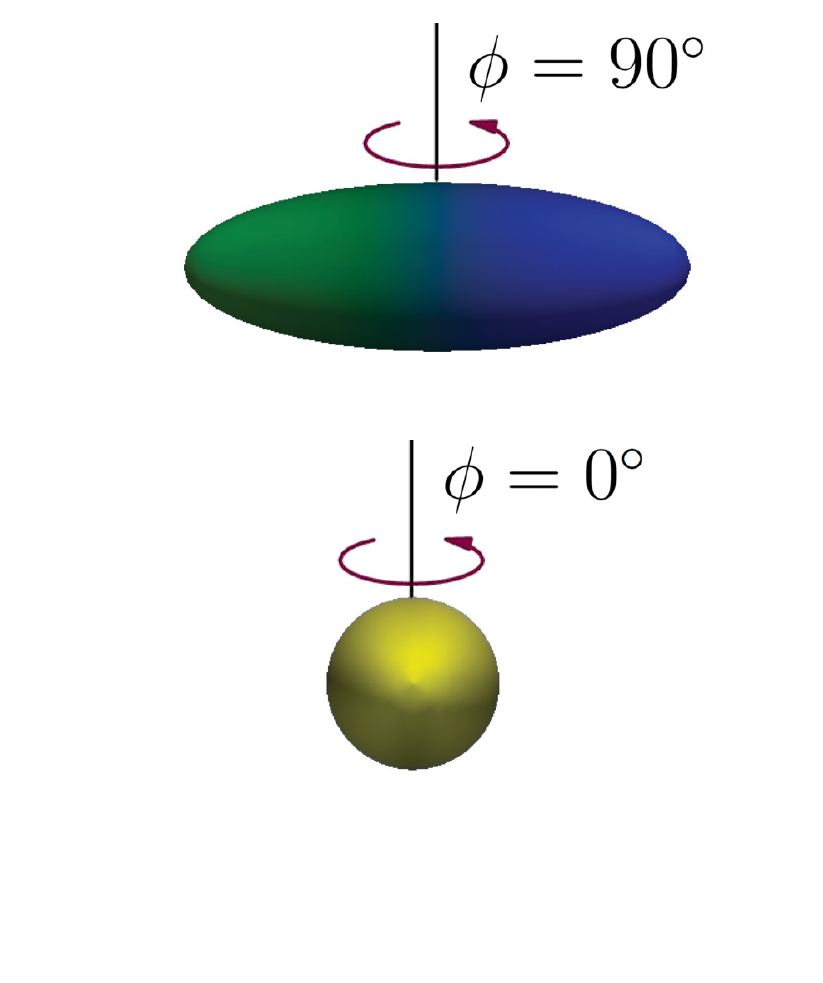} \hspace{-10pt}
\put(-33,-18){\footnotesize (e)} 
\vspace{15pt}
\caption{\label{fig:17} Initial particle configuration for all studied cases a) $1$, b) $1/3$, c) $3$-$0^{\circ}$, d) $3$-$45^{\circ}$ and e) $3$-$90^{\circ}$ projected in the $y-z$ plane, with the $z$ parallel to gravity. $\phi$ is the angle between the symmetric axis of the prolate particle and the $x$-direction, normal to the page.} 
\end{figure}

 The two particles start from rest and with their stable orientation (major-axis perpendicular to the falling direction). 
The initial orientation of the spheroids, defined by the direction of the symmetry axis, are given in table~\ref{table:1}  for all $5$ cases under investigation. For oblate and spherical particles, the orientation is in the vertical direction whereas  it is in the horizontal plane for prolates. 
Figure~\ref{fig:17} shows the initial position and orientation.  
 For prolate particle pairs, among all possible initial conditions, we vary the relative angle between the projection of the major axis in a plane perpendicular to gravity. 
 Three cases with angles of $0^{\circ}$, $45^{\circ}$ and $90^{\circ}$ are investigated, see figure~\ref{fig:17} c)-e) where colors are used only for a better visualization. The initial position of the centre of the leading particle, denoted as $P_1$, is set to $0.5L_x$, $0.5L_y$ and $0.8L_z$, where  $L_x$, $L_y$ and $L_z$ are the dimensions of the numerical domain. The trailing particle, denoted as $P_2$, is above $P_1$ at a vertical distance between the particle surfaces equal to $D_{eq}$. An offset of $0.1D_{eq}$ is introduced in the horizontal direction ($y-$direction for the sake of clarity) to trigger the DKT \cite{Breugem2012,Feng1994}. 
 To be able to detect the particle interactions, a relatively high resolution of $48$ grid cells per equivalent diameter is chosen. The boundary conditions and the dimensions of the computational domain are those used for a single sedimenting particle, except in the gravity direction where the length is reduced to $L_z= 45D_{eq}$.

\begin{figure}[t]
\centering
\includegraphics[width=0.9\linewidth]{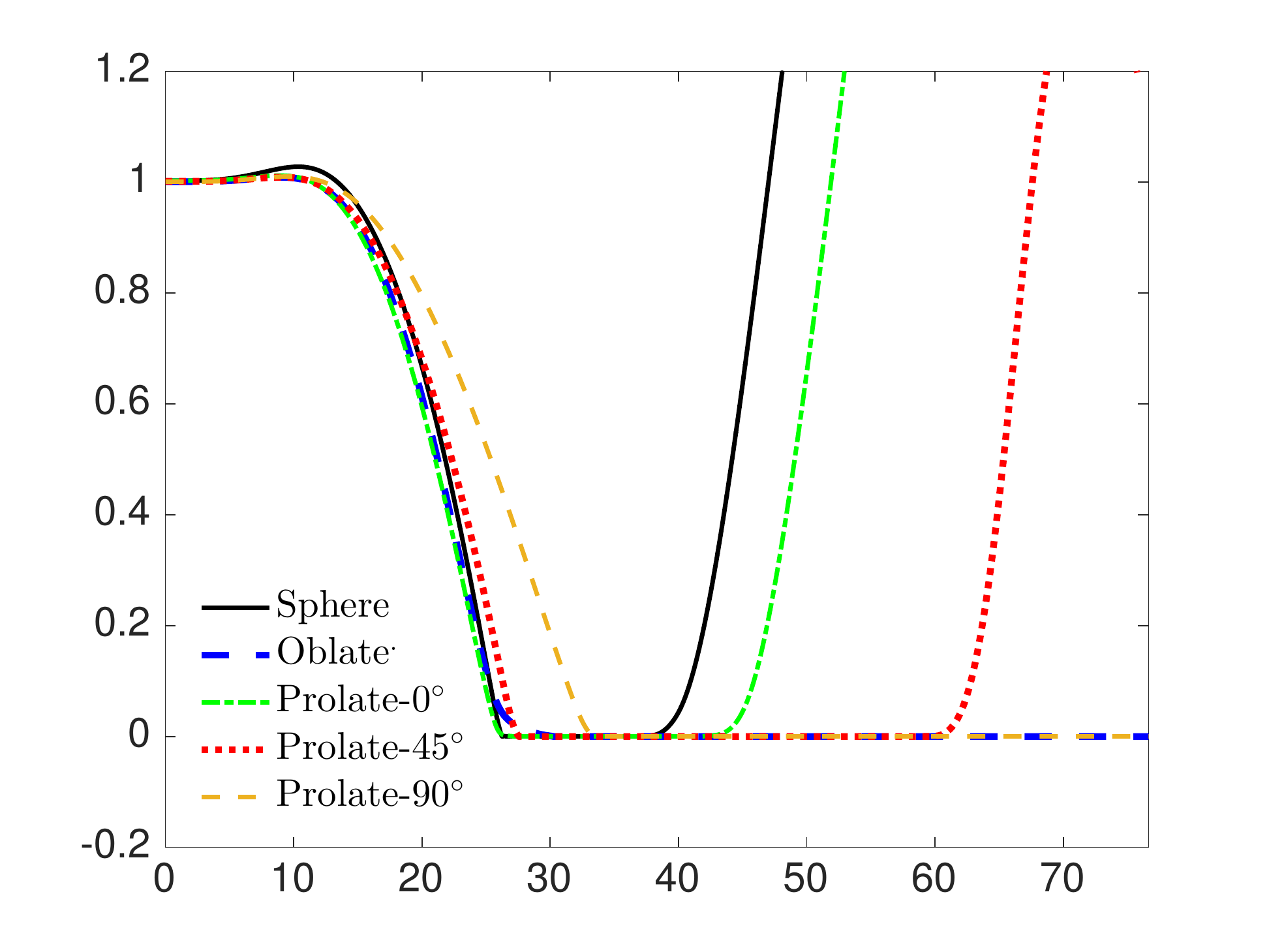}
\put(-310,110){\rotatebox{90}{ \Large $d_{12}$}}
\put(-160,0){{\Large $t^*$}}
\vspace{-5pt}
\caption{\label{fig:18} 
Nearest distance $d_{12}$ between the surfaces of the particle pairs, normalized by the equivalent diameter $D_{eq}$, versus non-dimensional time $t^*$ for all studied cases.} 
\end{figure}

The time history of the nearest distance between the two settling particles is reported in figure~\ref{fig:18}: the particle shape indeed alters the DKT and the tumbling disappears in some cases. %
For the oblate pair and the prolate with $90^{\circ}$ angle between the major axes of the two particles, the tumbling phase disappears and the particles continue in contact until they hit the bottom wall. 
The time duration of the drafting and kissing phases 
are listed in table~\ref{table:2} together with the increase
of the maximum vertical velocity of the trailing particle 
with respect to the case of an isolated particle.

\begin{table}[t]
\centering
\begin{tabular}{l | *{6}{c}}
Case & $1$ & $1/3$ & $3$-$0^{\circ}$ & $3$-$45^{\circ}$ & $3$-$90^{\circ}$ \\
\hline
Velocity Increase of $P_2$ & $46.68\%$ & $51.06\%$ & $49.64 \%$ & $48.19\%$ & $31.21\%$ \\ 
Drafting period & $26.25$  & $30.29$ & $27.32$ & $27.57$ & $33.91$ \\ 
Kissing period  & $10.16$ & $\infty$ & $13.27$ & $30.78$ & $\infty$ 
\end{tabular} 
\caption{\label{table:2} Increase of the maximum vertical velocity of the trailing particle, compared to an isolated particle, and time durations of the drafting and kissing phase for the cases considered. The values are reported in non-dimensional time $t^* = t \sqrt{D_{eq}/g}$.} 
\end{table}

\begin{figure}[t]
\centering
\includegraphics[width=0.9\linewidth]{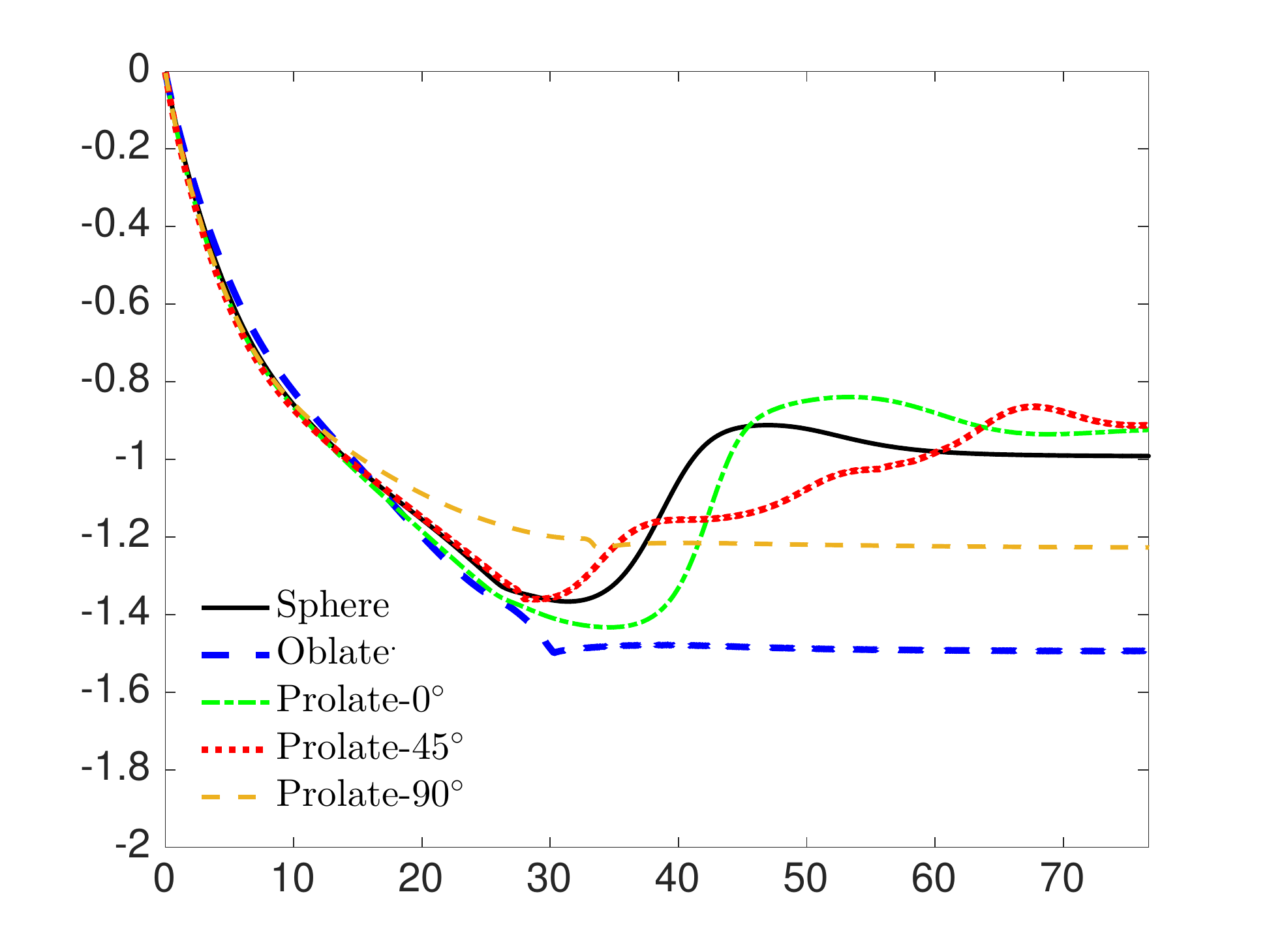}
\put(-310,90){\rotatebox{90}{ \Large $V_{avg}/ |u_{\,t}|$}}
\put(-160,0){{\Large $t^*$}}
\caption{\label{fig:19} 
Time evolution of the average vertical velocity of the two particles, normalized by the absolute value of terminal velocity of an isolated particle, for all cases studied.} 
\end{figure}

This increase of the velocity of the trailing particle 
depends on the overlap with the wake of $P_1$. 
For the cases denoted as $3$-$0^{\circ}$ and $3$-$90^{\circ}$, the particles preserve the angle between their major axes while drafting. 
For case the $3$-$45^{\circ}$, instead, $P_2$ starts rotating in the drafting phase, reducing the angle between the major axes of the two to about $12^{\circ}$. 
More details about the secondary motions of the particles are given later in this section. 
The reduced difference in the velocity of $P_2$ between cases $3$-$0^{\circ}$ and $3$-$45^{\circ}$ is thus due to the rotation of $P_2$, which reduces the relative angle and increases the overlap with the wake of $P_1$; therefore the two cases are similar in terms of overlap in the drafting phase. 
The velocity of the leading particle $P_1$ also increases as $P_2$ approaches owing to the lubrication forces between the particles. 
The duration of the drafting phase is the longest for case $3$-$90^{\circ}$ due to the minimum overlap between $P_2$ and the wake of $P_1$. 
This phase is also relatively long for the oblate case despite of the maximum increased velocity of $P_2$; we attribute this to the lubrication forces between the oblate particles just before the kissing phase.

Particle pair interactions affects the statistics of settling suspensions in the dilute regime, as shown by the intermittent behaviour reported in \cite{Fornari2016} for spherical particles.  The average vertical velocity of the two particles, normalized by the terminal velocity of an isolated particle, is therefore depicted  versus time in figure~\ref{fig:19}. 
For spheres, the average of the two particle velocities first increases to $\approx 1.4$ and then converges to $1$ as the particles starts the tumbling phase, meaning that the interaction between the two does not affect the vertical velocity after the particles move apart from each other. 
For the prolate particles, cases $3$-$0^{\circ}$ and $3$-$45^{\circ}$, the average vertical velocity converges to approximately $0.92$, lower than the terminal velocity of an isolated prolate particle at $Ga=80$. This reduction is caused by the change in the wake regime as helical vortices develop in the wake of the two particles, see figure~\ref{fig:14}c).
Interestingly, the average settling speed for the two cases without tumbling,
oblate particle pairs and case $3$-$90^{\circ}$,  
increases by $49\%$ and $22\%$ respectively.

\subsubsection{Oblate particles pairs}
\begin{figure}[p]
\centering
\includegraphics[width=0.4\linewidth]{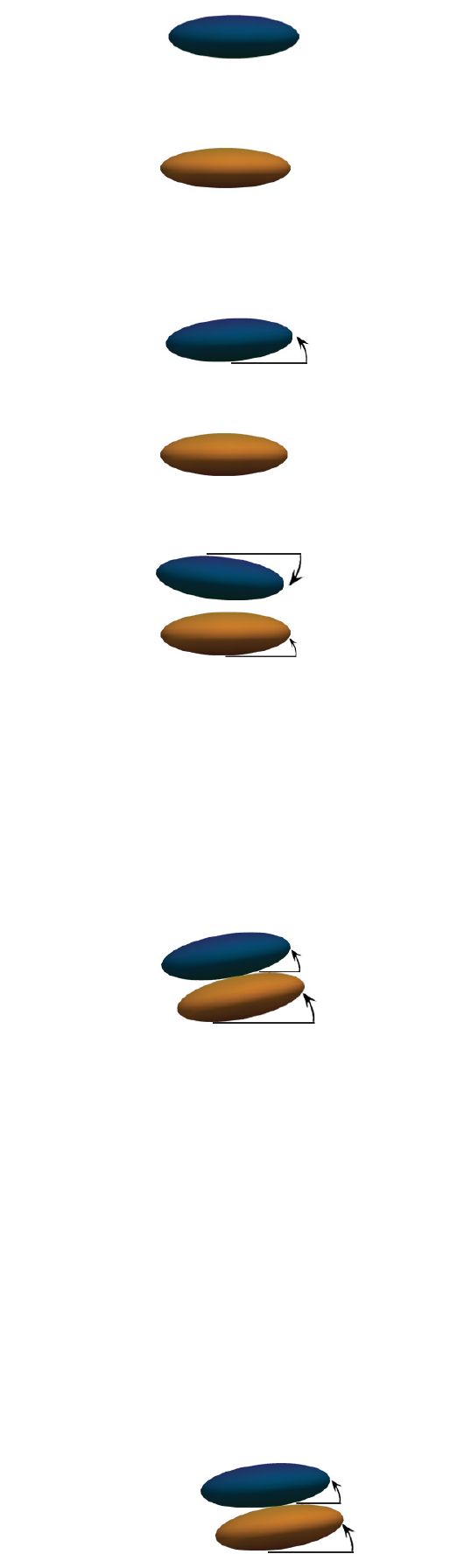}
\put(-40,467){{$\Theta = 0^{\circ}$}}
\put(-40,427){{$\Theta = 0^{\circ}$}}
\put(-40,370){{$\Theta = 3.7^{\circ}$}}
\put(-40,339){{$\Theta = 0^{\circ}$}}
\put(-40,302){{$\Theta = -5.6^{\circ}$}}
\put(-40,281){{$\Theta = 1^{\circ}$}}
\put(-40,185){{$\Theta = 9.2^{\circ}$}}
\put(-40,168){{$\Theta = 11.2^{\circ}$}}
\put(-25,21){{$\Theta = 6^{\circ}$}}
\put(-25,5){{$\Theta = 9.1^{\circ}$}}
\put(-160,15){{\Large $t^* = 54$}}
\put(-160,175){{\Large $t^* = 36$}}
\put(-160,295){{\Large $t^* = 25$}}
\put(-160,360){{\Large $t^* = 18$}}
\put(-160,445){{\Large $t^* = 0$}}
\vspace{-5pt}
\caption{\label{fig:20} 
Time sequence of the DKT process for oblate particle pairs with $\mathcal{AR}=1/3$ and $Ga=80$ at non-dimensional times $t^*=0$, $18$, $25$, $36$ and $54$. $\Theta$ denotes the angle with respect to the horizontal direction.} 
\end{figure} 

\begin{figure}[t]
\centering
\includegraphics[width=0.495\textwidth]{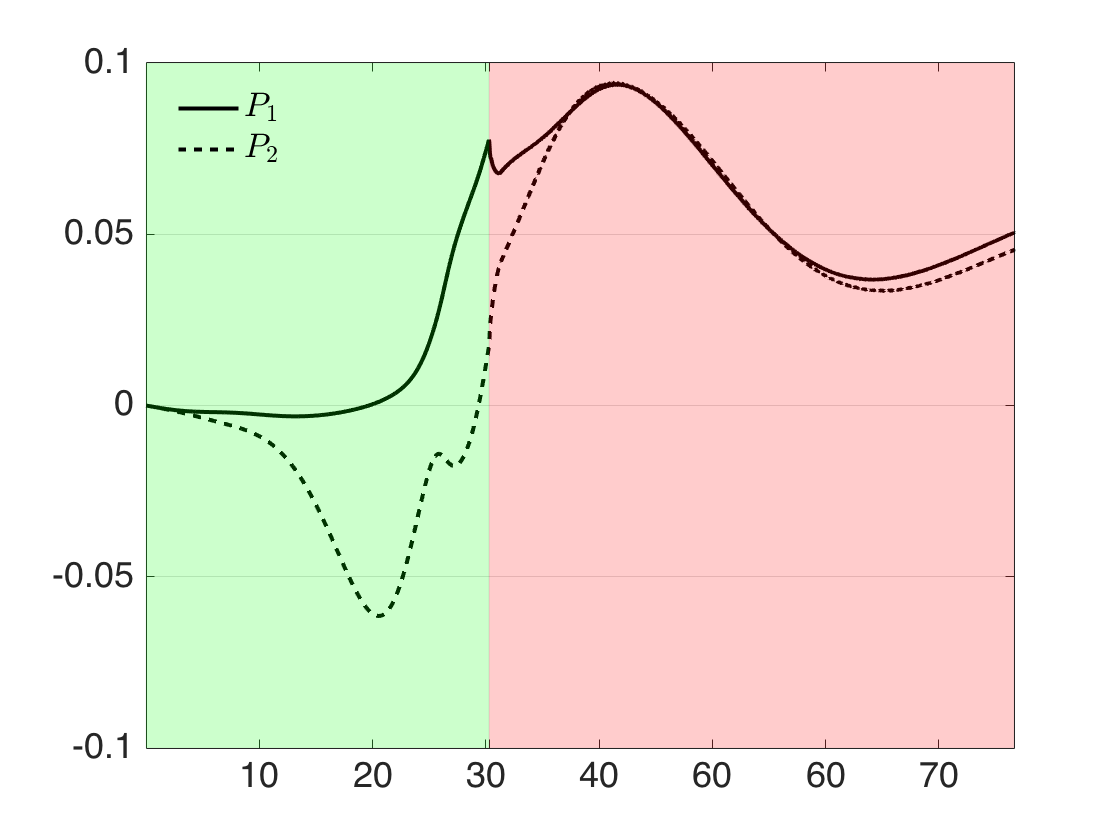} 
\includegraphics[width=0.495\textwidth]{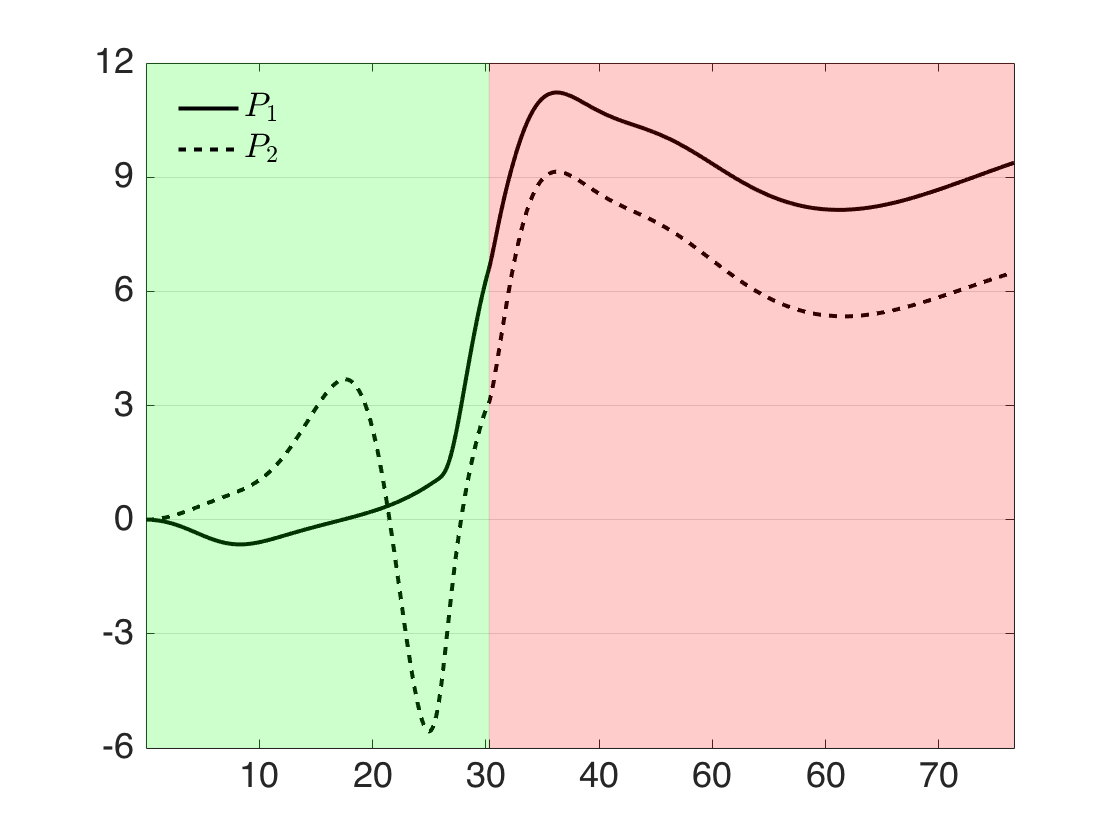}
\put(-345,52){\rotatebox{90}{$V_y/|u_{\,t}|$}}
\put(-168,58){\rotatebox{90}{$\Theta \left( {}^{\circ} \right)$}}
\put(-85,-2){{$t^*$}}
\put(-259,-2){{$t^*$}}
\put(-265,-20){\footnotesize (a)}
\put(-90,-20){\footnotesize (b)}
\caption{\label{fig:21}  a) The horizontal velocity $V_y$, normalized by the absolute value of terminal velocity of an isolated case and b) the horizontal ($y$) inclination angle of the oblate particle pairs with $\mathcal{AR}=1/3$. The drafting and kissing phase is shown by the light green and the pink background, respectively.} 
\end{figure}

We first recall that, unlike spherical particles, settling spheroids can resist horizontal motions by changing the orientation so that their broad-side becomes perpendicular to the velocity direction. They can also be re-oriented by an external torque thus drifting horizontally to balance the horizontal component of the drag force. 
Figure~\ref{fig:20} shows the DK(T, no tumbling in this case) process for oblate particles with aspect ratio $\mathcal{AR}=1/3$. 
The corresponding horizontal velocity and inclination, $\Theta$, defined in the $yz$ plane due to the symmetry of the problem and the initial offset in the $y$-direction,
are given as a function of non-dimensional time $t^*$ in figure~\ref{fig:21}a) and figure~\ref{fig:21}b). 

The trailing particle, $P_2$, initially located above and on the right hand side of $P_1$, experiences a torque originating from the drag difference on its right and left side,
which  results in a small positive $y$-inclination. 
With this orientation, $P_2$ gains horizontal velocity $V_y$ towards $P_1$. This motion forces the leading particle to drift in the same direction, to which the particle resists by tilting in the direction opposite to that of $P_2$, negative $\Theta$ in figure~\ref{fig:21}b ($t^*\approx10$). 
As a consequence, $P_2$ moves from the right  to the left side of $P_1$; at this point ($t^*\approx25$),  the particles experience the same oscillation but in the opposite direction and with higher lateral velocity and inclination due to the reduced distance between the two. 

The drafting phase, indicated by the light green background in figure~\ref{fig:21}, ends as $P_2$ finally reaches $P_1$ ($t^*\approx30$) and the particles fall in contact until they hit the bottom wall. 
The particles, once in contact, move with a vertical velocity larger than that of an isolated one ($\approx 1.5$ times), while experiencing two opposite torques that keep them attached and with a positive and nearly constant inclination. 
%

\subsubsection{Prolate particle pairs}

\begin{figure}[p]
\centering
\includegraphics[width=0.4\textwidth]{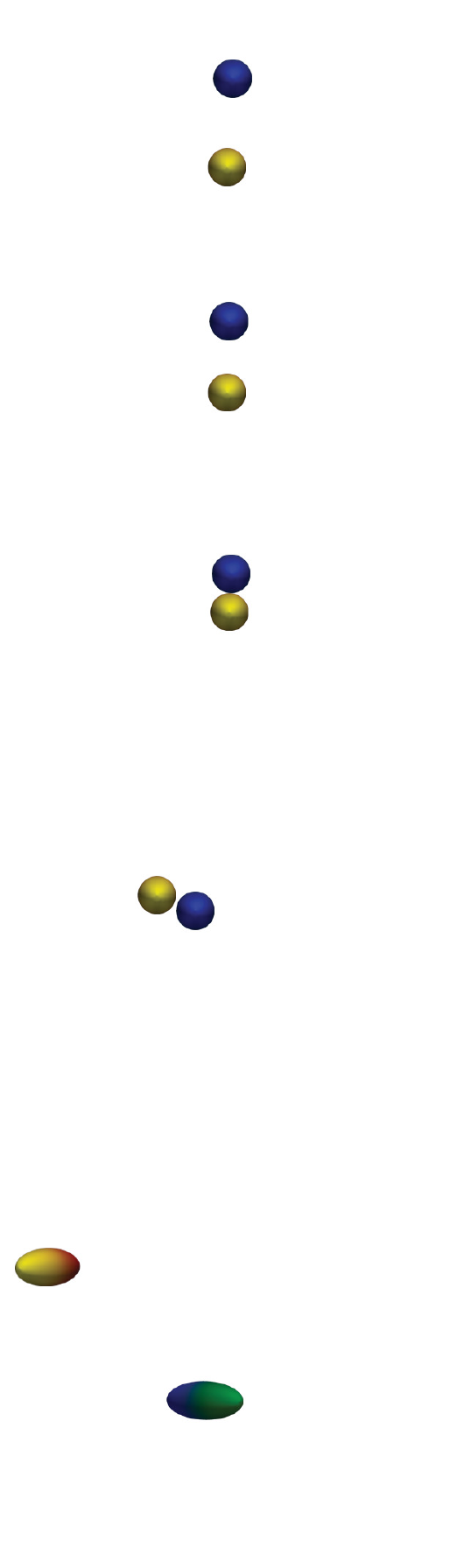} \hspace{-39pt} 
\put(-57,-5){\footnotesize (a)}
\includegraphics[width=0.4\textwidth]{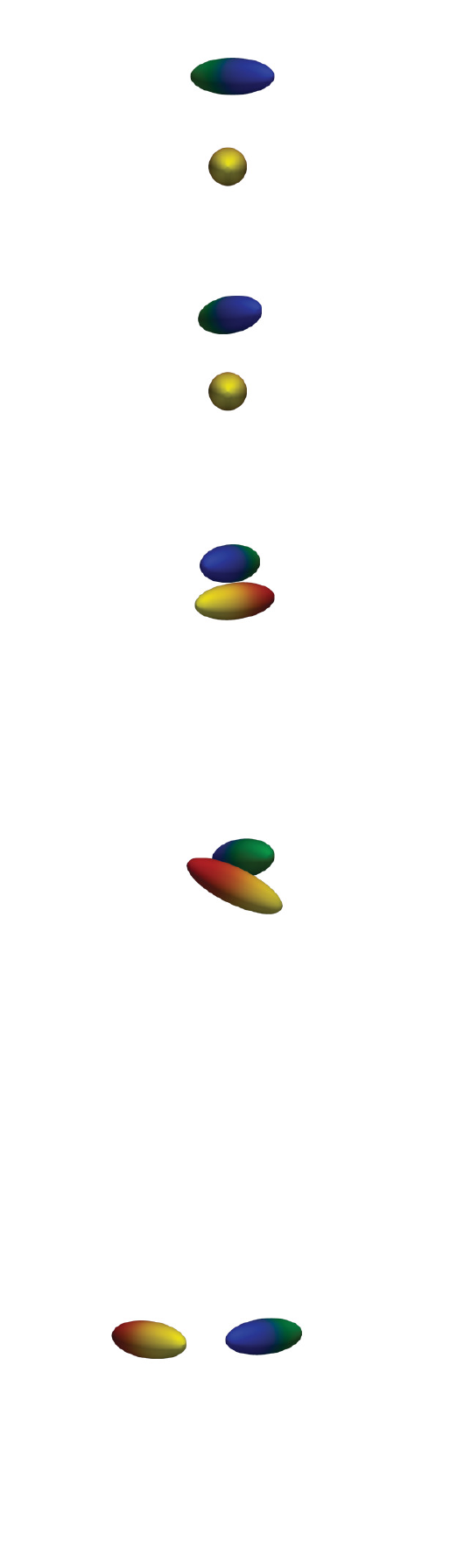} \hspace{-39pt}
\put(-38,-5){\footnotesize (b)}
\includegraphics[width=0.4\textwidth]{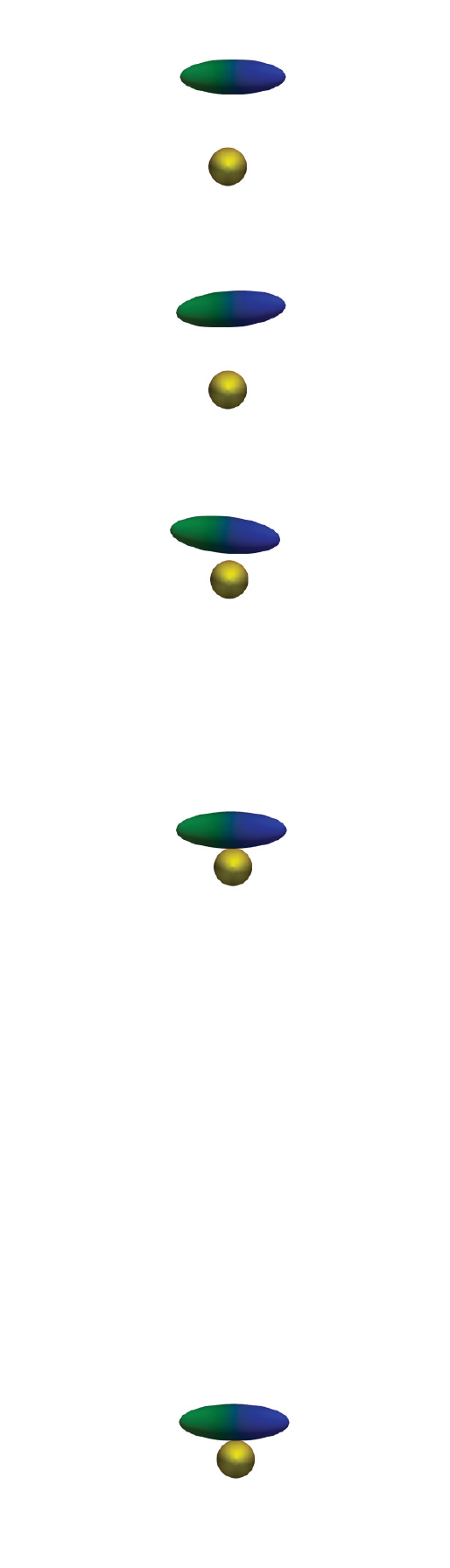} \hspace{-39pt}
\put(-36,-5){\footnotesize (c)}
\put(-330,55){{\Large $t^* = 76$}}
\put(-330,205){{\Large $t^* = 45$}}
\put(-330,295){{\Large $t^* = 30$}}
\put(-330,370){{\Large $t^* = 20$}}
\put(-330,440){{\Large $t^* = 0$}}
\caption{\label{fig:22} Sequences of a DKT process for prolate particle pairs with $\mathcal{AR}=3$ in the three cases a) $3$-$0^{\circ}$, b) $3$-$45^{\circ}$ and c) $3$-$90^{\circ}$ at non-dimesional times $t^*=0$, $20$, $30$, $45$ and $76$.} 
\end{figure}
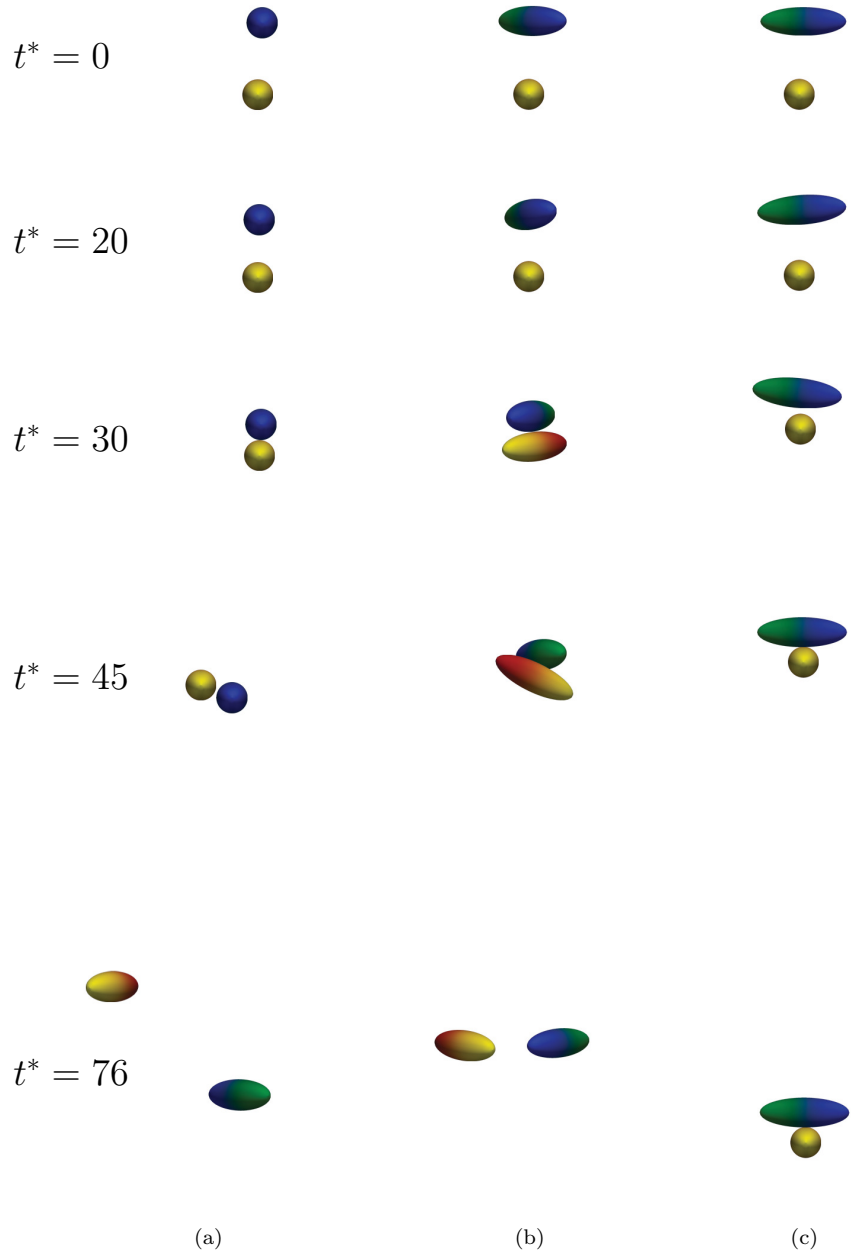

Prolate particles show different behaviours in the three studied cases; at $Ga=80$ and for $\mathcal{AR}=3$ they are in the unstable regime (see figure~\ref{fig:13}) 
where their motion and wake structure are sensitive to the interactions with other particles or ambient noise. 
The sequence of the DKT is displayed in figure~\ref{fig:22} for the three cases that  we discuss next separately. 

\begin{figure}[t]
\centering
\includegraphics[width=0.495\textwidth]{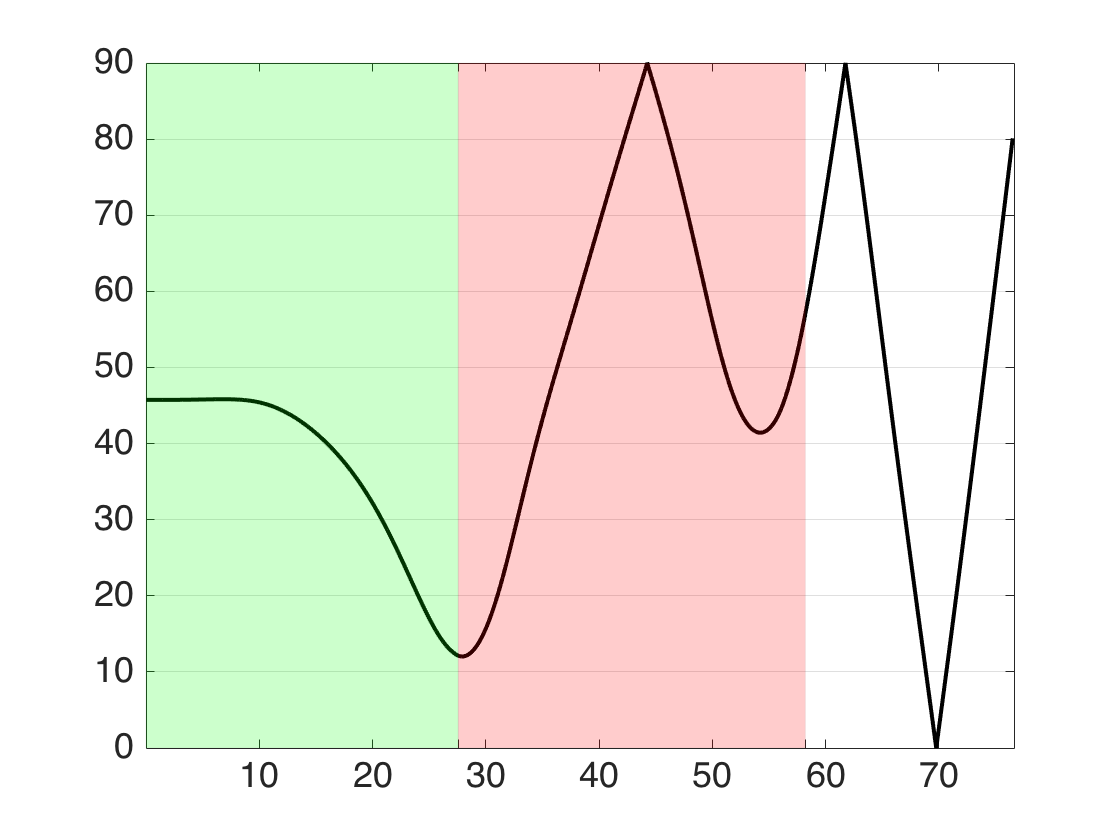} 
\includegraphics[width=0.495\textwidth]{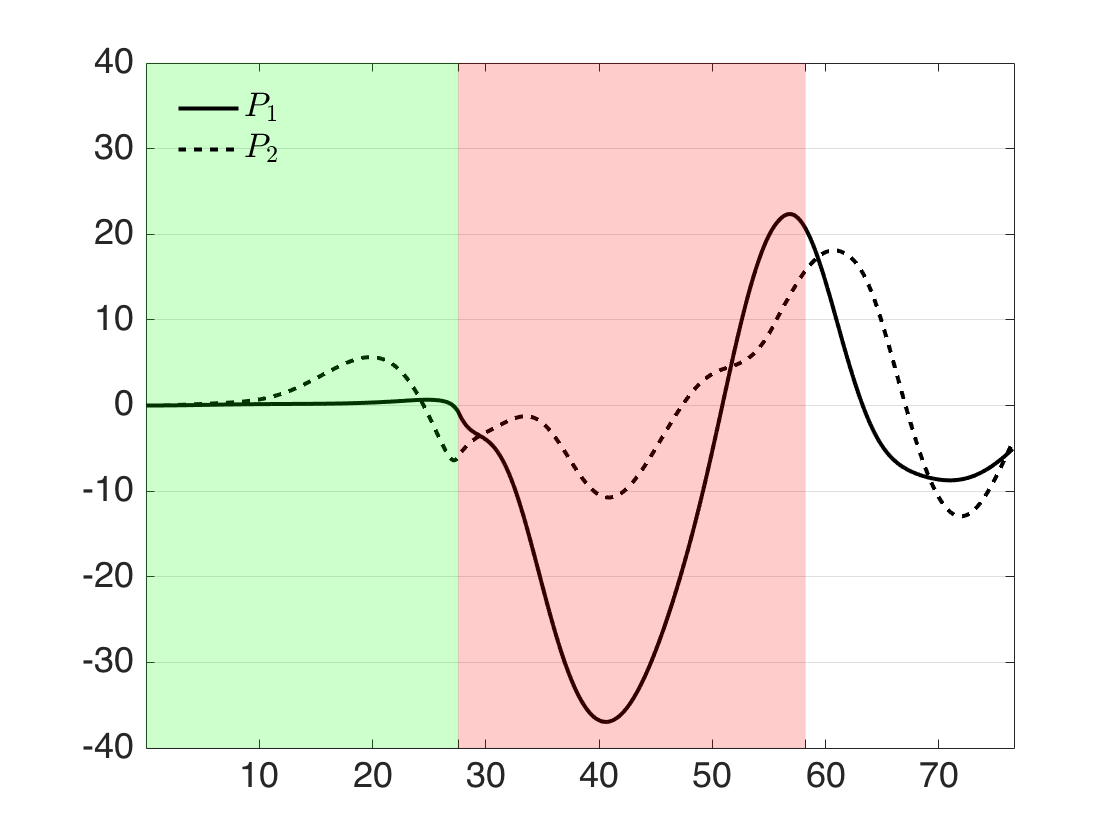}
\put(-345,53){\rotatebox{90}{$\phi_{rel}\left({}^{\circ}\right)$}}
\put(-168,58){\rotatebox{90}{$\Theta\left({}^{\circ}\right)$}}
\put(-85,-2){{$t^*$}}
\put(-259,-2){{$t^*$}}
\put(-265,-20){\footnotesize (a)}
\put(-90,-20){\footnotesize (b)}
\caption{\label{fig:23}  Time history of a) The relative angle $\phi_{rel}$ between the major axes of the prolate particles in the horizontal plane ($xy$) and b) the horizontal inclination angle of the prolate particle pair for the case $3$-$45^{\circ}$. The drafting and kissing phase is shown by the light green and the pink background, respectively.} 
\end{figure}

\paragraph{Case $3$-$0^{\circ}$ } $\,$ \\
In this case the particles are initially parallel,
figure~\ref{fig:22}a,
the DKT is  analogous to the case of spherical particles, just with a slightly longer duration of the drafting phase and an increase in the duration of the kissing phase of about 30\%. The particles start their rotation around the vertical ($z$) axis in the tumbling phase, with helical vortices appearing in their wake. These are triggered by the particle interactions at $Ga$ less than $100$, the critical value for an isolated particle. 

\paragraph{Case $3$-$45^{\circ}$ } $\,$ \\
In this case,  $P_2$ starts rotating already in  the drafting phase, thereby reducing the angle between the major axes of the two particles.
This is due to the torque that $P_2$ experiences in the low pressure regions behind the poles of $P_1$. Figure~\ref{fig:23}a reports the relative angle $\phi_{rel}$ between the major axes of the particles in the horizontal $xy$ plane:
the relative angle reduces to $12^{\circ}$ at the end of drafting phase. 
The motion of $P_2$ into the wake of $P_1$ triggers the particle rotation earlier than in case $3$-$0^{\circ}$.
The coupled rotation continues in the kissing phase, with a duration approximately twice that of two spheres (figure~\ref{fig:18}); this prevents $P_2$ from overtaking $P_1$ in the falling direction. The particles undergo a complex rotating motion, with a periodic horizontal inclination, indicated in figure~\ref{fig:23}b. 
This continues also in the tumbling phase, although with smaller values of the inclination angle, $\Theta$. 
The particle wake is characterised by helical vortices as those shown in figure~\ref{fig:14} for an isolated prolate. 
It should be noted here that the case $3$-$45^{\circ}$ can be taken as a model 
of the results pertaining 
larger initial vertical distances between the particles, when the rotating motion might have already begun in the drafting phase and the DKT becomes substantially  independent of the initial particle orientation.

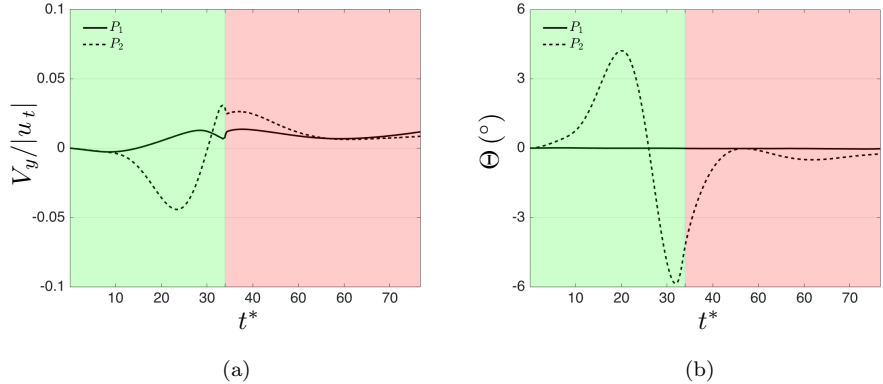
\begin{figure}[t]
\centering
\includegraphics[width=0.495\textwidth]{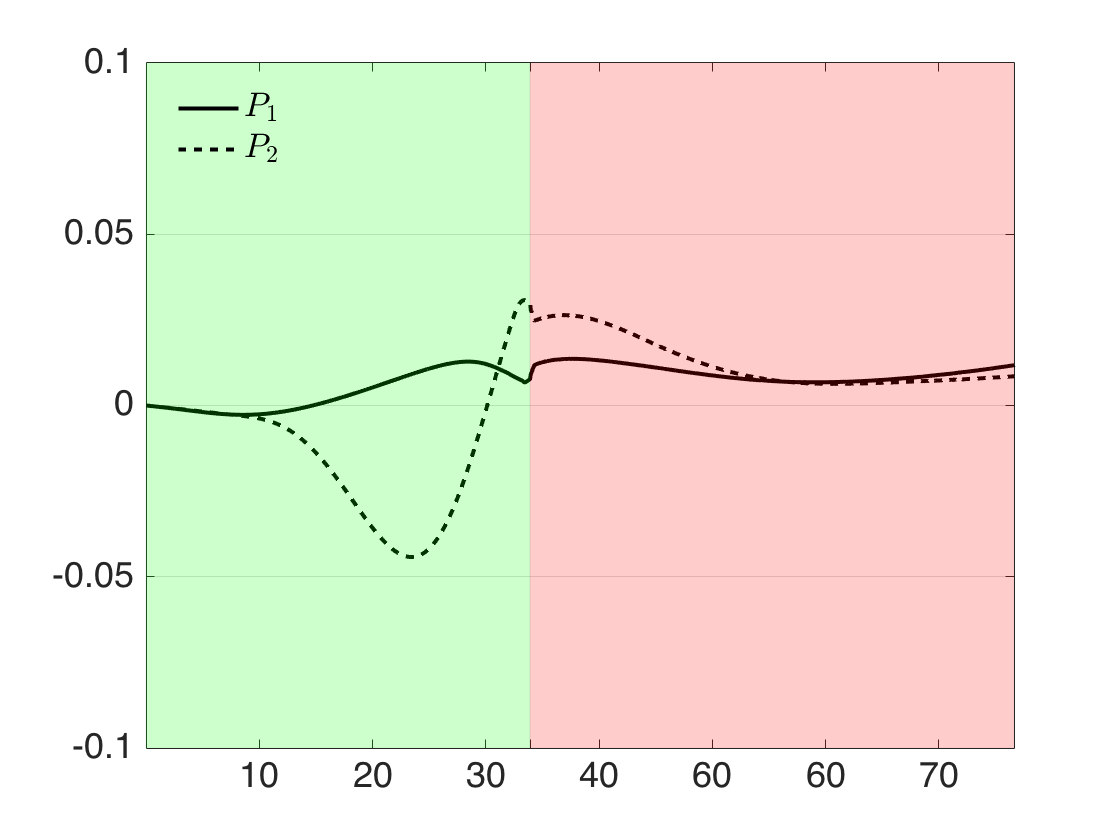} 
\includegraphics[width=0.495\textwidth]{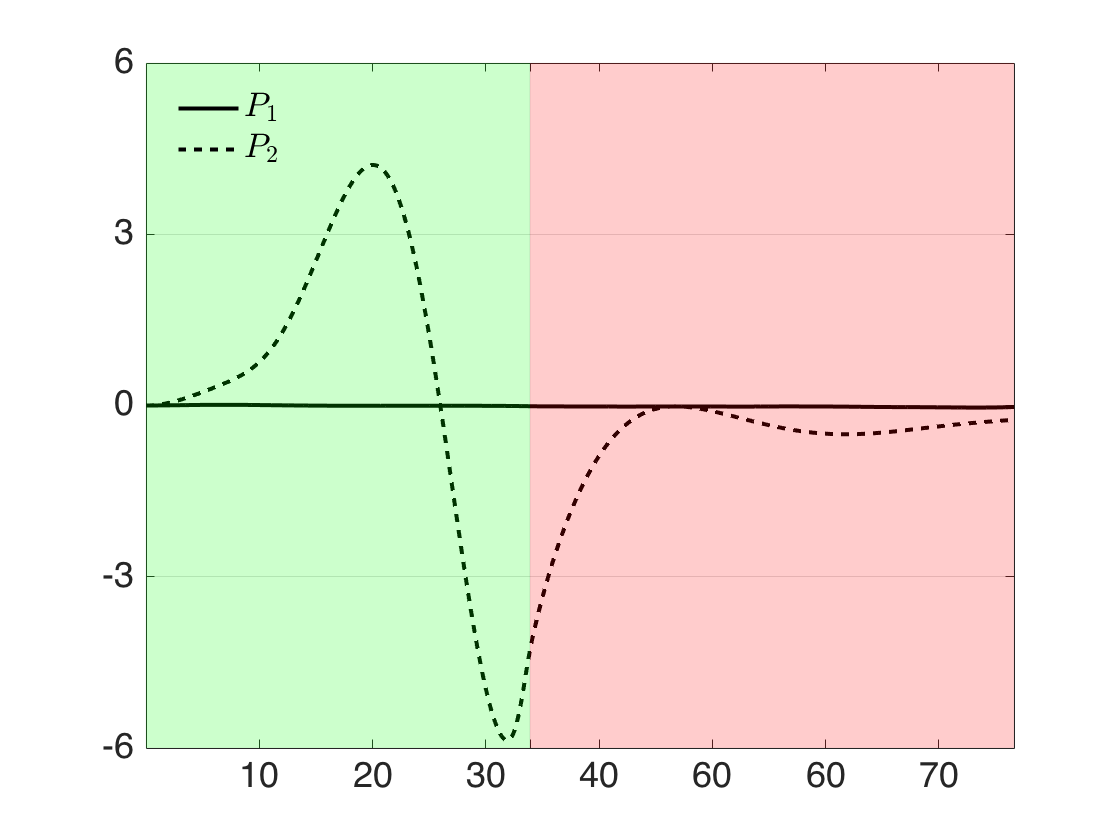}
\put(-345,52){\rotatebox{90}{$V_y/|u_{\,t}|$}}
\put(-168,58){\rotatebox{90}{$\Theta \left( {}^{\circ} \right)$}}
\put(-85,-2){{$t^*$}}
\put(-259,-2){{$t^*$}}
\put(-265,-20){\footnotesize (a)}
\put(-90,-20){\footnotesize (b)}
\caption{\label{fig:24}  a) Horizontal velocity, $V_y$ normalized by the absolute value of terminal velocity of an isolated particle, and b) the horizontal ($y$) inclination angle of the prolate particle pair for the case $3$-$90^{\circ}$. The drafting and kissing phase are shown by the light green and the pink background, respectively.} 
\end{figure}

\paragraph{Case $3$-$90^{\circ}$ } $\,$ \\
The DK(T, no tumbling in this case) is similar to the case of oblate particles  as $P_2$ experiences an inclination $\Theta$ with respect to the horizontal plane which reduces the initial horizontal offset in the drafting phase, see figure~\ref{fig:22}c.
Figure~\ref{fig:24} displays the particle horizontal velocity and inclination in time. $P_1$ does not experience any tilting 
 due to the symmetry in the $x$-direction. Thus, $P_2$ is attracted
 in the wake of $P_1$ at a lower velocity than in the case of oblate particles.
 The kissing phase continues until the particles hit the bottom wall, as for $\mathcal{AR}=1/3$ but without particle oscillation or rotation.  The vertical velocity 
 increases by approximately $22\%$ in the kissing phase, and, contrary to our expectations, the particle rotation is delayed until they are about to hit the bottom wall. 
 This observation can be explained by considering the formation of a new body, consisting of the two particles in a cross, which is more stable than an individual prolate particle.  

\subsection{The extent of collision domain for two sedimenting spheroids}           

The results above
show that in the case of non-spherical bodies, the trailing particle $P_2$ 
is attracted in the wake of the leading particle $P_1$;  the possibility to change its orientation gives $P_2$ an extra horizontal velocity.
In turn, $P_1$ also changes its orientation as $P_2$ approaches. 
%

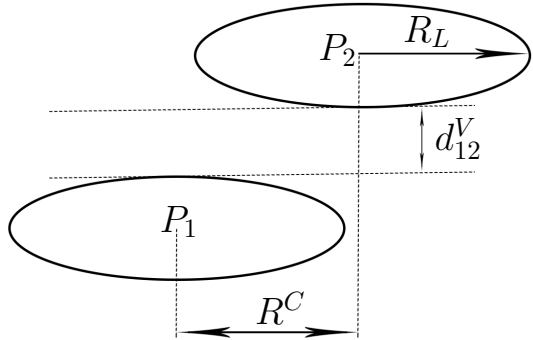
\begin{figure}[t]
\centering
\includegraphics[width=0.6\linewidth]{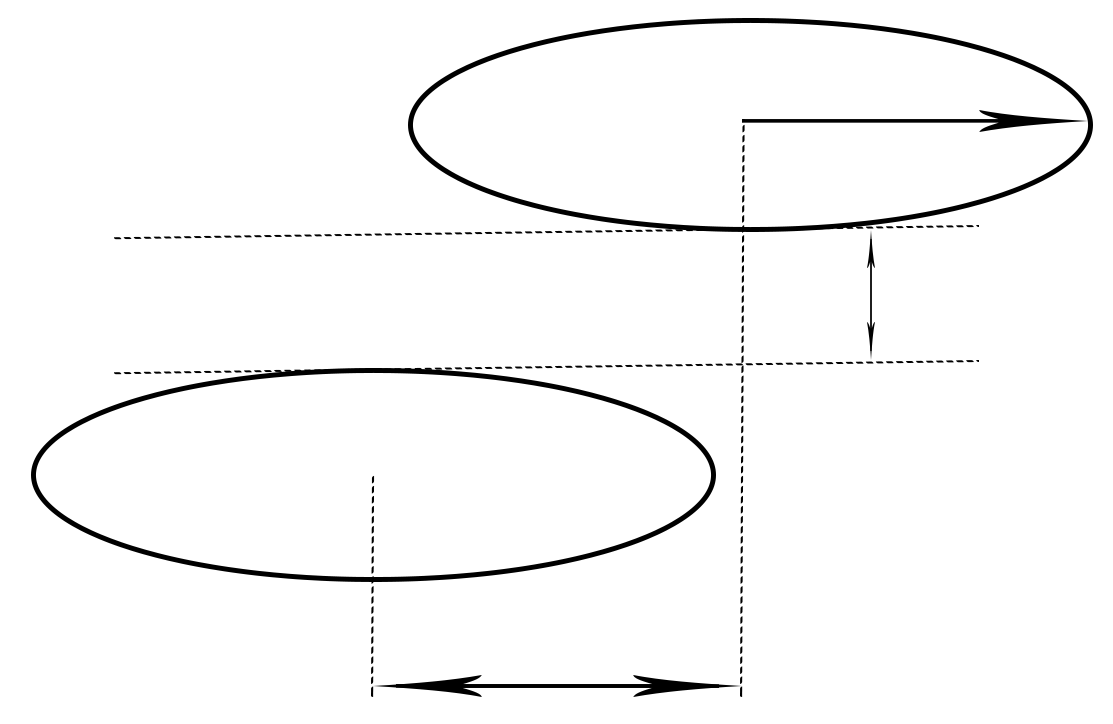}
\put(-108,10){{\Large $R^C$}}
\put(-40,76){{\Large $d^V_{12}$}}
\put(-52,117){{\Large $R_L$}}
\put(-144,45){{\Large $P_1$}}
\put(-84,110){{\Large $P_2$}}
\vspace{-5pt}
\caption{\label{fig:25} 
Schematic of initial conditions and parameters measuring the extent of collision.} 
\end{figure}

Motivated by these observations,  we speculate that the collision kernels may be significantly larger in the case of settling suspensions of non-spherical particles.
An attempt is therefore made to find the initial position from which 
two sedimenting spheroids with the same Galileo number and aspect ratios would eventually collide. 
To reduce the parameter space to be investigated by exploiting the symmetry of the problem we shall mainly focus on 
spherical and oblate particles. Indeed, in their stable configuration, these fall axisymmetrically, meaning that we can define on each horizontal plane above the leading particle $P_1$ 
 a circle with centre in $P_1$ and radius equal to the maximum distance to the centre of the trailing particle $P_2$ such that the two particles will collide. 
The collision between prolate particles, conversely, depends also on the initial relative orientation; however for sufficiently long vertical distances and sufficiently large $Ga$, the particles rotate along the vertical axis, creating an approximate symmetry in the horizontal direction. 

The extent of the collision area is computed by considering different vertical distances $d^V_{12}$ between the surfaces of the 
particles and computing the maximum horizontal distance between the particle centers for the collision to occur, $R^{C}_{max}$. Figure~\ref{fig:25} indicates the vertical distance $d^V_{12}$, the longer semi-axis of the 
spheroid $R_L$ and the collision radius $R^{C}$ whose maximum defines the collision area for each vertical distance.
For prolate particle pairs the 
collision area is found for an initial relative angle between the major axis of $45^{\circ}$  and a vertical distance $d^V_{12}= 2R_L$
such that they start rotating in the drafting phase. This is to make the outcome less dependent on the initial orientation.

\begin{figure}[t]
\centering
\includegraphics[width=0.9\linewidth]{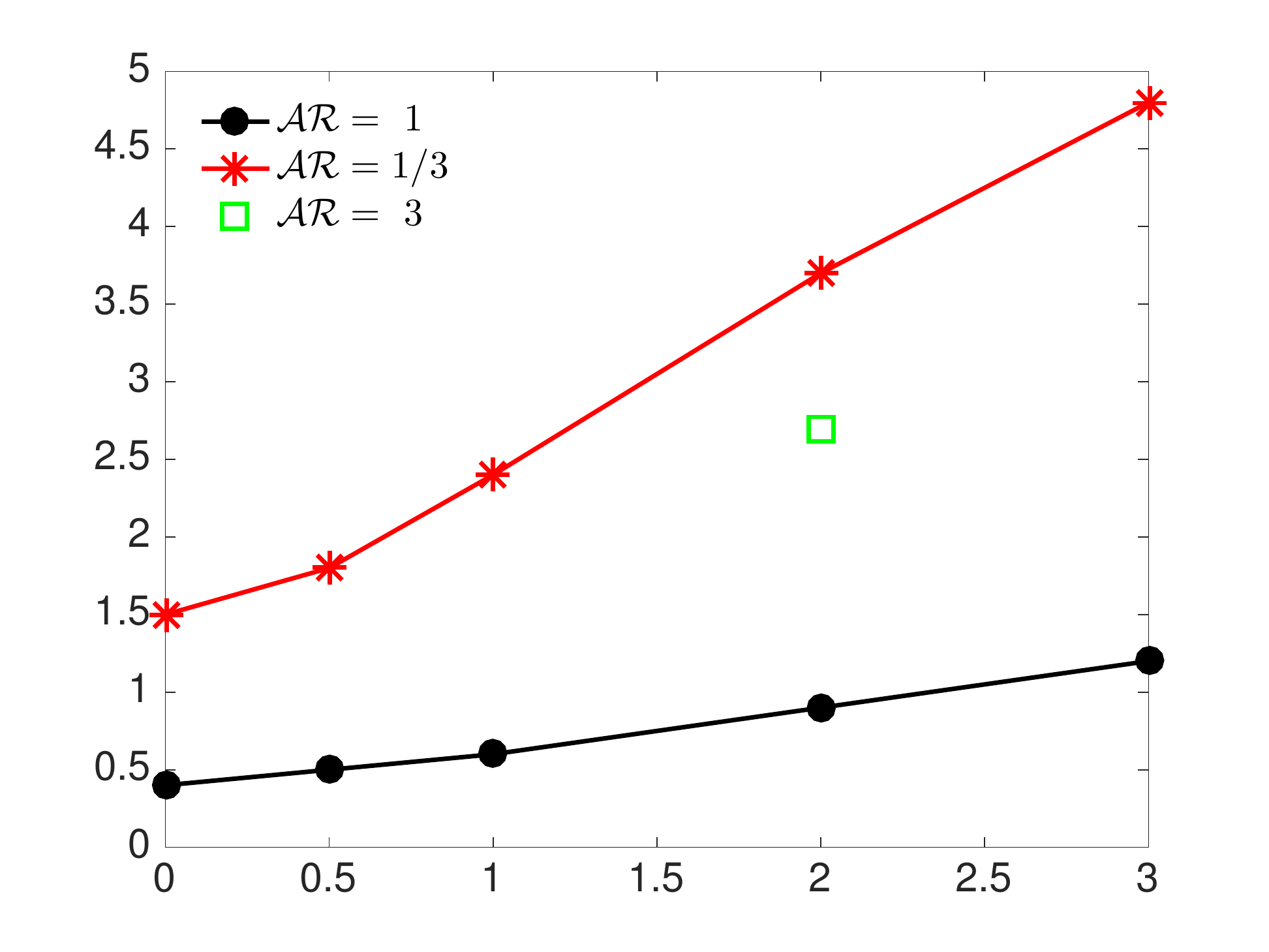}
\put(-310,80){\rotatebox{90}{ \Large $R^{C}_{max} \,\, / \,\,R_L$}}
\put(-175,-4){{\Large $d^V_{12} \,\, / \,\,R_L$}}
\vspace{5pt}
\caption{\label{fig:26} 
Maximum radius of the collision circle $R^{C}_{max}$ in the horizontal plane at different vertical distances $d^V_{12}$ between the surfaces of the two particles. The results are normalized here by the larger semi-axis of the spheroids $R_L$.} 
\end{figure}

 The results are shown in figure~\ref{fig:26}: each point in the figure is obtained with a series of simulations aiming to identify the occurrence of a collision for each initial vertical distance.
The 
collision domain is a diverging cone: the larger the initial vertical distance, the larger the horizontal distance over which the trailing particle can be attracted. Most importantly, we see that the collision area  is considerably larger (up to four times more) for oblate particles than for spherical ones. The maximum distance
for collision, $R^{C}_{max}$, is less than $1.5R_L$ for two spheres, meaning that a collision only happens if the particles overlap when projected on the horizontal plane  ($d^H_{12} < 
2R_L$); 
$R^{C}_{max}$ increases to approximately $5R_L$ for oblate particles when the vertical distance between the two particles is $3R_L$.
The data point for prolate particles reveals that $R^{C}_{max}$ is larger than for spheres and lower than for oblate particles.

\section{Final remarks} 
\label{Final remarks} 

A numerical codes is developed, based on the Immersed Boundary Method, to simulate suspensions of spheroidal particles.
The lubrication, collision and friction models used are presented here. These short-range interactions approximate the objects by two spheres with same mass and radius corresponding to the local surface curvature at the points of contact. We use  asymptotic analytical expression for the normal lubrication force  between unequal  spheres and a soft-sphere collision model with coulomb friction. The code is used to investigate the effect of particle shape on the sedimentation of isolated and particle pairs in a viscous fluid. The key observations can be summarised as follows:

\begin{itemize}
\item When examining the settling of an isolated particle, we find that the critical Galileo number $Ga_{cr}$ (based on the equivalent sphere dimeter $D_{eq}$) for the onset of secondary motions  decreases as the spheroid aspect ratio $\mathcal{AR}$ departs from 1. In particular, the critical $Ga$ decreases more for prolate particles for the same ratio between major and minor axis. 

\item For $Ga > Ga_{cr}$, oblate particles perform the so called zigzagging motion \citep{Mougin2006} whereas prolate particles 
rotate around the vertical (parallel to gravity) axis.

\item Different wake regimes are found for prolate particles with $\mathcal{AR} = 3$ (see figures~\ref{fig:13} and \ref{fig:14}) as we increase $Ga$. (i) steady axisymmetric wake ($Ga <  70$). (ii)  a rotating particle with four thread-like quasi-axial vortices in the wake ($70 < Ga < 100$). (iii) Helical vortices in the wake, associated with a reduction of the vertical velocity ($Ga > 100$).  Note that this last bifurcation is found to be sensitive to the level of ambient noise and the value of 100 is obtained with no noise and only 1 particle in the computational domain.

\item We also examine the Drafting-Kissing-Tumbling (DKT) of non-spherical particle pairs at $Ga=80$, starting with their stable orientation, i.e.\ the major axis orthogonal to gravity.
We find that  the tumbling phase disappears in the case of two oblate particles and when the prolates approach each other with their major axes almost orthogonal to each other (figures~\ref{fig:16}, \ref{fig:20} and \ref{fig:22}). 

\item In general, 
for non-spherical bodies, 
the trailing particle is more promptly attracted 
 (in terms of reducing horizontal distance between the centres) to the wake of the leading particle. 

\item We determine the volume behind the leading particle inside which the center of trailing particle should be for a collision to occur.
This collision domain is found to be considerably larger for oblate particles than for spherical particles. 
We also consider two prolates  at sufficiently long vertical distance so that they rotate in the drafting phase and the results can be seen as less dependent on the initial orientation. The distance at which collisions occur is found to be larger than for spherical particles and lower than oblate.    
\end{itemize}

The results of this study show that sedimenting spheroids are attracted towards each other from longer distances and stay in touch for considerably longer time after they collide than spheres. These two observations suggest that clustering in a suspension of sedimenting spheroids may be significantly larger than for spherical particles.
The next step would therefore be to examine collision kernels and clustering of non-spherical particles
in quiescent and turbulent environments and how the pair interactions studied here affect the global suspension behaviour.  

\section*{Acknowledgments}

This work was supported by the European Research Council Grant No. ERC-2013-CoG-616186, TRITOS. The authors acknowledge computer time provided by SNIC (Swedish National
Infrastructure for Computing) and the support from the COST Action MP1305: Flowing
matter.

 
\hypertarget{appA}{}
\appendix
\section{A simple model to predict $Re_t$ for spheroidal particles}

Here we propose a simple model to predict the terminal Reynolds number $Re_t$ for spheroidal particles at low Galileo numbers. This model assumes that for oblates, spheres and prolate particles the steady flow (wake) regime is similar. In this model the $Re$-dependent model of Abraham (1970) \cite{Abraham1970} for perfect sphere is employed to calculate the drag coefficient $C_d$. The assumption is that for sufficiently small Galileo number, the main effect of a change in spheroid aspect ratio (with respect to a perfect sphere) is the change in the frontal surface area, while $C_d$ remains the same when defining the terminal Reynolds number based on the equivalent sphere diameter:

\begin{equation}
 C_d = \left ( {\sqrt{\frac{24}{Re_t}} + 0.5407} \right)^2 
\label{eq:Cdrag}  
\end{equation}

The relation between the terminal Reynolds number $Re_t$, Galileo number $Ga$ and the aspect ratio $\mathcal{AR}$ is given below for oblate and prolate spheroids.

\paragraph{Oblate spheroids} $\,$ \\
A simple force balance, using the drag coefficient $C_d$ results in following equations for an oblate spheroid: 
\begin{equation}
\label{eq:OblateDragModel}  
\frac{1}{2} C_d \pi b^2 \rho_f u_t^2  = \frac{1}{6} (\rho_p - \rho_f) \pi D_{eq}^3 g \, ,   
\end{equation} 
where $\pi b^2$ is the projected surface area in direction of gravity when the particle falls with its stable orientation (major-axis perpendicular to the gravity direction), which can be written in term of $D_{eq}$ as
\begin{equation}
\label{eq:projectedAreaOb}  
\pi b^2 = \frac{1}{4} \pi D_{eq}^2 \mathcal{AR}^{-2/3} \, .  
\end{equation} 
 Substituting eq.~\ref{eq:projectedAreaOb} in \ref{eq:OblateDragModel} results in a relation between $Re_t$, $Ga$ and  $\mathcal{AR}$:
\begin{equation}
\label{eq:FinalDragModelOb}  
Re_t^2 + 18.12 Re_t^{1.5} + 82.09 Re_t - 4.56 Ga^2 \mathcal{AR}^{2/3} = 0  \, .  
\end{equation} 

\paragraph{Prolate spheroids} $\,$ \\ 
The same force balance holds for prolate spheroids when the frontal surface area $\pi a b$, written in terms of $D_{eq}$ as
\begin{equation}
\label{eq:projectedAreaPr}  
\pi a b = \frac{1}{4} \pi D_{eq}^2 \mathcal{AR}^{1/3} \, .  
\end{equation}
Upon substitution of eq.~\ref{eq:projectedAreaPr} in the force balance,  the final relation between $Re_t$, $Ga$ and  $\mathcal{AR}$ reads 
\begin{equation}
\label{eq:FinalDragModelPr}  
Re_t^2 + 18.12 Re_t^{1.5} + 82.09 Re_t - 4.56 Ga^2 \mathcal{AR}^{-1/3} = 0  \, .  
\end{equation}

\section*{References}

\bibliography{mybibfile}

\end{document}